\documentclass[traditabstract]{aa}

\newcommand{\lsim}{\lower 2pt \hbox{$\, \buildrel {\scriptstyle
<}\over {\scriptstyle \sim}\,$}}  \newcommand{\gsim}{\lower 2pt
\hbox{$\, \buildrel {\scriptstyle >}\over {\scriptstyle \sim}\,$}}
\usepackage{appendix}
\usepackage{amsmath}
\usepackage{graphicx}
\usepackage{txfonts}
\usepackage{natbib}
\usepackage{longtable}
\usepackage{rotating}
\usepackage{lscape}
\usepackage{float}
\usepackage{subfig}
\usepackage{caption}
\usepackage{multirow}
\usepackage{array}
\usepackage{fixltx2e}
\usepackage{hyperref}
\usepackage{breakurl}

\bibpunct{(}{)}{;}{a}{}{,} 

\newcommand{\Pal}{Pa$\alpha$ at 1.876${\mu}$m}
\newcommand{\Brdl}{Br$\delta$ at 1.945${\mu}$m}
\newcommand{\Brgl}{Br$\gamma$ at 2.166${\mu}$m}

\newcommand{\Ha}{H$\alpha$}
\newcommand{\Hb}{H$\beta$}
\newcommand{\Hg}{H$\gamma$}
\newcommand{\Pa}{Pa$\alpha$}
\newcommand{\Brd}{Br$\delta$}
\newcommand{\Brg}{Br$\gamma$}

\newcommand{\LTIR}{L$_{\rm IR}$}
\newcommand{\Lsolar}{L$_{\odot}$}
\newcommand{\Msolar}{M$_{\odot}$}
\newcommand{\Av}{A$_{\rm V}$}
\newcommand{\Reff}{R$_{\rm eff}$}
\newcommand{\reff}{r$_{\rm eff}$}
\newcommand{\rcore}{r$_{\rm core}$}
\newcommand{\Si}{$\Sigma_{\rm SFR}$}

\newcommand{\SFRHa}{SFR$\rm_{H\alpha}$}
\newcommand{\SFRPa}{SFR$\rm_{Pa\alpha}$}
\newcommand{\SFRMIR}{SFR$\rm_{24\mu m}$}
\newcommand{\SFRTIR}{SFR$\rm_{\rm IR}$}

\newcommand{\LMIR}{L$\rm_{24\mu m}$}
\newcommand{\LPa}{L$\rm_{Pa\alpha}$}
\newcommand{\LHa}{L$\rm_{H\alpha}$}

\newcommand\nodata{ ~$\cdots$~ }

\newcounter{subfig}
\newcounter{fake_fig}

\newcolumntype{x}[1]{%
>{\raggedleft\hspace{0pt}}p{#1}}%

\newcolumntype{z}[1]{%
>{\raggedright\hspace{0pt}}p{#1}}%

\begin{document}

\title{VLT-SINFONI sub-kpc study of the star formation in local LIRGs and ULIRGs}

\subtitle{Analysis of the global $\Sigma_{\rm SFR}$ structure and characterisation of individual star-forming clumps\thanks{Final data products are available in electronic form at the CDS via anonymous ftp to \url{cdsarc.u-strasbg.fr (130.79.128.5)} or via \url{http://cdsweb.u-strasbg.fr/cgi-bin/qcat?J/A+A/}}}

\author{J. Piqueras L\'opez\inst{1}
\and L. Colina\inst{1}
\and S. Arribas\inst{1}
\and M. Pereira-Santaella\inst{1}
\and A. Alonso-Herrero\inst{2}
}
\offprints{Javier Piqueras L\'opez\\ {\tt piqueraslj@cab.inta-csic.es} \smallskip}

\institute{
Centro de Astrobiolog\'ia (INTA-CSIC), Ctra de Torrej\'on a Ajalvir, km 4, 28850, Torrej\'on de Ardoz, Madrid, Spain
\and Instituto de F\'isica de Cantabria, CSIC-UC, Avenida de los Castros S/N, 39005 Santander, Spain
}

\date{  }

\abstract{We present a two-dimensional study of  star formation at kiloparsec and sub-kiloparsec scales of a sample of local ($z<0.1$) Luminous (10) and Ultraluminous (7) Infrared Galaxies (U/LIRGs), based on near-infrared VLT-SINFONI integral field spectroscopy (IFS). We obtained integrated measurements of the star formation rate and star formation rate surface density, together with their 2D distributions, based on \Brg\ and \Pa\ emission. In agreement with previous studies, we observe a tight linear correlation between the star formation rate (SFR) derived from our extinction-corrected \Pa\ measurements and that derived from \emph{Spitzer} 24\,$\mu$m data, and a reasonable agreement with SFR derived from \LTIR. We also compared our \SFRPa\ values with optical measurements from \Ha\ emission and find that the \SFRPa\  is on average a factor $\sim$3 larger than the \SFRHa, even when the extinction corrections are applied.

Within the angular resolution and sizes sampled by the SINFONI observations, we found that LIRGs have a median-observed star formation rate surface density of $\rm\Sigma_{\rm LIRGs}^{\rm obs}=1.16$\,\Msolar\,yr$^{-1}$\,kpc$^{-2}$, and $\rm\Sigma_{\rm LIRGs}^{\rm corr}=1.72$\,\Msolar\,yr$^{-1}$\,kpc$^{-2}$ for the extinction-corrected distribution. The median-observed and the extinction-corrected \Si\ values for ULIRGs  are $\rm\Sigma_{\rm ULIRGs}^{\rm obs}=0.16$\,\Msolar\,yr$^{-1}$\,kpc$^{-2}$ and $\rm\Sigma_{\rm ULIRGs}^{\rm corr}=0.23$\,\Msolar\,yr$^{-1}$\,kpc$^{-2}$, respectively. These median values for ULIRGs increase up to 1.38\,\Msolar\,yr$^{-1}$\,kpc$^{-2}$ and 2.90\,\Msolar\,yr$^{-1}$\,kpc$^{-2}$, when only their inner regions, covering the same size as the average FoV of LIRGs, are considered. For a given fixed angular sampling, our simulations show that the predicted median of the \Si\ distribution increases artificially with distance, a factor $\sim$2-3 when the original measurements for LIRGs are simulated at the average distance of our ULIRGs. This could have consequences on any estimates of the star formation surface brightness in high-z galaxies, and consequently on the derivation of the universality of star formation laws at all redshifts.

We identified a total of 95 individual star-forming clumps in our sample of U/LIRGs, with sizes that range within $\sim$60--400\,pc and $\sim$300--1500\,pc, and extinction-corrected \Pa\ luminosities of $\sim$10$^{5}$--10$^{7}$\,\Lsolar\ and $\sim$10$^{6}$--10$^{8}$\,\Lsolar\ in LIRGs and ULIRGs, respectively. The \Si\ of the clumps presents a wide range of values within 1--90\,\Msolar\,yr$^{-1}$\,kpc$^{-2}$ and 0.1--100\,\Msolar\,yr$^{-1}$\,kpc$^{-2}$ for LIRGs and ULIRGs. Star-forming clumps in LIRGs are about ten times larger and thousands of times more luminous than typical clumps in spiral galaxies, which is consistent with expected photon-bounded conditions in ionized nebulae that surround young stellar clusters.  Clumps in ULIRGs have sizes similar ($\times$0.5--1) to those of high-z clumps, having \Pa\ luminosities similar to some high-z clumps, and  about 10 times less luminous than the most luminous high-z clumps identified so far. This could be an indication that the most luminous giant clumps in high-z star-forming galaxies are forming stars with a higher surface density rate than low-z compact ULIRGs. We also observed a change in the slope of the L-r relation, from $\rm\eta=3.04$ of local samples to $\rm\eta=1.88$ from high-z observations. A likely explanation is that most luminous galaxies are interacting and merging, and therefore their size represents a combination of the distribution of the star-forming clumps within each galaxy in the system plus the additional effect of the projected distance between the galaxies. As a consequence, this produces an overall size that is larger than that of individual clumps, or galaxies (for integrated measurements)}

\keywords{Galaxies:general - Galaxies:evolution - Galaxies: structure - Galaxies:star formation - Infrared:galaxies - Infrared: ISM - ISM: HII regions}

\authorrunning{J. Piqueras L\'opez et al.:}
\titlerunning{VLT-SINFONI sub-kpc study of  star formation in local LIRGs and ULIRGs}
\maketitle 

\section{Introduction}

Star formation (SF) processes play a central role in understanding how galaxies evolve and build up their mass. Over the last decade, owing to the advent of new and powerful facilities like the \emph{Hubble Space Telescope} (HST), \emph{Spitzer} or \emph{Herschel}, together with new observational information from ground-based optical, infrared (IR) and radiotelescopes, it has been possible to obtain new insights into the physical mechanisms that govern the formation of stars at all scales. Those processes spawns from the pc scales of individual molecular clouds, kpc scales of galactic disks to the Mpc scales of gas flows onto disks or satellite objects from the intergalactic medium (see \citealt{KennicuttJr:2012ey} for a review). However, the hierarchy and coupling of the different mechanisms at different scales are not yet well understood.

Being able to characterise the SF processes in galaxies is crucial to understanding the so-called main sequence of star-forming galaxies, i.e. the tight correlation observed with the specific star-formation rate (sSFR, star formation rate per unit of stellar mass, M$\star$) with the stellar mass (see, e.g. \citealt{Daddi:2007ApJ670}, \citealt{Elbaz:2007jn}, \citeyear{Elbaz:2011ix}, \citealt{Rodighiero:2011jx}, \citealt{Wuyts:2011ApJ738}). The slope and normalisation of the sSFR-$\rm M\star$ relation play a central role in the growth of galaxies and in the evolution of their mass function \citep{Rodighiero:2011jx}. In addition, spatially resolved studies of the formation of stars are essential when answering whether the empirical Kennicutt-Schmidt law (KS, \citealt{Schmidt:1959bp,Schmidt:1963ApJ137}, \citealt{Kennicutt:1998p36}) is governed by local processes or by the global dynamics of the systems.

In this context of the cosmological evolution of the SF, the importance of luminous (LIRGs, $10^{11}$\Lsolar$<$\LTIR$< 10^{12}$\Lsolar) and ultraluminous (ULIRGs, $10^{12}$\Lsolar$<$\LTIR$< 10^{13}$\Lsolar) infrared galaxies has been well established in the last few decades. U/LIRGs start to play a significant role in the SF cosmic history of the Universe beyond $\rm z\sim1$ (see, e.g. \citealt{PerezGonzalez:2005ApJ630}). Previous studies based on \emph{Spitzer} and \emph{Herschel} concluded that the relative contribution of ULIRGs to the star formation rate (SFR) density of the Universe increases with redshift, and may even be the dominant component at $\rm z\geq2$ (\citealt{PerezGonzalez:2005ApJ630}, \citealt{Magnelli:2011eb}, \citeyear{Magnelli:2013ib}). These IR studies of large samples of bright IR galaxies also reveal that the dominant contribution to the SFR density is in the form of obscured SF, and that its contribution increases with redshift (see, e.g. \citealt{Magnelli:2013ib}).

The dominant power source of LIRGs and ULIRGs is likely to be extended SF activity in the low-luminosity objects, whereas the contribution of AGN increases with bolometric luminosity (\citealt{Farrah:2007ApJ667}, \citealt{Nardini:2008MNRAS385}, \citeyear{Nardini:2010p405}, \citealt{Yuan:2010eo}, \citealt{PereiraSantaella:2011fw}, \citealt{2011ApJ...730...28P}, \citealt{Alonso-Herrero:2012p744}). In particular, local U/LIRGs, although rare in the local Universe (see, e.g. \citealt{LeFloch:2005ApJ632}), are valuable candidates for studying extreme cases of compact SF, coeval AGN, and their attendant feedback processes in a great amount of detail, given their proximity.

The study of local LIRGs and ULIRGs using near-IR integral field spectroscopic techniques presents many advantages. In particular, it is possible to disentangle the 2D distribution of the SF using high spatial resolution, and characterise spatially-resolved individual star-forming regions. The use of near-IR wavelengths also enables us to analyse those dust-enshrouded SF regions that are partially or completely obscured in the optical. 

In this work, we present a detailed 2D study of the extinction-corrected, global star formation rate (SFR) and the sub-kpc structure of the star formation rate surface density (\Si) in a local sample of LIRGs and ULIRGs, and briefly discuss the effect of the spatial sampling on the \Si\ measurements. Besides the spaxel-by-spaxel approach, we also analyse in detail the properties of individual star-forming clumps, in terms of their size and \Si, and compare the results with local and high-z clumps from other samples.

The paper is organised as follows. In Sections \ref{section:sample} and \ref{section:observations}, we briefly describe the sample, observations, and data-reduction process, which are detailed in Paper I. Section~\ref{section:analysis_sfr} contains a summary of the SFR tracers and calibrations used and details the procedures for obtaining the \Si\ maps and characterising the star-forming clumps. The results and analysis of the \Si\ maps and distributions, as well as the individual regions, are presented in Sect.~\ref{section:results_sfr}, whereas Sect.~\ref{section:summary_sfr} includes a brief summary of the main results. Throughout this work we consider H$_{0}=$70\,km\,s$^{-1}$\,Mpc$^{-1}$, $\rm\Omega_{\rm \Lambda}$ = 0.70, $\rm\Omega_{\rm M}$ = 0.30.

\section{The sample}
\label{section:sample}

\begin{table}[t]
\caption{The SINFONI sample}
\tiny
\centering
{\setlength{\tabcolsep}{1.8pt}
\begin{tabular}{cccccc}
\hline
\hline
     ID1      &   ID2   & z &   D$_{\rm L}$       &     Scale    & log (\LTIR/\Lsolar) \\
Common  &   IRAS  &    &(Mpc)    &(pc/") &  \\
     (1)       &   (2)     &    (3)   &    (4)      &      (5)     &        (6)      \\
\hline
\object{IRAS 06206-6315} & \object{IRAS 06206-6315} & 0.092441& 425 & 1726 & 12.31\\
\object{NGC 2369} & \object{IRAS 07160-6215} & 0.010807& 48.6 & 230 & 11.17 \\
\object{NGC 3110} & \object{IRAS 10015-0614} & 0.016858& 78.4 & 367 & 11.34 \\
\object{NGC 3256} & \object{IRAS 10257-4338} & 0.009354 & 44.6 & 212 & 11.74 \\
\object{ESO 320-G030} & \object{IRAS 11506-3851} & 0.010781& 51.1 & 242 & 11.35\\
\object{IRAS 12112+030}5 & \object{IRAS 12112+0305} & 0.073317& 337 & 1416 & 12.38 \\
\object{IRASF 12115-4656} & \object{IRAS 12115-4657} & 0.018489 & 84.4 & 394 & 11.10 \\
\object{NGC 5135} & \object{IRAS 13229-2934} & 0.013693& 63.5 & 299 & 11.33 \\
\object{IRAS 14348-1447} & \object{IRAS 14348-1447} & 0.083000& 382 & 1575 & 12.41 \\
\object{IRASF 17138-1017} & \object{IRAS 17138-1017} & 0.017335& 75.3 & 353 & 11.42 \\
\object{IRAS 17208-0014} & \object{IRAS 17208-0014} & 0.042810& 189 & 844 & 12.43 \\
\object{IC 4687} & \object{IRAS 18093-5744} & 0.017345 & 75.1 & 352 & 11.44 \\
\object{IRAS 21130-4446} & \object{IRAS 21130-4446} & 0.092554& 421 & 1712 & 12.22 \\
\object{NGC 7130} & \object{IRAS 21453-3511} & 0.016151 & 66.3 & 312 & 11.34 \\
\object{IC 5179} & \object{IRAS 22132-3705} & 0.011415& 45.6 & 216 & 11.12 \\
\object{IRAS 22491-1808} & \object{IRAS 22491-1808} & 0.077760& 347 & 1453 & 12.23 \\
\object{IRAS 23128-5919} & \object{IRAS 23128-5919}& 0.044601& 195 & 869 & 12.04 \\
\hline
\hline
\end{tabular}}
\tablefoot{Col. (3): redshift from the NASA Extragalactic Database (NED). Cols. (4) and (5): Luminosity distance and scale from Ned Wright's Cosmology Calculator \citep{Wright:2006p4236} given h$_{0}$ = 0.70, $\Omega_{\rm M}$ = 0.7, $\Omega_{\rm M}$ = 0.3. Col. (6): \LTIR (8--1000$\mu$m) calculated from the IRAS flux densities $f_{12}$, $f_{25}$, $f_{60}$, and $f_{100}$ \citep{Sanders:2003p1433}, using the expression given in \cite{Sanders:1996p845}. A complete version of this table can be found in \cite{Piqueras2012A&A546A}}
\label{table:sample}
\end{table}

The sample is a subset of a larger sample ($\sim$70 sources) of local LIRGs and ULIRGs, as described in \cite{Arribas:2008p4403}, that covers the whole range of LIRGs and ULIRGs luminosities and their different morphological classes.

The SINFONI sample comprises a representative set of 17 sources that are divided in two subsamples of 10 LIRGs and 7 ULIRGs, and that cover the luminosity range of log(\LTIR/\Lsolar)$=11.10-12.43$ (Table~\ref{table:sample}). The mean redshifts of the LIRGs and ULIRGs subsets are $z_{\rm LIRGs}=0.014$ and $z_{\rm ULIRGs}=0.072$ and their mean luminosities are log(\LTIR/\Lsolar)$=11.33$ and log(\LTIR/\Lsolar)$=12.29$, respectively. For a detailed description of the sample, see \cite{Piqueras2012A&A546A} (Paper~I).

\section{Observations and data reduction}
\label{section:observations}

The seeing-limited observations of the sample were made between April 2006 and July 2008 during the periods 77B, 78B, and 81B with SINFONI \citep{Eisenhauer:2003p8484} on the VLT. We made use of the K-band (1.95--2.45\,$\mu$m) 0\farcs125$\times$0\farcs250\,pixel$^{-1}$ configuration of the instrument, which yields a FoV of 8"x8" by a two-dimensional 64x64 spaxel frame for each single exposition. Owing to the jittering process and the different pointings used in some of the sources, the effective FoV of the observations extends beyond that from $\sim$9"$\times$9" to $\sim$12"$\times$12". In terms of physical scales, the FoV typically covers  a size of $\sim3\times3$\,kpc for the LIRGs and $\sim12\times12$\,kpc for the ULIRGs subsample, with a resolution of $\sim$0.63\,arcsec full-width-at-half-maximum (FWHM) that corresponds, on average, to $\sim$0.2\,kpc and $\sim$0.9\,kpc for LIRGs and ULIRGs, respectively. The spectral resolution of the K-band observation is typically R$\sim$3300, and the average  FWHM measured from the OH sky lines is $6.4\pm0.6$\,\AA\ with a dispersion of 2.45\,\AA\ per pixel. For further details on the observations, the criteria used to select the different pointings, and integration times, see Paper I.

For the reduction of the data, we used the ESO pipeline ESOREX (version 3.8.3) and our own IDL routines for the flux calibration. On the individual
frames we applied the usual calibration corrections of dark subtraction, flat fielding, detector linearity, geometrical distortion, wavelength calibration, and subtraction of the sky emission. Afterwards, the individual cubes from each exposures were combined into a single data cube or into a final mosaic for those sources with several pointings.

We implemented some modifications in our reduction and calibration process that differ from Paper I. Besides using a newer version of ESOREX, we included additional steps in the process to reduce the spectral and spatial noise of the individual observations, and to match the background of the individual cubes before coadding the final data cube. For the calibration of the individual cubes, we also reduced the aperture used to measure the integrated flux of the standard star, from 5$\sigma$ to 3$\sigma$ of the best 2D Gaussian fit of its collapsed image. Although the flux measurements could be slightly affected, this change enabled us to reduce significantly the noise in the atmospheric transmission curves  in some  cases. As a consequence of the new calibration, we observed some differences in the relative calibration between H- and K-band data cubes, compared with previous published measurements. In the worst case scenario, these differences could be as high as 20\% on each band, whereas the uncertainties of the absolute flux calibration is, on average, below $\leq15$\% in both bands. Throughout this work, we have assumed a conservative 15\% systematic error for all flux measurements, added in quadrature to the statistical error obtained from the fitting.

As described in Paper I, we fitted a single Gaussian profile to the emission lines to obtain the maps of the \Pa, \Brg, and \Brd\ lines. To maximise the S/N ratio over the entire FoV, the data were binned following a Voronoi tessellation, using the IDL routines described in \cite{Cappellari:2003p4908}. The average S/N of the maps varies from object to object and from line to line, and is typically between 15-25 for the brightest line (\Brg\ and \Pa\ in LIRGs and ULIRGs, respectively), and between 8-10 for the weakest (\Brd\ and \Brg\ for LIRGs and ULIRGs, respectively). The individual values for the S/N thresholds used for binning the maps can be found in Paper~I.

\section{Data analysis}
\label{section:analysis_sfr}

\subsection{Star formation measurements. Optical and infrared tracers}
\label{section:sfr_tracers}
One of the key issues regarding the measurement of the SFR is the calibration of the SFR indicators. The UV/optical/near-IR range indicators probe the SF by directly    measuring  the stellar light. The young and most massive stars produce considerable amount of ionising photons that ionise the surrounding gas which, owing to recombination processes, creates line emission cascades including the Balmer, Paschen, and Brackett series. The conversion from the flux of light into SFR is performed under the assumption of a particular stellar IMF, which has to be fully sampled (i.e. stars are formed in every mass bin), and the SF has to be roughly constant over the time scale probed by the specific emission used. In the present work, we focus on the hydrogen recombination tracers that relate the intensity of a particular emission line with the SFR through the ionising photon rate. In particular, for the \Ha\ line, we have the well-known calibration from \cite{Kennicutt:1998p36}:

\begin{align}
\label{eq:kennicuttHa}
 \rm SFR\,[M_{\odot} yr^{-1}] &= 7.9 \times 10^{-42} \times \rm L_{H\alpha}\,[erg\,s^{-1}],
\end{align}

\noindent which assumes a Salpeter IMF from 0.1 to 100\,\Msolar\ \citep{Salpeter:1955ApJ121} and solar abundances, and the star formation has to remain constant over $\sim6$\,Myr for the expression to be applicable. The variations of the calibration constant are $\sim15$\% due to variations of the electron temperature within $\rm T_e=5000-20000$\,K and almost negligible for electron density variations within the range $\rm n_e=10^2-10^6$\,cm$^{-3}$ \citep{Osterbrock:2006AGN2}.

It is well known that this expression yields higher values of  SFR than those based on other IMF like Kroupa \citep{Kroupa:2001MNRAS322} or Chabrier \citep{Chabrier:2003PASP115}. The conversion factor to transform the SFR calculated using these IMFs with respect to those considering Salpeter are 1.44 and 1.59 respectively (\citealt{Kennicutt:2009ApJ703}, \citealt{Calzetti:2007ApJ666}). From this expression, taking the recombination factors H$\alpha$ to \Pa\ and \Brg\ (T$=10,000$\,K and $\rm n_{\rm e}=10^4\,cm^{-3}$, case B; \citealt{Osterbrock:2006AGN2}) into account, we can directly obtain equivalent relations in terms of the \Brg\ and \Pa\ luminosities:

\begin{align}
\label{eq:kennicuttPa}
   \rm SFR\,[M_{\odot} yr^{-1}] &=  8.2 \times 10^{-40} \times \rm L_{Br\gamma}\,[erg\,s^{-1}]\\
\label{eq:kennicuttBrg}
   \rm SFR\,[M_{\odot} yr^{-1}] &=  6.8 \times 10^{-41} \times \rm L_{Pa\alpha}\,[erg\,s^{-1}]. 
\end{align}

Although these near-IR lines have the advantage of being less affected by dust attenuation, they are also progressively fainter and more sensitive to the physical conditions of the gas (i.e. density and temperature, \citealt{Calzetti:2012ux}). In particular, \Brg\ is $\sim$100 times fainter than \Ha, on average, and its luminosity changes to $\sim35$\% within $\rm T_e=5000-20000$\,K, whereas \Pa\ is $\sim7.5$ times fainter than \Ha\ and its luminosity varies $\sim25$\% within the same temperature range. The dependence on the density is, in both cases, less than a $\sim3$\%  \citep{Osterbrock:2006AGN2}.

Although hydrogen recombination lines and UV emission represent the most traditional SFR tracers \citep{Kennicutt:1998p36}, indicators based on dust-reprocessed stellar light have been widely used since the advent of highly-sensitive IR space telescopes like \emph{IRAS}, \emph{Spitzer} and \emph{Herschel}. Among all the available IR-continuum tracers, we focused on the \LTIR\ and monochromatic 24\,$\mu$m indicators to compare them with our SFR measurements from the near-IR hydrogen lines. In particular, we used the total-IR-based (TIR, \LTIR[8--1000$\mu$m]) SFR calibration derived by \cite{Kennicutt:1998p36}:

\begin{align}
\label{eq:kennicuttTIR}
 \rm SFR\,[M_{\odot} yr^{-1}] &= 4.5 \times 10^{-44} \times \rm L_{IR}\,[erg\,s^{-1}],
\end{align}

\noindent which assumes a continuous burst of star formation of age 10-100\,Myr, with a completely dust-enshrouded stellar population and dust heating that is fully dominated by young stars. The main disadvantage of this indicator is the need to obtain multiple measurements along the IR spectral energy distribution (SED) and/or perform extrapolations.

In this sense, the monochromatic SFR indicators, like the 24$\mu$m continuum, have the advantage of only requiring a single measurement. This tracer, as all the indicators in the mid-IR range, is based on the continuum emission dominated by a warm ($\rm T\ge50$\,K) dust component in thermal equilibrium and small-grain dust heated by the absorption of individual starlight photons \citep{Draine:2003di}. Among the multiple calibrations available in the literature, we used the calibration from \cite{Rieke:2009ApJ692}:
 
 \begin{align}
\label{eq:AAH24}
 \rm SFR\,[M_{\odot} yr^{-1}] &= 1.2\times 10^{-9} \rm L_{\rm 24\mu m}(7.76\times 10^{-11} \rm L_{\rm 24\mu m})^{0.048},
 \end{align}

\noindent where L$_{\rm 24\mu m}$ is expressed in [L$_{\odot}$]. Unlike the previous optical and TIR tracers, several calibrators of the SFR use a non-linear relation between the luminosity at 24\,$\mu$m and the SFR. These models predict that the 24\,$\mu$m luminosity increases proportionally faster than the SFR, owing to the increasing mean dust temperature. For a detailed description of the multiple 24$\mu$m calibrations (linear and non-linear), see \cite{Calzetti:2010ka}, and for a general review, see \cite{KennicuttJr:2012ey}.

\begin{table*}[h]
\caption{Integrated star-forming properties of the sample}
\centering
{\small
{\setlength{\tabcolsep}{3pt}
\begin{tabular}{ccx{0.8cm}@{ $\pm$ }z{0.8cm}x{0.8cm}@{ $\pm$ }z{0.8cm}x{0.8cm}@{ $\pm$ }z{0.8cm}x{0.7cm}@{ $\pm$ }z{0.7cm}x{0.7cm}@{ $\pm$ }z{0.7cm}x{0.7cm}@{ $\pm$ }l}

\hline
\hline
\noalign{\smallskip}
  Object  & $\rm R_{eff}$ &  \multicolumn{2}{c}{$\rm L_{Pa\alpha}^{obs}$}& \multicolumn{2}{c}{$\rm L_{Pa\alpha}^{corr}$}& \multicolumn{2}{c}{$\rm L_{24\mu m}$}& \multicolumn{2}{c}{$\rm \Sigma_{SFR}^{obs}$}& \multicolumn{2}{c}{$\rm \Sigma_{SFR}^{corr}$}&  \multicolumn{2}{c}{$\rm A_{V}$} \\
\noalign{\smallskip}
(1) & (2) &  \multicolumn{2}{c}{(3)}& \multicolumn{2}{c}{(4)}& \multicolumn{2}{c}{(5)}& \multicolumn{2}{c}{(6)}& \multicolumn{2}{c}{(7)}&  \multicolumn{2}{c}{(8)}\\
   \noalign{\smallskip}
\hline
IRAS06206-6315   &  2.5 &     4.1 &     0.6 &    11.0 &    1.7 &   \multicolumn{2}{c}{\nodata} & 0.53 &    0.08 &    1.42 &    0.22 &  7.4 & 2.2 \\
NGC2369                &  0.5 &    0.73 & 0.11 &     4.9 &     0.8 &  1.7   &   0.3  &   2.0 &     0.3 &    13.9 &     2.1 &   21 &   6 \\
NGC3110$^{\dag}$                &  1.9 &     2.4 &     0.4 &     5.0 &     0.8 &   2.4  &  0.4   & 0.93 &    0.14 &     2.0 &     0.3 &    8 &   2 \\
NGC3256$^{\ddag}$                &  1.0 &     3.7 &     0.6 &     6.7 &     1.0 &  10.1 &  1.5   &    3.0 &     0.5 &     5.4 &     0.8 &  6.6 & 2.0 \\
ESO320-G030        &  0.7 &    0.93 &  0.14 &    1.58 & 0.24 &    1.8 &   0.3  &    1.7 &     0.3 &     2.8 &     0.4 &  6.1 & 1.8 \\
IRAS12112+0305  &  2.6 &     3.8 &     0.6 &    11.0 &    1.7 &  14.5 &   0.5  & 0.46 &    0.07 &    1.35 &    0.20 &  8.0 & 2.4 \\
IRASF12115-4656 &  1.3 &     1.9 &     0.3 &     5.1 &     0.8 & 1.40  & 0.05 & 0.87 &    0.13 &     2.4 &     0.4 &   11 &   3 \\
NGC5135                &  0.5 &    1.35 & 0.20 &     3.1 &     0.5 &   3.2  &   0.5  &   4.1 &     0.6 &     9.4 &     1.4 &    9 &   2 \\
IRAS14348-1447   &  3.8 &     7.8 &     1.2 &       18 &        3 &  20.2 &   0.7  & 0.44 &    0.07 &    1.06 &    0.16 &  6.5 & 2.0 \\
IRASF17138-1017 &  0.6 &     1.7 &     0.3 &     3.3 &     0.5 &  3.60 &  0.09 &    4.5 &     0.7 &     8.9 &     1.3 &  7.6 & 2.3 \\
IRAS17208-0014   &  0.9 &     4.9 &     0.7 &    13.2 &    2.0 &   15.2 &   0.4 &   5.1 &     0.8 &    13.7 &     2.1 &  7.3 & 2.2 \\
IC4687                    &  1.1 &     4.4 &     0.7 &     8.3 &     1.3 &   3.7  &   0.6   &    3.1 &     0.5 &     5.8 &     0.9 &  7.2 & 2.2 \\
IRAS21130-4446   &  1.6 &     3.9 &     0.6 &     8.3 &     1.3 &   \multicolumn{2}{c}{\nodata} & 1.23 &    0.19 &     2.6 &     0.4 &  5.7 & 1.8 \\
NGC7130                &  1.0 &    1.62 &  0.25 &     5.4 &     0.8 &   3.3  &  0.5  &  1.24 &    0.19 &     4.1 &     0.6 &   13 &   4 \\
IC5179$^{\dag}$                    &  1.6 &     1.7 &     0.3 &     3.2 &     0.5 &  1.56 & 0.23 & 0.66 &     0.10 &    1.25 &  0.19 &  7.2 & 2.2 \\
IRAS22491-1808   &  1.7 &     2.3 &     0.4 &     5.6 &     0.9 &   17.7 &  0.4  & 0.65 &    0.10 &    1.56 &    0.24 &  6.6 & 2.0 \\
IRAS23128-5919   &  2.0 &    11.6&     1.7 &      30 &        4 &   18.6 &  0.5  &     2.4 &     0.4 &     6.3 &     0.9 &  7.2 & 2.2 \\
  \noalign{\smallskip}
\hline
\hline
\end{tabular}}}
\tablefoot{Integrated star-forming properties of the sample. Col. (2):  \Ha\ effective radius in [kpc] from \cite{Arribas:2012p1203}. Cols (3) and (4): Observed (3) and extinction-corrected (4) \Pa\ luminosities measured within \Reff, expressed in [$\rm \times10^{7}$\,\Lsolar]. The \Pa\ luminosities for the LIRGs are obtained from the \Brg\ fluxes using the case B recombination factor at T$=10,000$\,K and n$\rm_{\rm e}=10^4\,cm^{-3}$ \citep{Osterbrock:2006AGN2}. Col. (5): \emph{Spitzer}/MIPS 24\,$\mu$m luminosities from \cite{PereiraSantaella:2011fw} and archival data, in units of [$\rm \times10^{10}$\,\Lsolar]. Cols. (6) and (7): Observed (6) and extinction-corrected (7) \Si\ in [\Msolar\,yr$^{-1}$\,kpc$^{-2}$]. Col. (8): \Av\ in magnitudes. All the uncertainties are calculated by a \emph{bootstrap} method of $\rm N=300$ simulations, added in quadrature to the 15\% systematic error from the absolute flux calibration.
$^{\dag}$Due to the limited FoV of the observations, in these objects the H$\alpha$ effective radius is greater than our FoV and the luminosities should be considered lower limits.
$^{\ddag}$ Since the main nucleus of NGC 3256 was not observed, we centred the aperture in the centre of the FoV so the measurements might be inaccurate (see Paper~I).}
\label{table:reff_table}
\end{table*}

\begin{table*}
\caption{Statistics of the SFR distributions}
\centering
\resizebox{.9\textwidth}{!}{
{\setlength{\tabcolsep}{5pt}
\begin{tabular}{cccccccccc}
\hline
\hline
\noalign{\smallskip}
  Object & \large$\Sigma_{\rm obs}^{\rm median}$ & \large$\Sigma_{\rm corr}^{\rm median}$ & \large$\Sigma_{\rm obs}^{\rm mean}$ &\large$\Sigma_{\rm corr}^{\rm mean}$ & \large$\Sigma_{\rm obs}(\rm P_{5})$ & \large$\Sigma_{\rm obs}(\rm P_{95})$ & \large$\Sigma_{\rm corr}(\rm P_{5})$ & \large$\Sigma_{\rm corr}(\rm P_{95})$ & \large A$_{\rm V,median}$ \\
  \noalign{\smallskip}
  (1) & (2) & (3) &(4)  & (5) & (6) & (7) & (8) & (9) & (10) \\
\noalign{\smallskip}
\hline
IRAS06206-6315 &      0.1 &      0.2 &     1.2 (0.6) &      1.5 (2.3) &     0.1 &      1.5 &      0.1 &      4.5 &  6.4 \\
NGC2369 &      1.4 &      3.9 &     5.0 (2.5) &     13.8 (21.5) &     0.3 &      7.8 &      0.6 &     60.7 & 18.5 \\
NGC3110 &      1.0 &      1.6 &     2.2 (0.8) &      2.6 (2.6) &     0.3 &      3.0 &      0.5 &      7.6 &  8.8 \\
NGC3256$^{\dag}$ &      2.4 &      3.4 &     6.6 (3.1) &     10.6 (8.1) &     0.5 &      9.0 &      0.5 &     18.6 &  4.9 \\
ESO320-G030 &      1.4 &      1.8 &     3.0 (1.1) &      3.6 (3.3) &     0.4 &      3.8 &      0.5 &      8.7 &  5.1 \\
IRAS12112+0305 &      0.2 &      0.3 &     1.8 (1.0) &      2.8 (3.4) &     0.1 &      2.2 &      0.1 &      6.4 &  6.4 \\
IRASF12115-4656 &      0.7 &      1.0 &     1.6 (0.6) &      2.9 (3.3) &     0.4 &      2.1 &      0.5 &      7.5 &  7.9 \\
NGC5135 &      1.5 &      2.3 &     6.2 (2.6) &      8.6 (6.8) &     0.3 &      7.7 &      0.4 &     19.7 &  8.0 \\
IRAS14348-1447 &      0.1 &      0.2 &     1.9 (0.9) &      3.2 (3.2) &     <0.1 &      1.6 &      0.1 &      4.8 &  6.3 \\
IRASF17138-1017 &      1.5 &      2.8 &     9.4 (4.3) &     12.6 (8.3) &     0.4 &     12.5 &      0.5 &     23.3 &  6.6 \\
IRAS17208-0014 &      0.2 &      0.3 &     2.9 (2.0) &      4.3 (6.4) &     0.1 &      5.1 &      0.1 &     12.6 &  5.4 \\
IC4687 &      1.9 &      3.3 &     5.7 (2.8) &      8.7 (5.4) &     0.4 &      7.9 &      0.5 &     16.0 &  5.4 \\
IRAS21130-4446 &      0.2 &      0.3 &     2.4 (1.0) &      1.7 (2.7) &     <0.1 &      3.0 &      <0.1 &      6.1 &  4.2 \\
NGC7130 &      0.6 &      0.8 &     4.9 (3.5) &      4.9 (13.4) &     0.3 &      5.3 &      0.3 &     17.4 &  5.4 \\
IC5179 &      0.7 &      0.7 &     5.2 (1.7) &      2.8 (5.3) &     0.3 &      3.0 &      0.3 &      6.1 &  4.1 \\
IRAS22491-1808 &      0.1 &      0.2 &     1.1 ( 0.6) &      1.2 (1.7) &     <0.1 &      1.6 &      0.1 &      4.6 &  6.4 \\
IRAS23128-5919 &      0.2 &      0.2 &     4.0 (3.0) &      4.6 (9.8) &     <0.1 &      3.5 &      0.1 &      8.1 &  6.2 \\
\hline
LIRGs &      1.2 &      1.7 &     2.1 (2.6) &      4.5 (8.8) &     0.3 &      6.8 &      0.4 &     17.0 &  5.3 \\
ULIRGs &      0.2 &      0.2 &     0.7 (2.0) &      1.6 (6.6) &     <0.1 &      2.8 &      0.1 &      6.8 &  6.5 \\
            
\hline
\hline
\end{tabular}}}
\tablefoot{Statistics of the star formation rate surface density, \Si, distributions. Cols. (2) and (3):  Median \Si\ values of the observed (2) and extinction-corrected (3) spaxel-by-spaxel distributions. Cols. (4) and (5): Weighted mean \Si\ values of the observed (4) and extinction-corrected (5) distributions. The standard deviation of each distribution is shown in brackets. Cols. (6) to (9): 5th and 95th percentiles of the distributions. Col. (10): Median \Av\ from the spaxel-by-spaxel extinction distributions \citep{PiquerasLopez:2013hx}. All the quantities are expressed in [\Msolar\,yr$^{-1}$\,kpc$^{-2}$], except \Av, which is expressed in magnitudes.
$^{\dag}$ The main nucleus of NGC 3256 was not observed (see Paper~I).}
\label{table:distrib_table}
\end{table*}

\subsection{Integrated measurements of the star formation rate}

We obtained integrated measurements of the SFR and \Si\ by stacking the spectra of the individual spaxels within the H$\alpha$ effective radius from \cite{Arribas:2012p1203}. The spectra of each individual spaxel were previously derotated, i.e. corrected from the intrinsic large-scale velocity field, to prevent from beam smearing effects, and improve the S/N of the lines. The lines were then fitted to a Gaussian profile, accounting for the instrumental broadening using the OH sky line at 2.190\,$\mu$m (see Paper~I, for details).

The SFR and \Si\ values were derived using the \Brg\ and \Pa\ emission lines for LIRGs and ULIRGs, respectively, and Eqs.~\ref{eq:kennicuttPa} and \ref{eq:kennicuttBrg}. These values were then corrected from extinction using the \Av\ measurements from \cite{PiquerasLopez:2013hx} (Paper~II). These extinction values are obtained from the \Brg/\Brd\ and \Pa/\Brg\ line ratios for LIRGs and ULIRGs, respectively, and the extinction law described in \cite{Calzetti:2000p2349} ($\rm\alpha_{\rm Br\gamma} = 0.096$ and $\rm\alpha_{\rm Pa\alpha} = 0.145$, see Paper~II). Calzetti's law has to be applied to derive first the extinction for the stellar continuum and then obtain the extinction towards the ionized gas using the empirical ratio \emph{f} = E(B--V)$\rm_{stars}$/E(B--V)$\rm_{gas}$ = 0.44,  from \cite{Calzetti:2000p2349}. This might lead to significant differences in the \Av\ measurements when compared with other galactic extinction laws like \cite{Fitzpatrick:1986ApJ307} or \cite{Cardelli:1989ApJ345}, specially in the ultraviolet and optical. {In addition, the value of this ratio, \emph{f}, is under debate, and it might not be considered as a constant for all type of galaxies and redshifts (e.g. \citealt{Kashino:2013ev}, \citealt{Koyama:2015gz}, \citealt{Pannella:2015gk}). In the near infrared, all these extinction laws are very similar and differences are $\leq0.02\times$\Av\ at \Pa, \Brd,\ and \Brg\ wavelengths. However, the use of a different value of \emph{f} might have an impact when comparing near-IR measurements of the SFR with those derived from \Ha\ measurements at high-z (see Sect.~\ref{section:clumps_highz}). We have investigated this effect by comparing the extinction-corrected \Ha\ luminosities obtained assuming a higher value of \emph{f} = 0.7 \citep{Kashino:2013ev}. For moderate A$\rm_{H\alpha}$ values (0.7--2.1 mag, see \citealt{Genzel:2011ApJ733} and \citealt{Swinbank:2012ApJ760}), the extinction-corrected \Ha\ luminosities are $\sim$20--50\% lower than those derived assuming a ratio of \emph{f} = 0.44. Although this effect could be important for high values of A$\rm_{H\alpha}$, the use of a different value of \emph{f} would not have a critical impact on the conclusions of the present work. We have then assumed that the canonical value of $f=0.44$ from \cite{Calzetti:2000p2349}, derived from a sample of low-z starburst galaxies}. Observed and extinction-corrected measurements of the \Si, velocity dispersion, and \Av\ for each object are listed in Table~\ref{table:reff_table}.

\subsection{Spatially-resolved star formation rate at sub-kpc scales}
\label{section:sfr}

\begin{figure*}[t]
\begin{center}
\resizebox{\hsize}{!}{\includegraphics[angle=0]{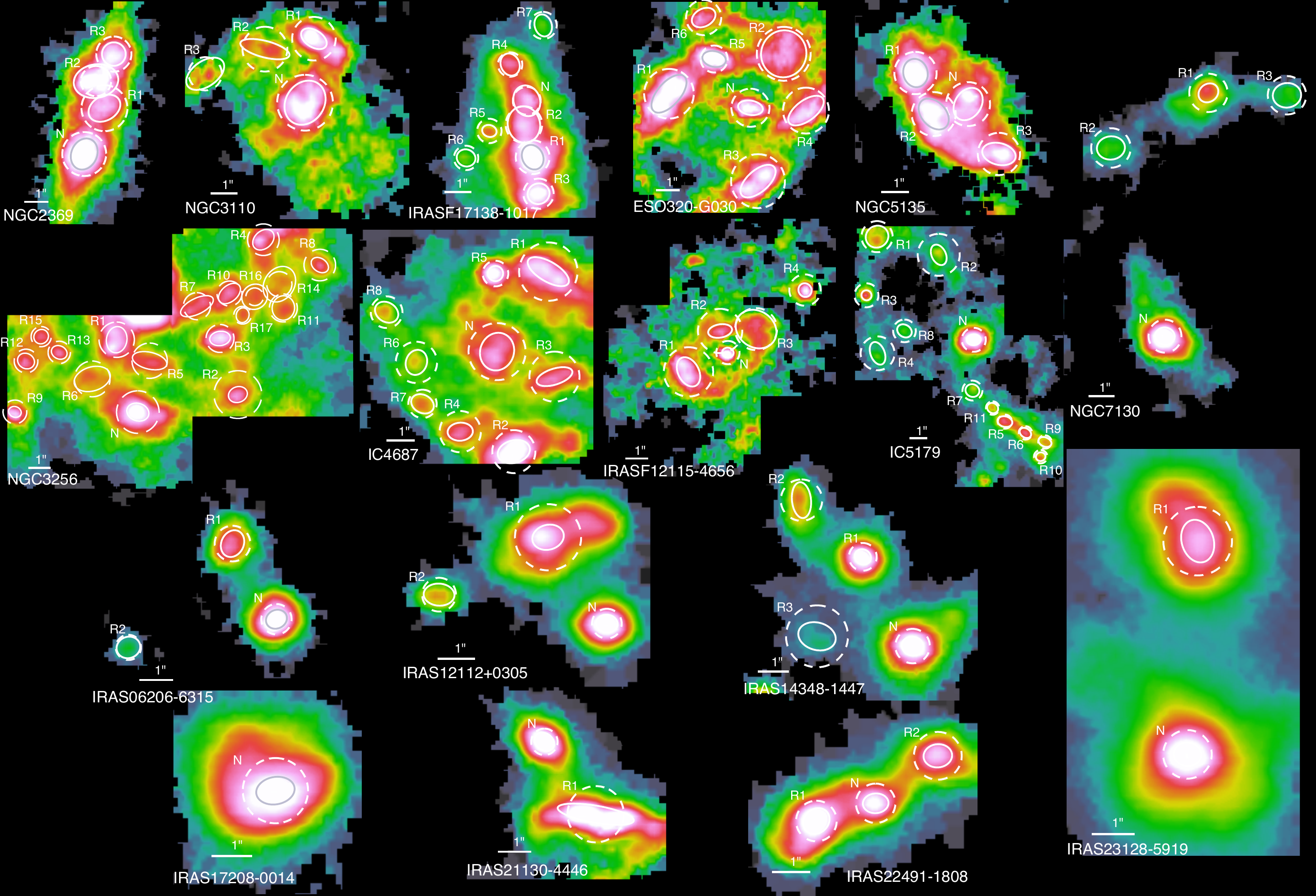}}
\caption{Maps of the observed \Brg\ (LIRGs) and \Pa\ (ULIRGs) emission. The angular scale of each object is indicated by a white horizontal bar that marks 1 arcsec. The nucleus of each object (defined as the brightest region in the K-band continuum, see Paper~I) is labeled as `N', whereas the remaining regions are tagged as `R\#' by decreasing total observed flux (see Table~\ref{table:regions_table}). The white (grey) ellipses over each individual region are the 1$\sigma$ fitting to the 2D gaussian profile, whereas the white (grey) dashed circles have r$\rm_{eff}$ radius. The colour scheme is logarithmic and autoscaled.}
\label{sfr_maps}
\end{center}
\end{figure*}

To obtain the extinction-corrected values of the SFR and \Si\ for the maps and spaxel-by-spaxel distributions, we applied the \Av\ correction presented in Paper~II. As outlined in Paper II, the correction is made on a spaxel-by-spaxel basis on those spaxels where the weakest line (\Brd\ and \Brg\ in LIRGs and ULIRGs, respectively) has been detected above a S/N threshold of 4. In those spaxels, where this point-to-point correction is not available, we have used a flux-weighted median value of \Av.

The individual SFR uncertainties are obtained directly from the 1$\sigma$ errors in the flux and in the extinction measurements. As described in Paper~II, the \Av\ uncertainties are derived from the error in the line fluxes using Monte Carlo simulations, with the main advantage that uncertainties take into account both the photon noise, and the errors from the line fitting or from an inaccurate continuum determination. Since the \Av\ measurements are highly sensitive to the S/N of the weakest line of the ratio, we observed an artificial increase of the corrected SFR in those regions with low surface brightness, specially in the ULIRGs. In regions that are typically the external regions of the sources, the uncertainties in the \Av\ measurements reach up to 70-80\%, and are directly propagated to the SFR measurements.

Figures~\ref{figure:LIRG} and \ref{figure:ULIRG} in Appendix~\ref{sec:tables&figures} show the two-dimensional structure of the extinction-corrected \Si, together with the \Av\ map, and \Brg\ and \Pa\ emission maps for LIRGs and ULIRGs, respectively. The statistics of the spaxel-by-spaxel distributions (i.e. median and percentiles) are shown in Table~\ref{table:distrib_table}. The detection limit of the emission maps  varies slightly from object to object. We found an average sensitivity limit of $\rm10^{-18}\,erg\,s^{-1}\,cm^{-2}$ per spaxel on the \Brg\ and \Pa\ maps in LIRGs and ULIRGs, respectively. Considering the closest LIRG and ULIRG, this flux threshold yields to \Si\ observed values of $\sim$0.3\,\Msolar\,yr$^{-1}$\,kpc$^{-2}$ and $\sim$0.03\,\Msolar\,yr$^{-1}$\,kpc$^{-2}$ per spaxel in LIRGs and ULIRGs, respectively, which are different owing to the use of different tracers. Therefore, we hereafter assume these values as our detection limits in the \Si\ maps and spaxel-by-spaxel distributions.

\subsection{Star-forming clumps. Identification and sizes}
\label{section:region_size}

\begin{figure}[]
\begin{center}
\resizebox{\hsize}{!}{\includegraphics[angle=0]{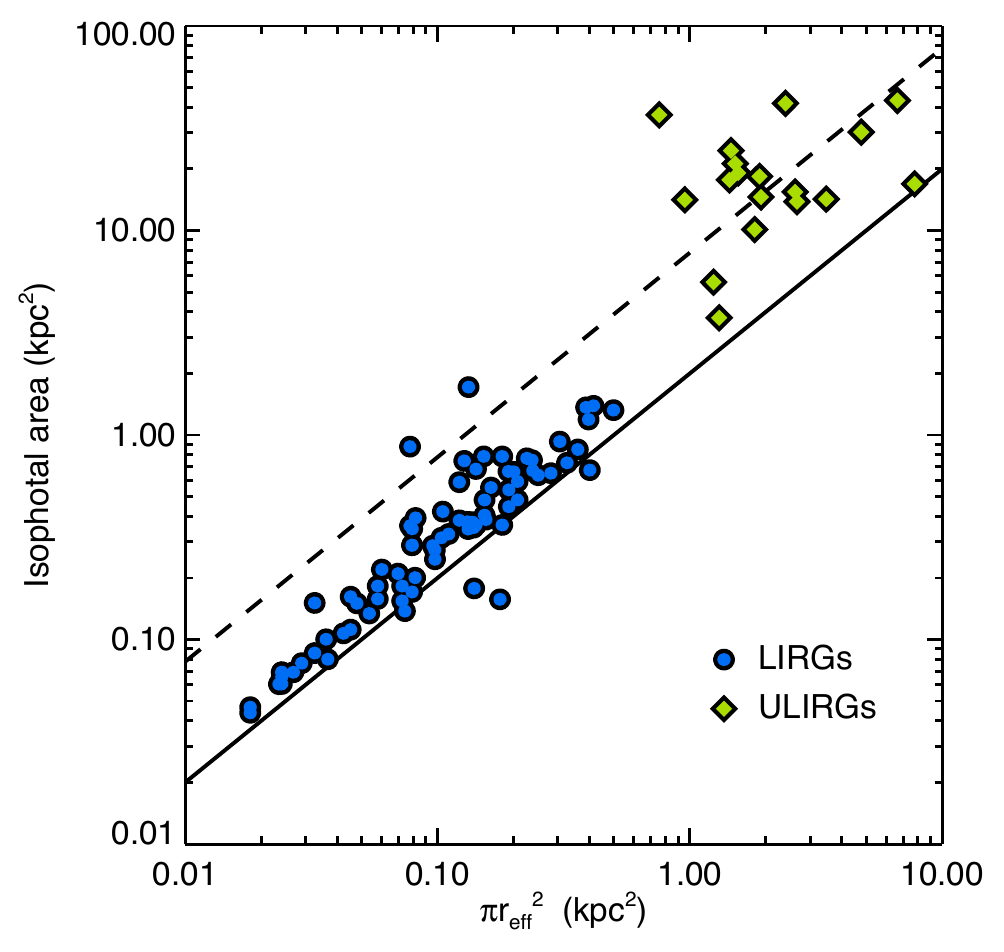}}
\caption{Comparison between the isophotal area (region enclosed by an isophote of $\rm10^{-18}\,erg\,s^{-1}\,cm^{-2}$) and the circular area obtained from the r$\rm_{eff}$ of the individual star-forming regions in LIRGs (circles) and ULIRGs (diamonds). The thick black line represents the ratio between both quantities for a region with a uniform flux distribution, whereas the dashed line represents the same ratio for a region with a symmetric 2D gaussian light profile and a total flux of 10$\rm^{-13}$\,$\rm erg\,s^{-1}\,cm^{-2}$.}
\label{comp_radii}
\end{center}
\end{figure}

As shown in Figs.~\ref{figure:LIRG} and \ref{figure:ULIRG}, the spatial distribution of the SF is clumpy with diffuse as well as compact regions of high-surface brightness. We  identified a total of 95 of these individual star-forming regions/complexes. Figure~\ref{sfr_maps} shows the \Brg\ (LIRGs) and \Pa\ (ULIRGs) maps of the objects of the sample, where the individual regions have been overplotted. Characterising the size of an individual star-forming clump is a key issue when accurately measuring important properties of the region, such as its luminosity, velocity dispersion, and \Si. Two parameters are widely employed to characterise the size of an emitting region, i.e. the effective radius or half-light radius, \reff\ \citep{Kennicutt:1979ApJ228}, and the `core' radius, \rcore\ \citep{Sandage:1974ApJ190}. 

The \reff\ is traditionally defined as the radius that encloses half of the total flux of the emitting region, assuming circular symmetry, although it could be generalised in terms of an isophotal region with arbitrary shape that encloses a total flux greater than a defined fraction of light. Depending on the flux threshold, the intrinsic flux distribution of the region and the sensitivity of the observations, this radius can represent either the core of the region or also include the surrounding diffuse inter-clump emission. The presence of tails, halos, or smeared background emission when two regions are close together increase \reff\ and, therefore, bias the measurements. This could be particularly important when comparing low- and high-redshift samples, since the contribution of the surrounding background could be significantly higher locally than at higher redshift.  

The so-called core method consists of fitting the light profile of the region using a particular analytic function, typically a 2D Gaussian function. One of the advantages of this method is that it does not depend on any particular flux or surface brightness threshold, and it is less sensitive to the local background. This method is widely used for high-redshift data where the regions are almost unresolved, and it seems suitable for a direct comparison between low- and high-redshift observations. However, the main limitation of the method is that it assumes a light profile, typically Gaussian, that might not accurately reproduce  the intrinsic profile of resolved clumps, particularly in local samples where the spatial resolution of the observations is  typically of tenths of parsecs. For further discussion on both methods see \cite{Wisnioski:2012MNRAS422}.

To obtain the \reff\ of the individual clumps, we considered a flux threshold of $\rm10^{-18}\,erg\,s^{-1}\,cm^{-2}$ to calculate the total flux of each region. This flux limit corresponds typically to a S/N ratio of $\sim2$ in the \Brg\ and \Pa\ lines per spaxel for LIRGs and ULIRGs, respectively. For these isophotal regions, we estimate the \reff\ using the A/2 method described in \cite{Arribas:2012p1203}, i.e. the \reff\ is obtained as $\rm r_{eff} = \sqrt{\rm A/\pi}$, where A is the area covered by the minimum set of spaxels that accounts for half of the total flux of the region.

We show a comparison between the area of the isophotal region ($\rm A_{iso}$) and the circular area derived from the \reff\ ($\rm A_{eff}$) in Fig.~\ref{comp_radii}. This relationship gives us an idea of how compact or extended each individual region is. Following on from this, we also plotted the expected ratio of both areas for an ideal region with an uniform flux distribution, that represents the extreme case of an extended region with diffuse background emission,  and a mock region with a symmetric 2D Gaussian light profile that represents a typical unresolved, compact star-forming clump. The region with a constant flux distribution represents the lower limit for the ratio, since $\rm A_{iso} \sim 2\times A_{eff}$. On the other hand, the relationship between both areas for the mock region would depend on the total flux of the clump, the FWHM of the 2D gaussian profile, and our limiting flux threshold. We considered an unresolved clump with a total flux of 10$\rm^{-13}$\,$\rm erg\,s^{-1}\,cm^{-2}$ that corresponds to the total flux of a typical star-forming clump of our sample, with a FWHM of $\sim$0.63\,arcsec and a detection limit of $\rm10^{-18}\,erg\,s^{-1}\,cm^{-2}$. Most of the clumps from the LIRG subsample lay close to the lower limit of $\rm A_{iso} \sim 2\times A_{eff}$, whereas the regions observed in the ULIRG subsample present higher ratios, similar to those of more compact distributions. Although it could be argued that this might respond to intrinsic differences between the structure of star-forming clumps in LIRGs and ULIRGs, the differences in the spatial resolution and FoV of our observations, between both LIRG and ULIRG subsamples, means that this conclusion is not straight-forward, and could be a direct consequence of the distinct diffuse background contribution in both subsamples.

To calculate \rcore, we fitted each individual region to a 2D Gaussian profile with the local background level as a free parameter. The final value of the radius is obtained as a quadratic mean of the widths of the Gaussian function, $\rm \sigma_{x'}$  and $\rm \sigma_{y'}$, where $\rm x'$ and $\rm y'$ are the canonical axes of the ellipse.

\subsection{Star-forming clumps.  Line luminosity and extinction correction}
\label{section:region_analysis}

\begin{figure*}[t]
\begin{center}
\resizebox{1.\hsize}{!}{\includegraphics[angle=0]{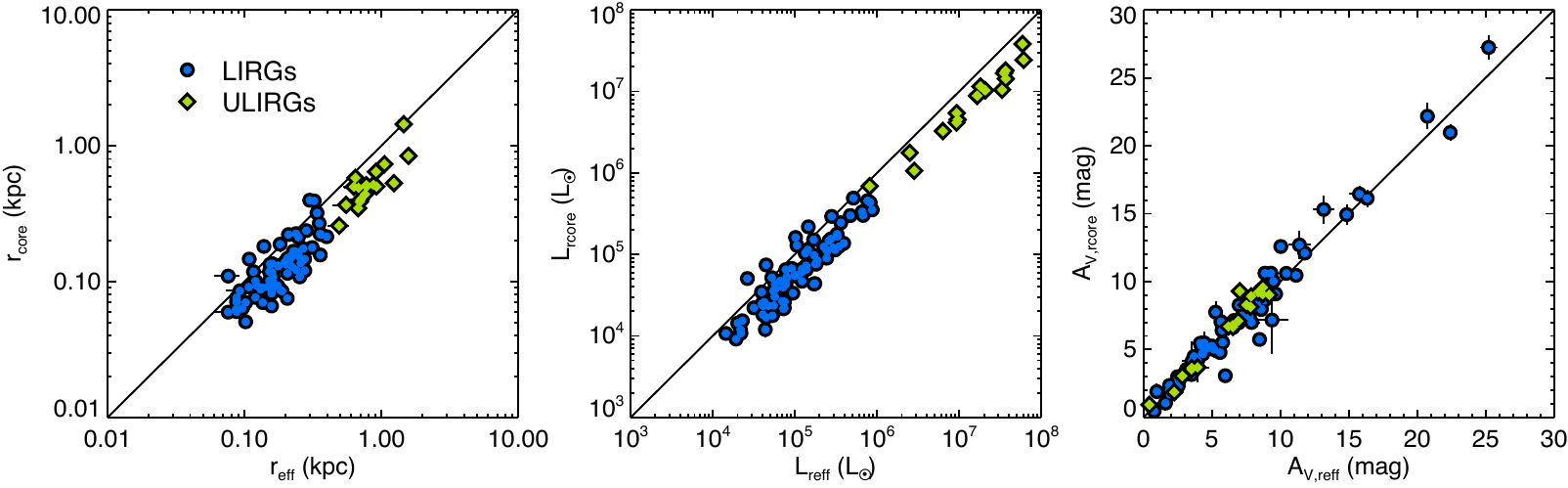}}
\caption{Comparison between the radius (left), observed line luminosity (centre), and visual extinction (right) of individual clumps in LIRGs (blue) and ULIRGs (green), measured using the effective radius and core radius methods. {Uncertainties at 1$\rm \sigma$ level are represented as black lines}. The thick black lines represent a one-to-one ratio in all the panels.}
\label{regions_circ_vs_ell}
\end{center}
\end{figure*}

We measured the line luminosity and visual extinction of each clump from the stacked spectra of the individual spaxels within the circular area defined by \reff\ and by the ellipse given by $\rm \sigma_{x'}$  and $\rm \sigma_{y'}$. The measurements are performed using the same procedure as for the whole galaxy, we fitted a Gaussian profile to the derotated stacked spectrum of each region, accounting for the instrumental broadening, and used a spaxel-by-spaxel correction to account for the extinction. The resulting values are shown in Table~\ref{table:regions_table}.

The measurements of the SFR and \Si\ could also be performed directly over the emission maps. We checked whether the method, i.e. fitting the stacked spectrum or synthetic aperture photometry of the emission map, would bias the measurements. We found that the values of the luminosity obtained directly from the maps are slightly larger than those obtained from the line fitting, although the differences are less than $\sim10\%$. There is also a small dependence with the luminosity/size of the region, those regions with small sizes/less numbers of spaxels show the largest differences ($\lsim10\%$), whereas the more extended regions present differences between both measurements of less than $\sim5\%$ owing to the larger number of spaxels within the region.

Figure~\ref{regions_circ_vs_ell} shows a comparison between the luminosity, radius, and visual extinction of the clumps measured using the \reff\ and \rcore\ methods. We found that the sizes of the clumps measured using \reff\ are typically larger than those obtained by the core method, with differences of $\sim30\%$ on average, and hence, the luminosities extracted using the effective radius method yield values larger than  the 
 corresponding ones for the \rcore\ measurements. The differences in the area of the region obtained by both methods are typically of  50-60\% or less, and are translated to differences in luminosity of $\sim50\%$ on average. We observed that the \Av\ values are less sensitive to the method used for characterising the size of the clumps, and found a close one-to-one correlation between the values measured using the effective radius and the core radius, with differences lower than $\sim20\%$ in \Av.

\section{Results and discussion}
\label{section:results_sfr}

\subsection{Integrated star formation: Optical vs near-IR star formation rates.}
\label{section:optical_vs_ir}

We compared our \Brg\ and \Pa\ SFR measurements with those derived from our own IFS-based \Ha\ measurements from \cite{GarciaMarin:2009p8459} and \cite{RodriguezZaurin:2011A&A527}. Although the SFR values for our LIRG subsample were derived from the \Brg\ luminosity, these were translated to \SFRPa, assuming a \Brg\ to \Pa\ line ratio given by the corresponding Case B recombination factor.

\cite{GarciaMarin:2009p8459} present optical IFS observations of 22 local ULIRG systems, and provided not only observed \Ha\ fluxes of three of our ULIRGs (IRAS~12112+0305, IRAS~14348-1447, and IRAS~17208-0014), but also spaxel-by-spaxel extinction-corrected measurements based on the \Ha/\Hb\ ratio, with an angular resolution of $\sim1$\arcsec per spaxel. For our LIRG subsample and the rest of the ULIRGs, we used the \Ha\ fluxes based on VIMOS IFS observations and \Av\ corrections from \cite{RodriguezZaurin:2011A&A527}. The \Av\ corrections for these objects are derived from the nuclear \Ha/\Hb\ ratios available in the literature, based on long-slit spectroscopic observations. To account for the extinction effects, these authors used the nuclear reddening spectroscopic measurements to correct the fraction of \Ha\ emission within a typical slit width ($\sim2$\,arcsec) and considered the remaining fraction of flux not affected by extinction. As discussed in \cite{RodriguezZaurin:2011A&A527}, although extinction effects are particularly important in the nuclear regions, this approach could  underestimate the extinction-corrected \SFRHa\ values, since, as shown in Figs.~\ref{figure:LIRG} and \ref{figure:ULIRG}, it is also frequent to find dusty, highly-obscured extranuclear \Ha-emitting regions in these objects.

 \begin{figure}[t]
\begin{center}
\resizebox{1.\hsize}{!}{\includegraphics[angle=0]{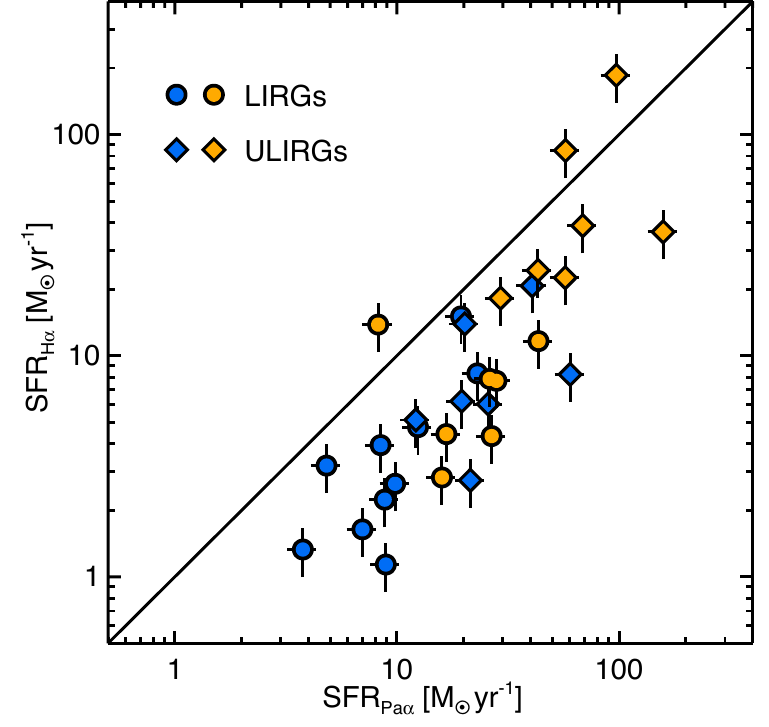}}
\caption{Comparison of star formation rates derived from \Ha\ and \Pa\ observed (blue) and extinction-corrected (yellow) luminosities for LIRGs (circles) and ULIRGs (diamonds). The solid black line represents a one-to-one ratio. The SFR has been corrected using its corresponding \Av\ correction, i.e. \SFRHa\ is corrected using optical measurements of \Av, and \SFRPa\ is corrected using the near-IR lines. The \Ha\ luminosities are extracted from \cite{GarciaMarin:2009p8459} and \cite{RodriguezZaurin:2011A&A527}.{Uncertainties at 1$\rm \sigma$ level are represented as black lines.}}
\label{sfr_palpha_vs_halpha_4p}
\end{center}
\end{figure}

Figure~\ref{sfr_palpha_vs_halpha_4p} shows the comparison between the optical and near-IR-based SFR values, for observed and extinction-corrected measurements. As expected, we found a mild correlation between the \SFRHa\ and \SFRPa\ values, although the \Pa\ measurements yield larger SFR values, even for extinction-corrected measurements. For the observed values, the \SFRPa\ values are on average $\sim3.6$ times larger than \SFRHa, although individual ratios range between from $\sim1.2$ to $\sim8$. When the extinction correction is applied, the \SFRPa/\SFRHa\ ratio decreases to $\sim2.9$ on average, while the individual factors are constrained between $\sim0.5$ and $\sim6$.

As mentioned in Paper~I, we found evidence of nuclear activity in four of the objects of the sample, i.e. IRASF 12115-4656, NGC5135, NGC7130, and IRAS 23128-5919, in terms of detection of [SiVI] coronal emission. Those objects, in particular the three LIRGs, present the highest \SFRPa/\SFRHa\ ratios. When these objects are removed from the sample, the average ratios of the  \SFRPa/\SFRHa\ becomes $\sim3.4$ and $\sim2.0$ for the observed and extinction-corrected values.

\subsection{Integrated star formation: Near-IR vs mid-IR star-formation rates}

\begin{figure*}[t]
\begin{center}
\resizebox{1.\hsize}{!}{\includegraphics[angle=0]{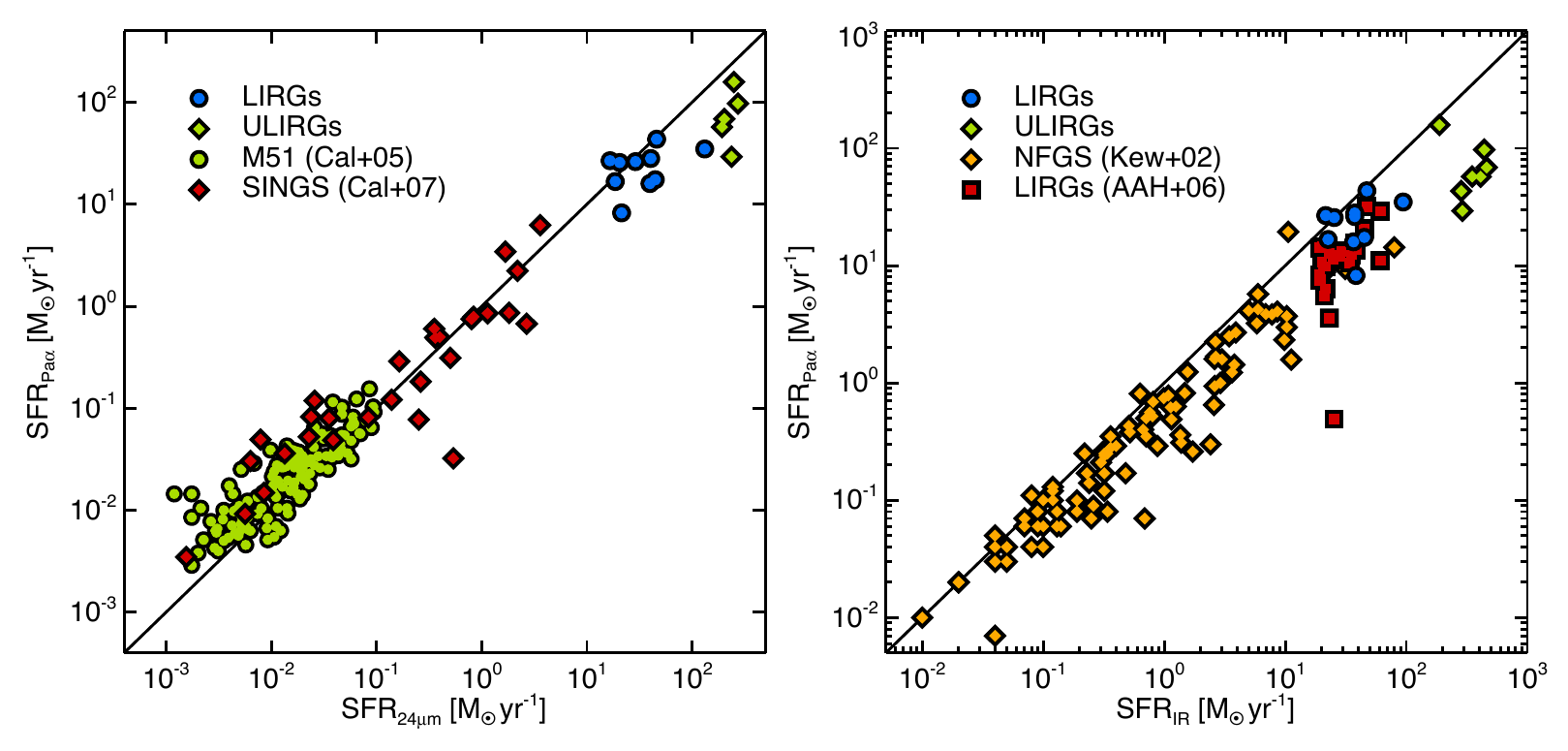}}
\caption{\emph{Left:} Comparison of the extinction-corrected \SFRPa\ with the monochromatic \SFRMIR. The blue circles and green diamonds correspond to our local LIRGs and ULIRGs, respectively. The green circles are data of M51 individual star-forming regions from \cite{Calzetti:2005ApJ633} and the red diamonds correspond to data from \cite{Calzetti:2007ApJ666} of SINGS galaxies. \emph{Right:} Comparison of extinction-corrected \Pa\ luminosity and \LTIR. Our LIRG and ULIRG samples are plotted as blue circles and green diamonds, respectively, orange diamonds are data from \cite{Kewley:2002AJ124} of normal galaxies from the NFGS, while red squares correspond to local LIRGs from \cite{AlonsoHerrero:2006p4703}. In both panels, the black lines correspond to a one-to-one ratio. {Uncertainties at 1$\sigma$ level of the \Pa\ measurements are smaller than the size of the symbols.}}
\label{sfr_24mum_ir}
\end{center}
\end{figure*}

Our extinction-corrected \SFRPa\ is compared with the \SFRTIR\ and \SFRMIR\ values derived using Eqs.~\ref{eq:kennicuttTIR} and \ref{eq:AAH24} (see Fig.~\ref{sfr_24mum_ir}),  and based on IRAS and 24$\mu$m fluxes that are already available \citep{PereiraSantaella:2011fw}. In addition, different sets of nearby galaxies and individual star-forming regions in M51 are also included to extend the comparison over more than four orders of magnitude in \SFRPa.

The  \Pa\ and 24\,$\mu$m tracers yield very similar SFR values characterized by a linear relation with some deviations at the low- and high-luminosity ends (Fig.~\ref{sfr_24mum_ir}, left panel). The deviations at the high-luminosity end have already been reported as being observed in previous works  using different calibrations and datasets (\citealt{AlonsoHerrero:2006p4703}, \citealt{Rieke:2009ApJ692} or \citealt{Calzetti:2007ApJ666}) and respond to the fact that the relation between \LMIR\ and the SFR is no longer linear, since the increase of starlight raises the dust temperature of large grains and hence the absorbed energy reradiated at 24\,$\mu$m. Besides this effect, the high-luminosity end (LIRGs and brighter) corresponds to objects where the star formation occurs in dusty environments with increasing density, where standard extinction corrections (e.g. based on hydrogen recombination line ratios) become less reliable and could lead to an underestimation of the true \Pa\ luminosity.

The comparison between the SFR derived from the \Pa\ and \LTIR\ luminosities (Fig.~\ref{sfr_24mum_ir}, right panel) shows a sub-linear relation with \SFRTIR\ values that are systematically higher than \SFRPa. In particular, for our sample of LIRGs and ULIRGs, the differences are up to a factor $\times2$. The origin of this systematic difference is not clear. It could be due to the fact that a fraction of the ionizing photons do not ionize the surrounding interstellar medium but are directly absorbed by the dust, i.e. the \SFRPa\ would then represent a lower limit to the true instantaneous SFR. On the other hand, \SFRTIR\ is sensitive to the overall radiation field produced by young massive ionizing stars, as well as older intermediate mass, non-ionizing stars, i.e. tracing SFR over longer time scales and, therefore, in a sense represent an upper limit to the true instantaneous SFR.  In summary, the comparison of the near- and mid-IR SFR tracers indicate that the extinction-corrected \Pa\ and the 24\,$\mu$m fluxes give  more consistent and more reliable values of the instantaneous SFR in our galaxies.

\subsection{Spatially resolved star formation. \Si\ distributions on (sub)kiloparsec scales}

\begin{figure*}[t]
\begin{center}
\resizebox{1.\hsize}{!}{\includegraphics[angle=0]{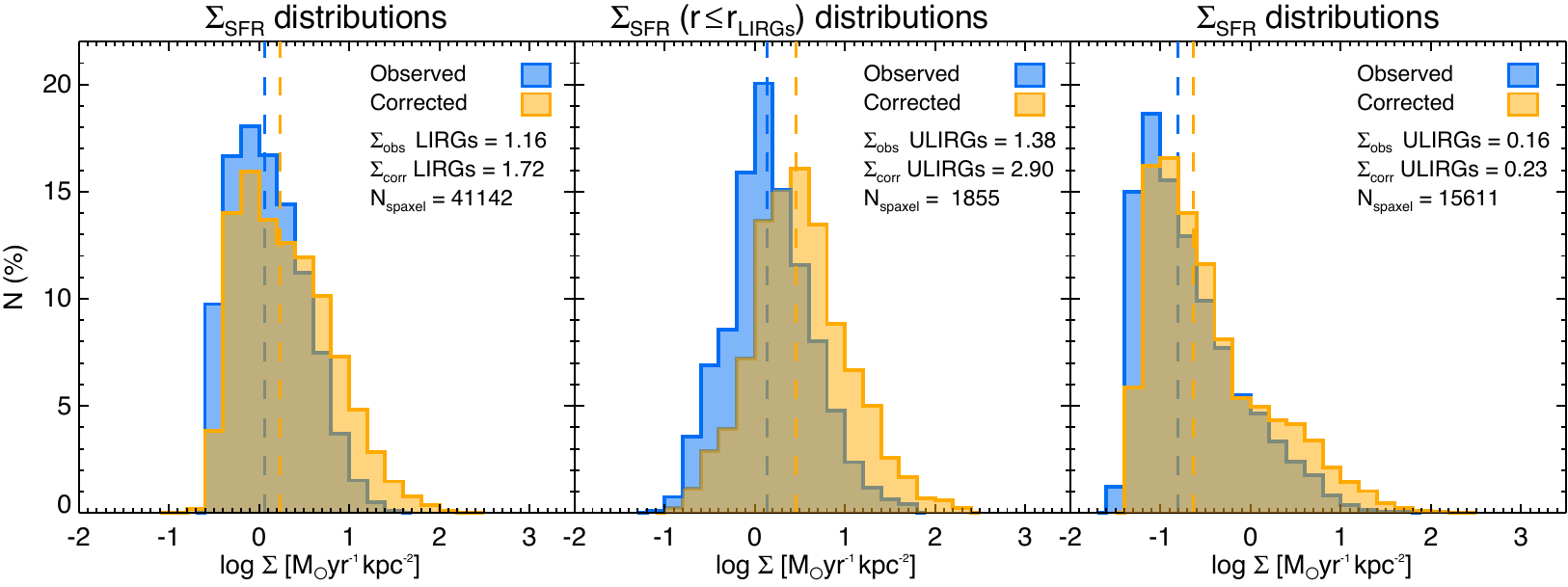}}
\caption{Observed and extinction-corrected spaxel-by-spaxel \Si\ distributions of the LIRG subsample (left), of the inner spaxels ($\rm r\leq r_{\rm LIRGs}$, see text for details) of the ULIRG subsample (centre), and the complete ULIRG distributions (right). The median \Si\ values and the total number of spaxels in each distribution are shown in the panels in units of [\Msolar\,yr$^{-1}$\,kpc$^{-2}$] and plotted as dashed vertical lines.}
\label{figure:crop_distributions_sfr_histograms}
\end{center}
\end{figure*}

The observed and extinction-corrected \Si\ distributions for the entire sample of LIRG and ULIRG \Si\ maps are presented in Fig.~\ref{figure:crop_distributions_sfr_histograms}, as derived from the spaxel-by-spaxel distribution for each galaxy (see Appendix~\ref{sec:tables&figures}  for the individual distributions). In addition, their respective median, weighted mean, and percentiles can be found in Table~\ref{table:distrib_table}.

In LIRGs, the median of the observed distribution is $\rm\Sigma_{\rm LIRGs}^{\rm obs}=1.16$\,\Msolar\,yr$^{-1}$\,kpc$^{-2}$, and increases to $\rm\Sigma_{\rm LIRGs}^{\rm corr}=1.72$\,\Msolar\,yr$^{-1}$\,kpc$^{-2}$ in the extinction-corrected distribution. The corrected distribution also becomes a $\sim50\%$ wider than the observed one. In ULIRGs, the median of the observed and corrected distributions are $\rm\Sigma_{\rm ULIRGs}^{\rm obs}=0.16$\,\Msolar\,yr$^{-1}$\,kpc$^{-2}$ and $\rm\Sigma_{\rm ULIRGs}^{\rm corr}=0.23$\,\Msolar\,yr$^{-1}$\,kpc$^{-2}$. 

As discussed in Paper~II, the difference in distance, and hence in angular resolution, is a key issue when comparing the results for LIRGs and ULIRGs, and therefore a direct interpretation is not straight-forward. The FoV of our SINFONI observations limits our analysis of the LIRGs to their innermost $\sim3$\,kpc, with a typical sampling of about 200\,pc. On the other hand, ULIRGs are located at larger distances and therefore the physical area covered by SINFONI FoV is about $\times16$ that for LIRGs, covering more external regions, but sampled with a lower spatial resolution of about 0.9\,kpc.

Due to the clumpy structure of the dust and ionized gas distributions, simulations have been performed to evaluate the effect of the poorer linear resolution in the derivation of \Si\ for the sample of LIRGs as the distance increases up to 900\,Mpc (see Appendix~\ref{sec:distance} for a detailed discussion). Simulations were performed for both observed and extinction-corrected \Si\ distributions assuming rebinning-only and a combination of rebinning plus additional smoothing due to PSF effects. The main conclusion is that there is a tendency for an increase of the \Si\ with the distance for as much as a factor of 2 (rebinning-only), while the increase is in general less relevant for the rebinning+PSF derived \Si. Thus, resolution effects do not explain the differences observed between LIRGs and ULIRGs. These are likely due to the different physical area covered by the SINFONI observations, excluding the lower surface brightness external regions for LIRGs. In fact, when comparing similar areas, the differences tend to decrease. 

We estimated an average radius, $\rm r_{LIRGs}=1.4\,kpc$, that correspond to the spaxel-weighted mean of the LIRG FoVs, and considered only those spaxels from the ULIRG distribution within this physical scales. The resulting observed and extinction-corrected \Si\ distributions are shown in the central panel of Fig.~\ref{figure:crop_distributions_sfr_histograms}. This set of the innermost spaxels of the ULIRG distributions that corresponds to the same physical regions sampled in the LIRGs, are $\sim10\%$ of the total in the ULIRG \Si\ distributions, and corresponds to those spaxels with the largest \Si\ values. When we consider only these values, the medians of the observed and extinction-corrected distributions reach up to 1.38\,\Msolar\,yr$^{-1}$\,kpc$^{-2}$ and 2.90\,\Msolar\,yr$^{-1}$\,kpc$^{-2}$, respectively, i.e. the median extinction-corrected \Si\ values for LIRGs and ULIRGs are within a factor of 2.

\subsection{Star-forming clumps in U/LIRGs. Line luminosities, sizes and \Si}
\label{section:clumps}

\begin{figure*}[t]
\begin{center}
\resizebox{1.\hsize}{!}{\includegraphics[angle=0]{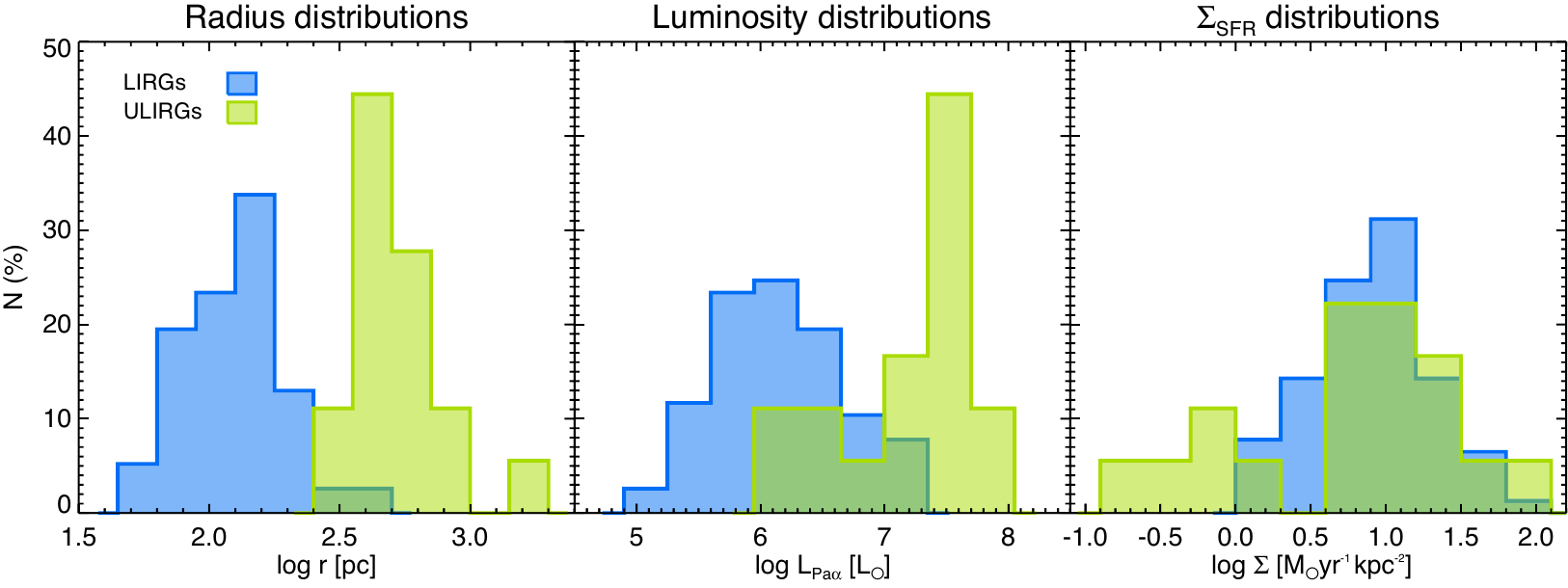}}
\caption{Distributions of the radius (left), extinction-corrected \Pa\ luminosities (centre), and extinction-corrected \Si\ (right) of the star-forming regions from LIRGs (blue) and ULIRGs (green).}
\label{figure:regions_distributions}
\end{center}
\end{figure*}

As explained before (see Secs.~\ref{section:region_size} and \ref{section:region_analysis}), the sizes, line luminosity and extinction-corrected \Si\ have been derived for a number of star-forming clumps identified in the \Brg\ and \Pa\ maps. The distribution of the radius, \LPa\ and \Si\ for the sample of  clumps is shown in Figure~\ref{figure:regions_distributions}. The clumps have sizes in the $\sim$60 to 400\,pc range for LIRGs and $\sim$300 to 1500\,pc for ULIRGs, whereas their extinction-corrected luminosities range correspond to $\sim$10$\rm^{5}$--10$\rm^{7}$\,\Lsolar\ and $\sim$10$\rm^{6}$--10$\rm^{8}$\,\Lsolar\ for LIRGs and ULIRGs, respectively. These line luminosities yield \Si\ values typically of 1--90\,\Msolar\,yr$^{-1}$\,kpc$^{-2}$ and 0.1--100\,\Msolar\,yr$^{-1}$\,kpc$^{-2}$ for LIRG and ULIRG clumps (see Table~\ref{table:regions_table} for the specific values). {As shown in Fig.~\ref{sfr_maps}, some of the smallest regions in the LIRG subsample, and some of the clumps in the ULIRG subset, are essentially unresolved, given the average spatial resolution of our dataset (see Secs.~\ref{section:observations} and \ref{section:region_size}).}

The lack of clumps with \Si\ below $\sim$1\,\Msolar\,yr$^{-1}$\,kpc$^{-2}$ and \LPa$\sim$10$\rm^{5}$\,\Lsolar\ in LIRGs could be explained in terms of observational biases. The observed regions from the LIRG subset come from the central kiloparsecs of the objects, and, as discussed in \ref{section:sfr}, our detection limit for the LIRGs is \Si$\sim$0.3\,\Msolar\,yr$^{-1}$\,kpc$^{-2}$ per spaxel. These effects bias our LIRG sample towards the detection of high-luminosity, high-\Si\ clumps.

A previous HST optical imaging study of star-forming clumps in LIRGs and ULIRGs  identified nearly 3 000 star-forming knots with a median radius of 32\,pc for the spatially-resolved knots, some with radii up to 200--400\,pc \citep{Miralles-Caballero:2011AJ142}. Subsequent ground-based optical integral field spectroscopy of the same sample, detected large H$\alpha$-emitting regions identified with star-forming clumps \citep{MirallesCaballero:2012bn}. Owing to the higher angular resolution of the HST images, the structure of these H$\alpha$ clumps consisted of few star-forming knots, with an average of 2.2 per clump.  However, the structure of these clumps is, in general, quite simple. In most cases, only one bright knot is observed within the area defined for the H$\alpha$ clump, or a centred bright knot plus a few fainter knots located at the border of the clump. Only a few clumps show a more complex knot structure, with several bright blue knots spread inside the area of the clump (see Fig.~2 and Table~1 in \citealt{MirallesCaballero:2012bn}). 

Except for internal extinction effects (i.e. different relative brightness), H$\alpha$ traces the same star-forming clumps identified in the \Brg\ and \Pa\ maps, therefore a similar clump-knot structure should be present in the \Pa\ clumps. This is confirmed by an H-band and \Pa\ study of nearby LIRGs based on HST imaging \citep{AlonsoHerrero:2006p4703}. Some of the galaxies in common with this sample (e.g. IC4687 or NGC7130) show that the \Pa\ clump is dominated by one knot and appears as more diffuse and extended (i.e. the effective radius of the \Pa\ clump larger than that of the knot).

In summary, the measured sizes and \Si\ of the clumps should be considered as upper and lower limits, respectively, with the real \Pa\ clumps more likely associated with an  individual, dominant, star-forming knot, and consequently being up to a factor 10 smaller in size, and a factor 100 larger in \Si. This could have important consequences when comparing the properties of the star-forming clumps in U/LIRGs with that of low-z spiral galaxies and high-z star-forming galaxies. Therefore the discussion in the following two sections is restricted to the \Pa\ radius-luminosity relations.

\subsection{Star-forming clumps in U/LIRGs: comparison with nearby galaxies}
\label{section:clumps_local}

\begin{table*}[ht]
\caption{Star-forming clumps. Low- and high-z samples}
\centering
\resizebox{1.\textwidth}{!}{\small
{\setlength{\tabcolsep}{5pt}
\begin{tabular}{cccccccc}
\hline
\hline
\noalign{\smallskip}
  Sample & SFR tracer & Av correction & Object & Scale & z & Instrument & Reference \\
  \noalign{\smallskip}
  (1) & (2) & (3) & (4)  & (5) & (6) & (7) & (8)\\
\noalign{\smallskip}
\hline
Planesas+97 & \Ha & \Ha/\Hb\ ratio & Circumnuclear star-forming rings, HII regions &  30--80\,pc & <0.005 & narrow band, NOT & (a) \\
Bastian+06 & \Hb & \Hg/\Hb\ ratio & Antennae, HII regions & $\sim100$\,pc & 0.00569 & VLT-VIMOS & (b) \\
Liu+13 & \Pa & Averaged$^{\rm1}$, \Ha/\Pa\ ratio & Nearby spiral galaxies, HII regions & $\sim20$\,pc & <0.003 & HST-NICMOS & (c) \\
Arribas+14 & \Ha & \Ha/\Hb\ ratio & Low-z U/LIRGs, HII regions & 200--400\,pc & <0.018 & VLT-VIMOS, INTEGRAL & (d) \\
Genzel+11 & \Ha & SED fitting$^{\rm2}$ & High-z galaxies, giant HII regions & 1.5--1.8\,kpc& 2.2--2.4 & AO VLT-SINFONI & (e) \\
AG+12 & \Ha & Averaged$^{\rm3}$, \Ha/\Hb\ ratio & SMG, integrated measurements & $\sim5$\,kpc &2.0--2.7 & NIFS, VLT-SINFONI & (f) \\
Swinbank+12 & \Ha & SED fitting$^{\rm2}$ & High-z disks, HII regions and integrated measurements & 0.9--1.0\,kpc&0.8--2.2 & AO VLT-SINFONI & (g) \\
Zanella+15 & \Hb & SED fitting$^{\rm4}$ & HII region & <0.5\,kpc & 1.987 & HST-WFC3 & (h) \\
\hline
\hline
\end{tabular}}}
\tablefoot{
Scale shown in Col. (5) corresponds to the typical physical resolution of the observations. $^{\rm1}$ Average extinction correction of $\rm A_{V} = 2.2$\,mag based on measurements from \cite{Calzetti:2007ApJ666}. $^{\rm2}$ Luminosities are corrected from extinction using stellar E(B--V) values from SED fitting. The extinction of the stars is then re-scaled to the extinction of the gas using the standard recipe A$_{\rm H\alpha}$ = 7.4\,E(B--V), where E(B--V)$\rm_{stars}$ = 0.44\,E(B--V)$\rm_{gas}$ from \cite{Calzetti:2000p2349}. $^{\rm3}$ Average extinction of $\rm A_{V} = 2.9$\,mag from \cite{Takata:2006ApJ651}. $^{\rm4}$ Luminosities are corrected from extinction using stellar E(B--V) values from SED fitting. The extinction of the stars is then re-scaled to the extinction of the gas using the recipe E(B--V)$\rm_{stars}$ = 0.83\,E(B--V)$\rm_{gas}$ from \cite{Kashino:2013ev}}
\tablebib{
(a)~\citet{Planesas:1997A&A325}; (b) \citet{Bastian:2006fo}; (c) \citet{Liu:2013vg}; (d) \citet{2014A&A...568A..14A}; (e) \citet{Genzel:2011ApJ733}; 
(f) \citet{Alaghband-Zadeh:2012MNRAS424}; (g) \citet{Swinbank:2012ApJ760}; (h) \citet{Zanella:2015ej}.
}
\label{table:samples}
\end{table*}

\begin{figure*}[t]
\begin{center}
\resizebox{1.\hsize}{!}{\includegraphics[angle=0]{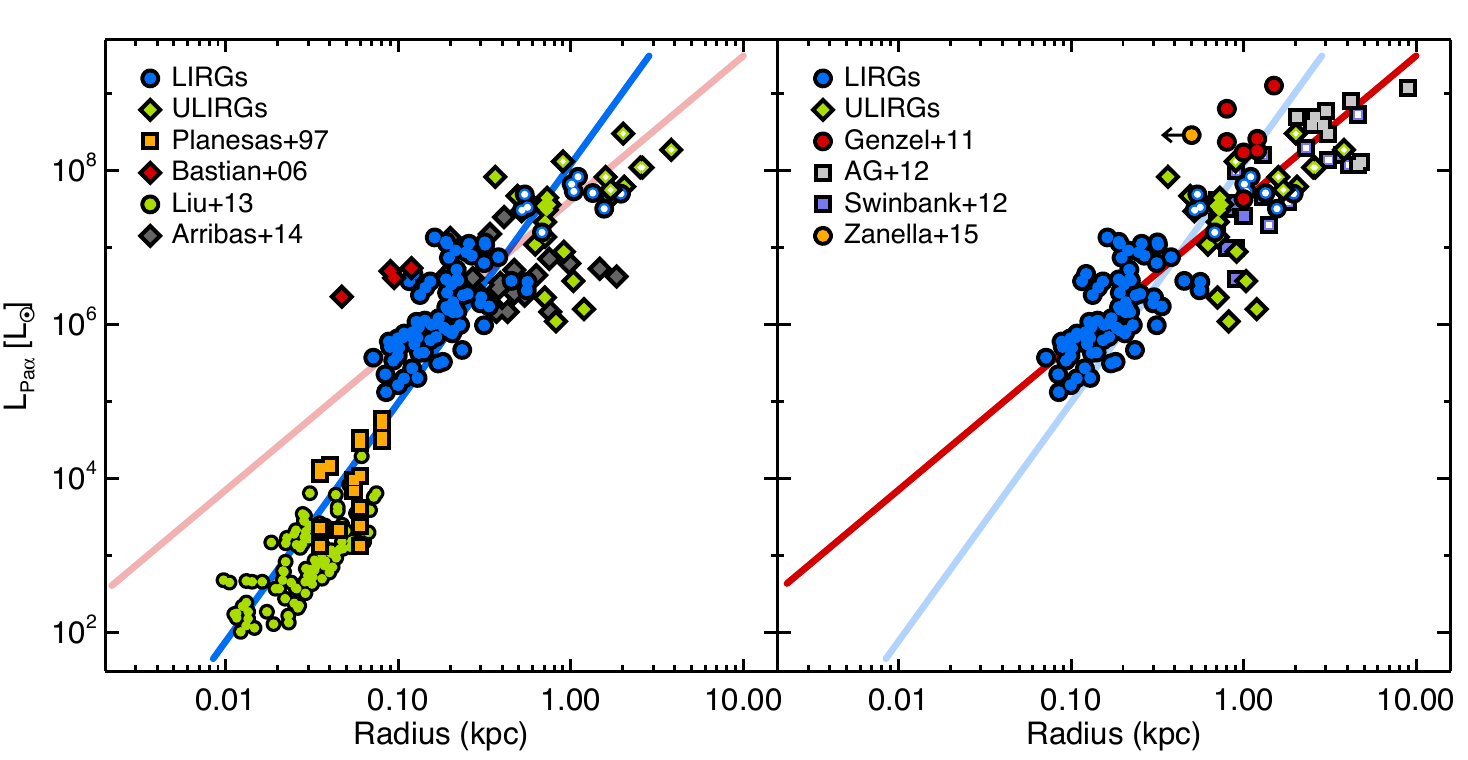}}
\caption{Dependence of the extinction-corrected \LPa\ with the radius of individual star-forming clumps and galaxies. The individual regions of LIRGs and ULIRGs are plotted as blue circles and green diamonds, respectively, whereas the integrated measurements of each object of our sample are marked using hollow symbols. {Uncertainties at 1$\sigma$ level of the LIRG and ULIRG measurements are smaller than the size of the symbols}. We show the comparison with local samples (left) and high-z samples (right). Further details on the local and high-z samples can be found in Table~\ref{table:samples}.
The blue and red lines correspond to power law fits $\rm L_{Pa\alpha} \sim r^{\eta}$ to our data of local LIRG and ULIRG clumps, together with the local samples ($\rm\eta=3.04$) and high-z points ($\rm\eta=1.88$), respectively.}
\label{sfr_rad_plots_corr}
\end{center}
\end{figure*}

\begin{figure}[t]
\begin{center}
\resizebox{1.\hsize}{!}{\includegraphics[angle=0]{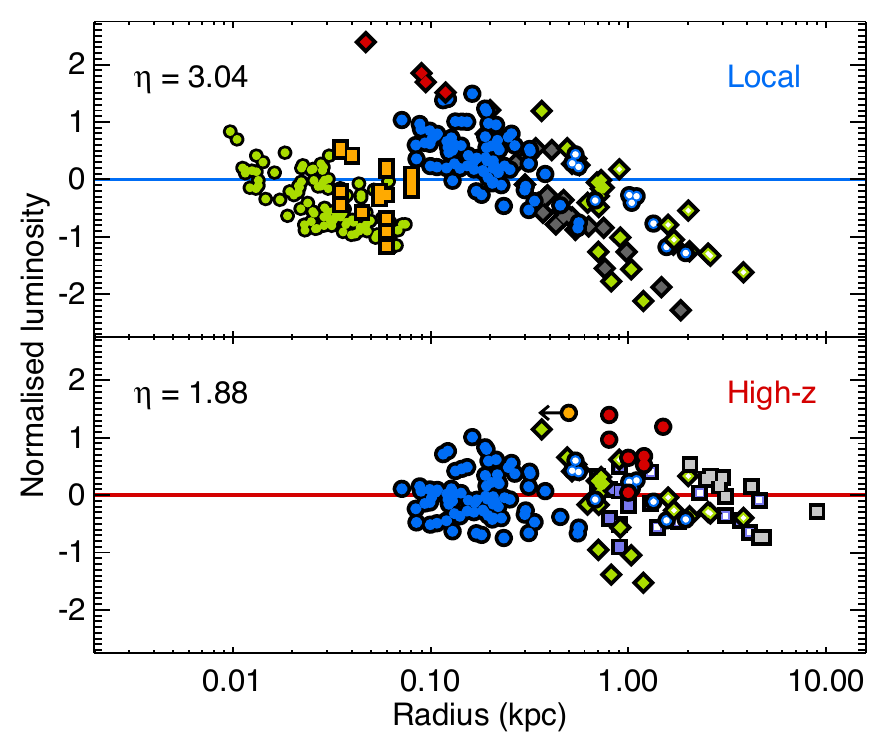}}
\caption{Normalised \LPa\ versus radius of individual star-forming clumps and galaxies, assuming the power-law fits from Fig.~\ref{sfr_rad_plots_corr} for U/LIRGs + local samples (top), and U/LIRGs + high-z samples (bottom). Symbols and colours are the same as in Fig.~\ref{sfr_rad_plots_corr}}
\label{sfr_rad_fits}
\end{center}
\end{figure}

We compared our measurements with different samples of star-forming clumps in nearby galaxies. Although most of the data from other samples correspond to \Ha\ measurements, we converted the extinction-corrected \Ha\ luminosities into \Pa\ luminosities using the Case B recombination factor at T$=10,000$\,K and n$\rm_{\rm e}=10^4\,cm^{-3}$ \citep{Osterbrock:2006AGN2} (\LHa/\LPa$\rm = 8.582$). Details of these different datasets taken from the literature can be found in Table~\ref{table:samples}. Since the complete sample from \cite{Liu:2013vg} consisted of $\sim$2 000 individual clumps, we randomly selected 100 individual clumps to obtain a subsample of similar size to the other samples. {To avoid any possible bias due to sampling selection, we repeated our analysis by extracting another 100 different subsamples of the \cite{Liu:2013vg} dataset, and checked that our results are not biased owing to selection effects}.

All the samples with the exception of \cite{Liu:2013vg} are based on \Ha\ measurements. As discussed in Sect.~\ref{section:optical_vs_ir}, even when the extinction corrections are applied, a factor  of $\sim$2--3 difference appears to exist between the \Ha- and \Pa-based SFR measurements in LIRGs and ULIRGs. This effect could be even larger when the correction from extinction is less accurate, such as object-averaged corrections, \Av\ measurements from different datasets, or SED fitting. Although the uncertainties of the extinction could be intrinsically large, the different methods used would yield additional sources of uncertainties when comparing the different datasets. To account for this effect, we used an additional average factor of 2.9 to translate the extinction-corrected \Ha\ luminosities into \Pa\ luminosities. The results are presented in Fig.~\ref{sfr_rad_plots_corr}.

The slope of the L-r relation is usually interpreted in terms of physical properties of the star-forming regions. If we assume that the regions are photon-bounded, they can be represented by a Str\"omgren sphere and their L-r relation is in the form $\rm L_{Pa\alpha} \sim r^{\rm\eta}$, with $\rm\eta=3$. The deviations from this model could result from a variety of factors, both physical and observational. If the regions are density-bounded, i.e. the hydrogen atoms recombine faster than they are ionised and some ionising photons escape, a shallower slope of the L-r is expected, since regions turn out to be less luminous at a given radius \citep{Wisnioski:2012MNRAS422}. \cite{Beckman:2000AJ119} present a detailed discussion of this transition and propose a luminosity threshold of \LPa$\gsim10^{5}$\,\Lsolar beyond which regions turn out to be density-bounded. Nevertheless, the exponent of this relation is subject to variations that could be unrelated to the intrinsic physical properties of the HII regions and are likely to be induced by observational biases as region-blending or low S/N ratio of the clumps. In addition, the different spatial resolution of the local and high-z observations might also contribute to change the slope of the L-r relation (\citealt{Scoville:2001AJ122}, \citealt{Liu:2013vg}). Despite this observational bias, the use of either \Ha\ or \Pa\ data to perform the analysis seems to have little impact on $\rm\eta$ \citep{Liu:2013vg}.

As shown in Fig.~\ref{sfr_rad_plots_corr}, if we compare the \Si\ and \LPa\ of our LIRG regions, we find that they have higher luminosity densities than clumps in local so-called normal galaxies, although they seem to be similar to the regions from the ongoing spiral-spiral merger of the Antennae \citep{Bastian:2006fo}. This is consistent with previous results (e.g. \citealt{AlonsoHerrero:2002p124}) who found that U/LIRGs show a large fraction of giant HII regions that are not just aggregations of normal HII regions. 

The blue line in the left panel of Fig.~\ref{sfr_rad_plots_corr} correspond to the power law fitting $\rm L_{Pa\alpha} \sim r^{\eta}$ to the local points, i.e. U/LIRGs and local samples, which yields an exponent of $\rm\eta=3.04${ and a correlation coefficient of $\rm r=0.967$}. This value is in very good agreement with what  would be expected for photon-bounded regions, represented by Str\"omgren spheres. However, there seems to be a clear change on the slope at r$\sim60$\,pc, which corresponds approximately to \LPa$\sim10^{5}$\,\Lsolar, and which coincides with the proposed luminosity threshold from photon- to density-bounded regions. Although the lack of low-luminosity regions in our sample precludes a more detailed analysis of this transition using a consistent dataset, the combination of the different samples points to a change in the regime around \LPa$\sim10^{5}$\,\Lsolar.

We combined our sample of clumps and galaxies with the more general sample from \cite{2014A&A...568A..14A} of local LIRGs and ULIRGs observed in the optical. When only LIRGs and ULIRGs are considered, star-forming regions seem to be significantly more luminous than predicted for their radius range and the resulting slope of the L-r relation yields $\rm\eta\sim2.02$ {(correlation coefficient of $\rm r=0.895$)}, which is therefore compatible with density-bounded regions.

\subsection{Star-forming clumps in U/LIRGs: comparison with high-z giant star-forming clumps and SMGs}
\label{section:clumps_highz}

We investigate how the observed properties of the star-forming clumps in U/LIRGs compare with those detected in high-z luminous star-forming galaxies. In particular, a comparison is made with the giant star-forming clumps and the integrated properties for a sample of sub-millimeter galaxies (SMG). The details for these high-z samples taken from the literature are summarized in Table~\ref{table:samples}. As discussed in Appendix~\ref{sec:distance} and previous sections, the spatial resolution of the data could be playing a role in deriving the observed properties. Although we explored this effect at low redshifts ($\rm z<0.2$), our dataset is not well suited to study and simulate the effect of distance at higher redshifts, given the limited FoV of our SINFONI observations. {Therefore, we restricted the comparison at clump level only for the AO-based samples of \citet{Genzel:2011ApJ733},  \cite{Swinbank:2012ApJ760}, and the z$\sim$2 star-forming clump of \cite{Zanella:2015ej}, while a comparison of the integrated properties of galaxies is considered only for our ULIRG sample, the integrated measurements of  \cite{Swinbank:2012ApJ760}, and the high-z SMG sample of Alaghband-Zadeh and collaborators (see Fig.~\ref{sfr_rad_plots_corr}, right panel).}  

Star-forming clumps in LIRGs appear as significantly ($\sim$3-20 times) smaller than the kpc-size clumps detected so far in high-z galaxies, and up to 1 000 times less luminous. On the other hand, clumps in ULIRGs have sizes similar ($\times$0.5--1) to those of high-z clumps, having \Pa\ luminosities similar to some high-z clumps \citep{Swinbank:2012ApJ760} and  about ten times less luminous than the most luminous high-z clumps identified so far \citep{Genzel:2011ApJ733}. These results suggest that the giant clumps in high-z star-forming galaxies appear to be forming stars with a higher surface density rate than ULIRGs, and in particular LIRGs. This could be the consequence of a higher molecular gas surface density in high-z galaxies, and therefore a higher star formation rate than expected, if the empirical Kennicutt-Schmidt law that seems valid on a galaxy scale also applies at the (sub)kpc scales in the highly luminous and dense environments of U/LIRGs and/or high-z galaxies. However, a recent analysis of the star-forming clumps in the low-z LIRG IC 4687, combining ALMA CO measurements with \Pa\ HST imaging on scales of 250\,pc \citep{PereiraSantaella:2016A&A587A}, shows a large scatter in the \Pa\ luminosity for a given mass of molecular gas, i.e. an effective break of the KS law at (sub)kpc scales in the high  (500--2500\,M$_{\sun}$ pc$^{-2}$) H$_2$ surface-density regions present in this galaxy. In practice, this indicates the coexistence of massive clumps of molecular gas that have been actively forming stars over the last few million years, with others that could form stars on  longer time scales but at a  much lower rate than expected. Moreover, some clumps with a relatively low molecular gas surface density appear to be forming stars at rates similar to those of the most dense clumps. {The slope of the L-r relation of the U/LIRGs and high-z clumps is $\rm\eta\sim1.88$, and a correlation coefficient of $\rm r=0.890$}. The large scatter in the \Pa\ luminosity among the high-z clumps (Fig.~\ref{sfr_rad_plots_corr}) could indicate a similar breaking of the KS law at high-z. The sample of high-z clumps is, however, still very limited and consists of a few clumps in luminous  (SFR $\sim$ 120--290\,M$_{\sun}$ yr$^{-1}$, \citealt{Genzel:2011ApJ733}), and fainter (SFR$\sim$16\,M$_{\sun}$ yr$^{-1}$, \citealt{Swinbank:2012ApJ760}) extended star-forming disks. A proper comparison of low- and high-z luminous star-forming galaxies with larger samples is required to investigate more deeply the differences of the star formation on (sub)kpc scales. 

The size and \Pa\ luminosity of star formation in low-z ULIRGs on galaxy scales is compared with that of high-z galaxies (Fig.~\ref{sfr_rad_plots_corr}), including extended disks \citep{Swinbank:2012ApJ760} and mergers \citep{Alaghband-Zadeh:2012MNRAS424}. Galaxies have sizes between 1 to 10\,kpc, from the most compact star-forming regions to interacting galaxies, where the global size is the combined luminosity of the system, and it is therefore affected by the projected distance of the galaxies involved in the interaction/merger and the distribution of the giant clumps within each of the galaxies.  Taking this into account, low-z mergers (i.e. local ULIRGs) appear to be slightly more compact than the high-z mergers \citep{Alaghband-Zadeh:2012MNRAS424} although the number of high-z sources is too small to derive firmer conclusions.

\section{Summary}
\label{section:summary_sfr}

We presented a detailed 2D study of star formation in a representative sample of local LIRGs and ULIRGs, based on VLT-SINFONI IFS K-band observations. The data sample the inner $\sim3\times3$\,kpc and $\sim12\times12$\,kpc in LIRGs and ULIRGs, respectively, with an average linear resolution of $\sim$0.2\,kpc and $\sim$0.9\,kpc (FWHM), respectively. The analysis of the SFR and \Si\ is performed using a spaxel-by-spaxel extinction correction, based on measurements of the \Brg/\Brd\ and \Pa/\Brg\ line ratios for LIRGs and ULIRGs, respectively (Paper~II).

The integrated near-IR based SFR of the objects of our sample has been compared with optical, mid-IR and \LTIR-based measurements. We found that the observed \SFRHa\ and \SFRPa\ values differ by a factor of $\sim$3.6 on average, and that the difference decreases slightly when the extinction correction is applied, up to \SFRHa/\SFRPa$\sim$2.9. In agreement with previous studies, we observed a tight linear correlation between the extinction-corrected \SFRPa\ and the 24\,$\mu$m measurements from \emph{Spitzer} and a reasonable agreement with SFR values from \LTIR, with differences of up to factors 2-3.

We obtained 2D \Si\ maps and spaxel-by-spaxel distributions of  individual galaxies (Figs.~\ref{figure:LIRG} and \ref{figure:ULIRG}). We found that, in a significant fraction of the objects ($\sim$57\,\%), the corrected \Si\ peaks either at the main or secondary nucleus of the systems, and that the extinction correction typically increases the median of the distributions by $\sim$50\,\%. Within the angular resolution and  sizes sampled by the SINFONI observations, we found that the LIRG subsample has median observed star formation surface brightness of $\rm\Sigma_{\rm LIRGs}^{\rm obs}=1.16$\,\Msolar\,yr$^{-1}$\,kpc$^{-2}$ and $\rm\Sigma_{\rm LIRGs}^{\rm corr}=1.72$\,\Msolar\,yr$^{-1}$\,kpc$^{-2}$ for the extinction-corrected distribution. The median observed and the extinction-corrected \Si\ values for ULIRGs  are $\rm\Sigma_{\rm ULIRGs}^{\rm obs}=0.16$\,\Msolar\,yr$^{-1}$\,kpc$^{-2}$ and $\rm\Sigma_{\rm ULIRGs}^{\rm corr}=0.23$\,\Msolar\,yr$^{-1}$\,kpc$^{-2}$, respectively. These median values for ULIRGs increase by up to 1.38\,\Msolar\,yr$^{-1}$\,kpc$^{-2}$ and 2.90\,\Msolar\,yr$^{-1}$\,kpc$^{-2}$ when only their inner regions, covering the same size as the average FoV of LIRGs, are considered.

The spatial sampling (i.e. physical scale per spaxel) of the observations shapes not only the \Av\ structure (Paper~II) but also the \Si\ distributions. Our simulations of the \Si\ maps of the LIRGs at increasing distances show that the predicted median of the \Si\ distributions is artificially increased by the poorer sampling of the maps. At the average distance of the ULIRG subsample (328\,Mpc), the computed median of the LIRG observed and extinction-corrected \Si\ distributions is a factor $\sim$2--3 larger than the original measurements. This has consequences for any estimates of  star formation surface brightness in high-z galaxies and, consequently, on the derivation of the universality of star formation laws at all redshifts.

We identified a total of 95 individual star-forming clumps in the \Brg\ and \Pa\ emission maps of LIRGs and ULIRGs. These regions present a range of sizes from $\sim$60 to 400\,pc and from $\sim$300 to 1500\,pc in LIRGs and ULIRGs, with \Pa\ luminosities of $\sim$10$^{5}$--10$^{7}$\,\Lsolar\ and $\sim$10$^{6}$--10$^{8}$\,\Lsolar, respectively. The \Si\ of the clumps presents a wide range of values within 1--90\,\Msolar\,yr$^{-1}$\,kpc$^{-2}$ and 0.1--100\,\Msolar\,yr$^{-1}$\,kpc$^{-2}$ for LIRGs and ULIRGs.

Star-forming clumps in LIRGs are about ten times larger and a thousand times more luminous than typical clumps in spiral galaxies, consistent with expected photon-bounded conditions in ionized nebulae that surround young stellar clusters.  Clumps in ULIRGs have sizes similar ($\times$0.5--1) to those of high-z clumps, having \Pa\ luminosities similar to some high-z clumps \citep{Swinbank:2012ApJ760} and  about 10 times less luminous than the most luminous high-z clumps identified so far \citep{Genzel:2011ApJ733}. This could be an indication that the most luminous giant clumps in high-z star-forming galaxies could be forming stars with a higher surface density rate than low-z compact ULIRGs.

We also observed a change in the slope of the L-r relation, from $\rm\eta=3.04$ of local samples to $\rm\eta=1.88$ from high-z observations. A likely explanation is that most luminous galaxies are interacting and mergers and, therefore, their size represents a combination of the distribution of the star-forming clumps within each galaxy in the system, plus the additional effect of the projected distance between the galaxies, therefore producing an overall larger size  than that of individual clumps or galaxies (for integrated measurements).

\appendix
\section{The effect of the linear resolution on the \Si\ measurements. Implications for high-z galaxies}
\label{sec:distance}

\begin{figure*}[t]
\begin{center}
\resizebox{\hsize}{!}{\includegraphics[angle=0]{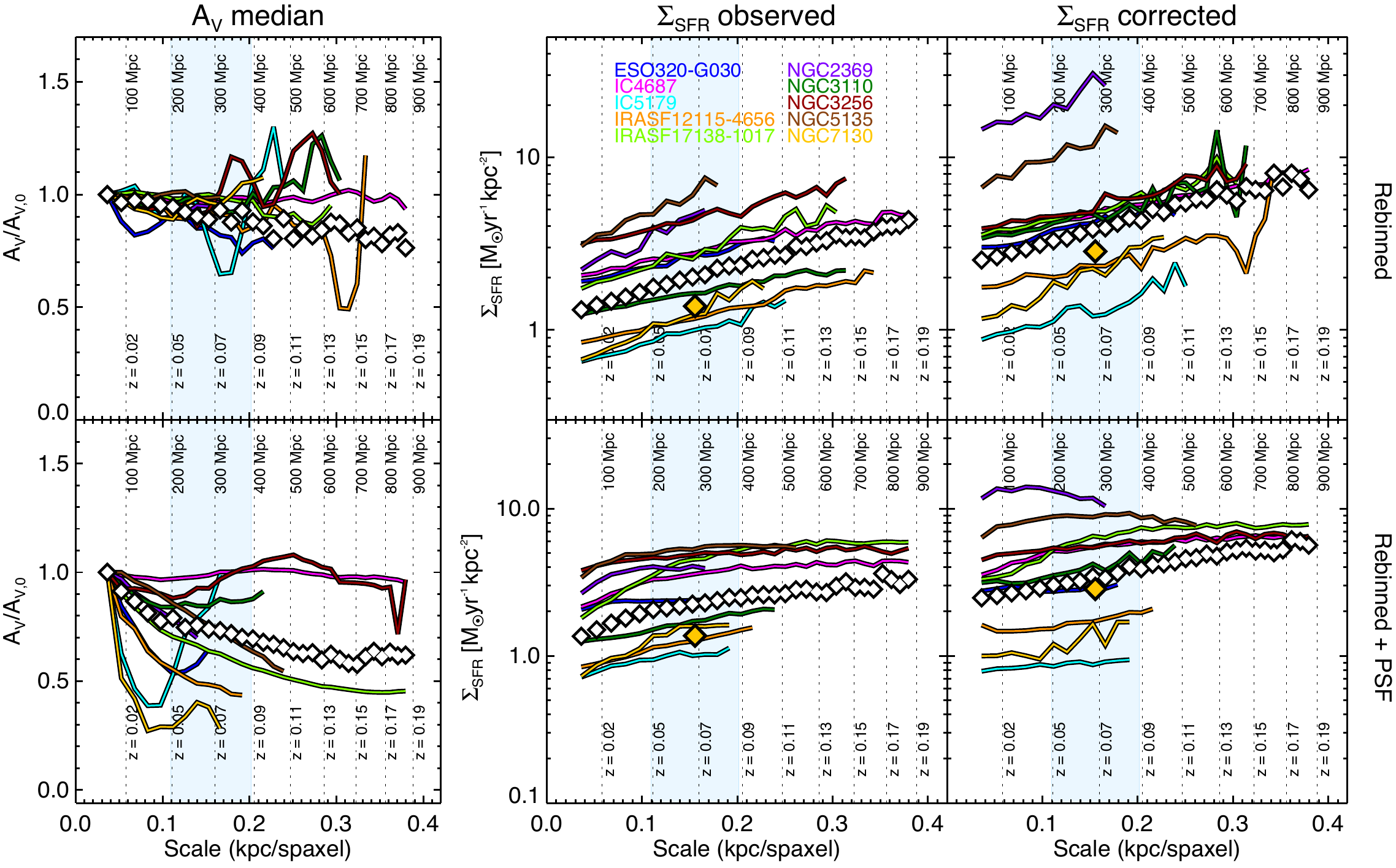}}
\caption{Evolution of the computed median \Av\ and \Si\ of the spaxel-by-spaxel LIRG distributions, as a function of distance/scale. \emph{Upper panels} correspond to  simulations where the maps are only rebinned, whereas \emph{lower panels} show the results from rebinning and smearing the maps as a result of the effect of seeing (see text for details). Colour lines represent the evolution of the median of each individual distribution, whereas white diamonds show the median of the whole LIRG distribution. The median \Si\ values of the ULIRG distributions within $\rm r_{LIRGs}=1.4\,kpc$ are also plotted as yellow diamonds, for reference.}
\label{sim_indiv_00}
\end{center}
\end{figure*}

Owing to the clumpy structure of the dust observed in LIRGs and ULIRGs at sub-kiloparsec and kiloparsec scales, the measurement of  internal extinction, and subsequent derived quantities, could be affected by the physical scale of the observations at increasing distance (Paper~II). Fig.~\ref{figure:LIRG} shows that the morphology of the ionized gas within the inner $\sim$3$\times$3\,kpc in LIRGs is mainly in the form of nuclear star formation rings, emission associated with the nucleus of the galaxy, and bright HII regions with typical sizes of a few hundred parsecs (see also \citealt{AlonsoHerrero:2006p4703}), some of them barely resolved in our observations. In addition, the \Av\ maps show that dust seems to be distributed in a patchy and clumpy structure at the same sub-kiloparsec scales. As shown in Fig.~\ref{figure:ULIRG}, ULIRGs show a more compact morphology of both dust and ionized gas phases, with unresolved structures beyond $\lsim$1\,kpc, which suggests that the reduced linear resolution of the observations might play a role in shaping the gas and dust distributions.

To probe how the differences in the pixel scale might bias the \Si\ measurements, we simulated the \Brg\ and \Brd\ maps of the LIRGs at increasing distances. We performed two different sets of simulations, the first set considers only the effect of the decreasing spatial sampling of the maps, whereas the second set also takes the smearing effect owing to the seeing into account. In the first set of simulations, we rebinned each individual map as if it was observed at increasing distances with the same instrument set-up. The simulated maps are therefore sampled by a decreasing number of spaxels with a constant angular resolution of 0$\farcs$125 per spaxel, as the original maps. The rebinning process is performed using the basic equations of angular diameter distance and luminosity distance and assuring that the total luminosity within the FoV is conserved. In the second set of simulations, before the rebinning process, we considered the effect of the seeing to obtain more realistic emission maps that could be directly compared with our SINFONI observations of the ULIRG subsample. The resulting maps presents the same spatial sampling and scale as those from the first set of simulations, but with a $\sim$0.63\,arcsec (FWHM) seeing as in the original observations. {Finally, in both sets of simulations, we have removed those spaxels with flux below $\rm10^{-18}\,erg\,s^{-1}\,cm^{-2}$from our maps, which corresponds to the detection limit of our original observations (see Sect.~\ref{section:sfr}).}

Once the simulated maps are built, we followed the same procedure as described in Sect.~\ref{section:analysis_sfr} to obtain a spaxel-by-spaxel \Av\ correction, and finally an extinction-corrected \Si\ distribution. As discussed in Paper~II, we typically sample  the innermost $\sim$3$\times$3\,kpc of the LIRGs, i.e. $\sim$\Reff. Since the average distance of the ULIRG subsample is $\times$10 larger, the whole FoV of the LIRGs are equivalent to $\sim$1" of the ULIRG FoV, i.e. $\sim$8\,spaxels. These limitations and the lack of information about the external regions of the LIRGs makes the extrapolation to larger distances to be not straight-forward.

The evolution of the median values of the simulated \Si\ observed and corrected distributions is shown in Fig.~\ref{sim_indiv_00}. We observe a clear increment of the median \Si\ with increasing distance in both the observed and corrected distributions, up to $\sim$90\% at 328\,Mpc, the average distance of the ULIRG subsample. We observe that, in most  cases, the predicted median of the LIRG distribution is slightly higher than the median of the ULIRG distribution. However, there is  good agreement between both quantities when the extinction correction is applied and the PSF effects are taken into account. Although the simulations are a first-order approximation, this result suggests that the \Si\ measurements of the ULIRGs may be affected by such distance effects and a further investigation is needed to characterise these effects in detail.

\section{Figures and tables}
\label{sec:tables&figures}

\clearpage

\setcounter{fake_fig}{\value{figure}}
\refstepcounter{figure}
\renewcommand{\thefigure}{\Alph{section}.\arabic{figure}\alph{subfig}}
\label{figure:LIRG}

\setcounter{figure}{0}
\setcounter{subfig}{1}

\begin{figure*}[t]
\begin{center}
\begin{tabular}{c}
\includegraphics[angle=0, width=0.93\hsize]{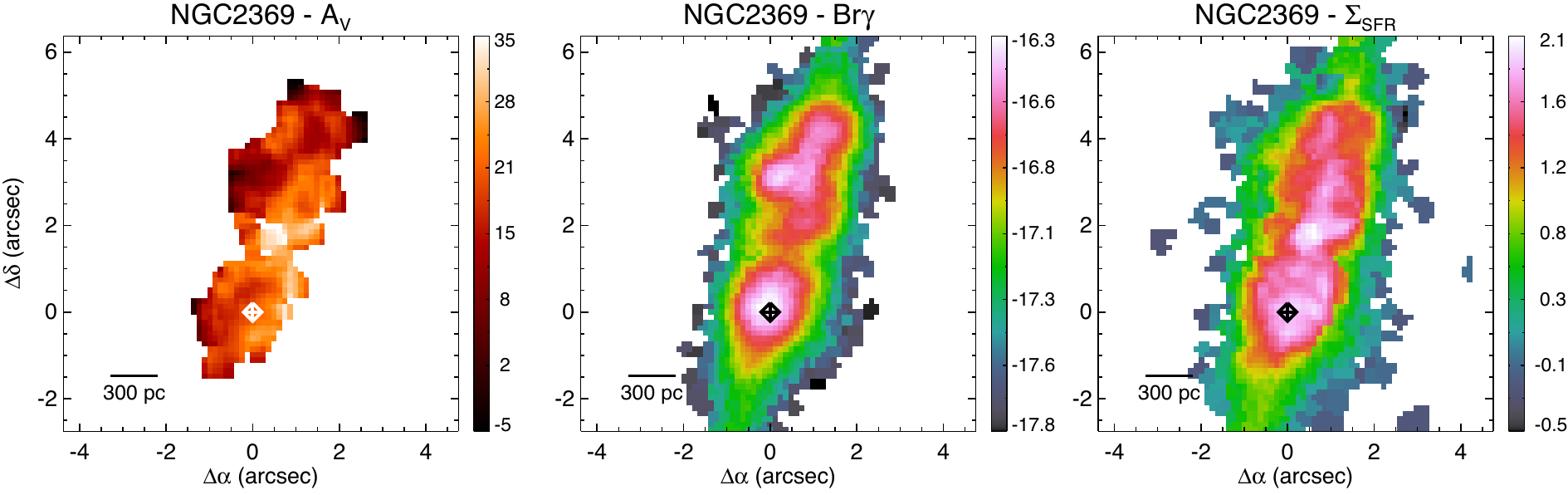} \\
\includegraphics[angle=0, height=.24\hsize]{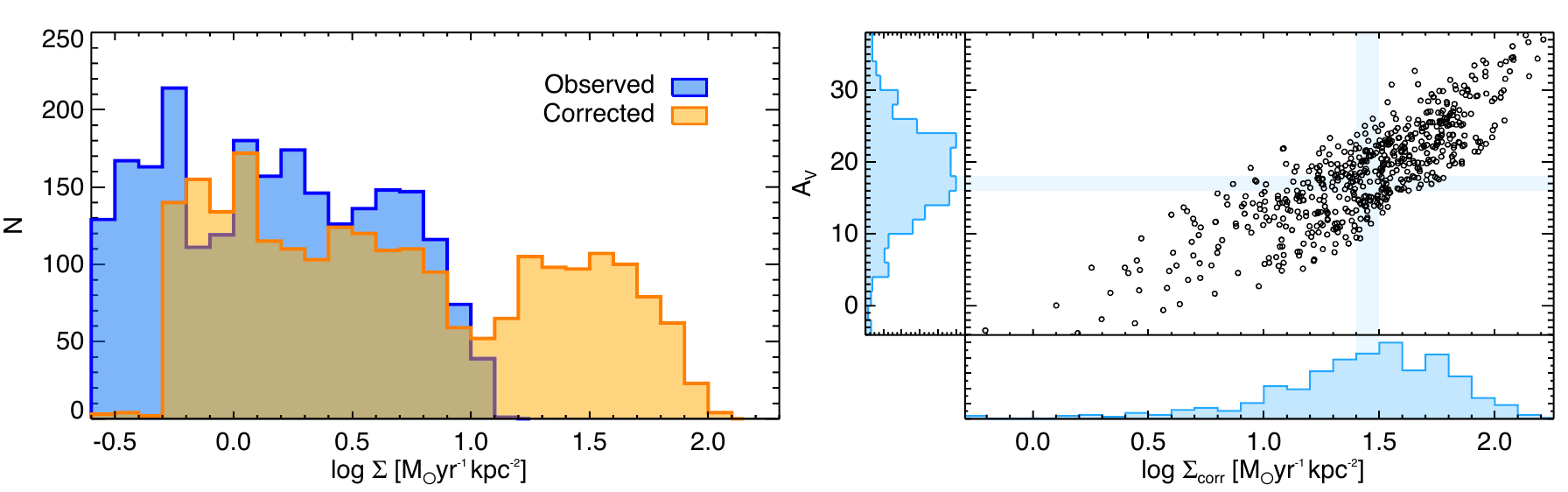} \\
\end{tabular}
\end{center}
\caption{\tiny \object{NGC 2369}. Top panels show the \Av\ map derived from the \Brgl\ and \Brdl\ line ratio, the observed maps of the \Brg\ emission, together with the star formation rate surface density (\Si) map, corrected from extinction. Units are [mag], log[$\rm erg\,s^{-1}\,cm^{-2}$], and [$\rm M_{\sun}\,yr^{-1}\,kpc^{-2}$], respectively. The nucleus and \Brg\ peak are indicated with a plus sign (+) and a diamond ($\Diamond$), respectively. The main nucleus is defined as the brightest spaxel of the SINFONI K band image (Paper~I), and the \Brg\ (\Pa) peak corresponds to the brightest spaxel of the corresponding emission map. Bottom left panel shows the observed (blue histogram) and the corrected-from-extinction (yellow histogram) \Si\ spaxel-by-spaxel distributions. The relationship between the corrected \Si\ values and the \Av\ is shown in the bottom right panel only for those points with a spaxel-by-spaxel correction of the extinction. The blue histograms show the projected distribution onto each axis and are arbitrarily normalised, whereas the blue lines are the median of each distribution.}
\label{figure:NGC2369}
\end{figure*}

\addtocounter{figure}{-1}
\addtocounter{subfig}{1}
\begin{figure*}[!hb]
\begin{center}
\begin{tabular}{c}
\includegraphics[angle=0, width=0.93\hsize]{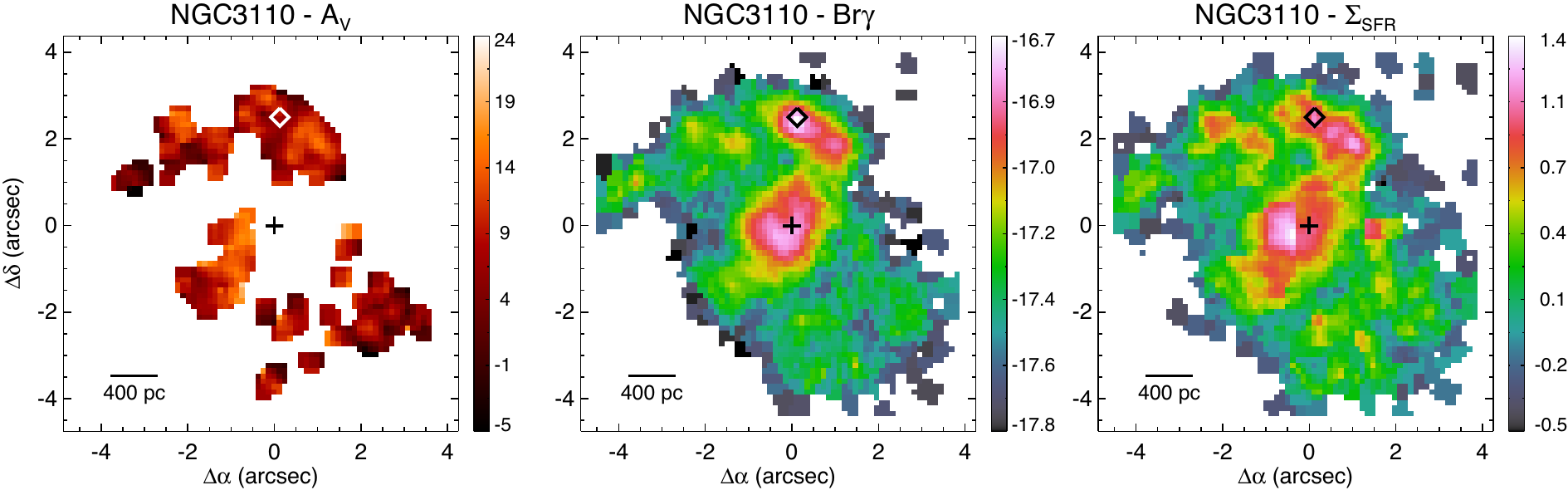} \\
\includegraphics[angle=0, height=.24\hsize]{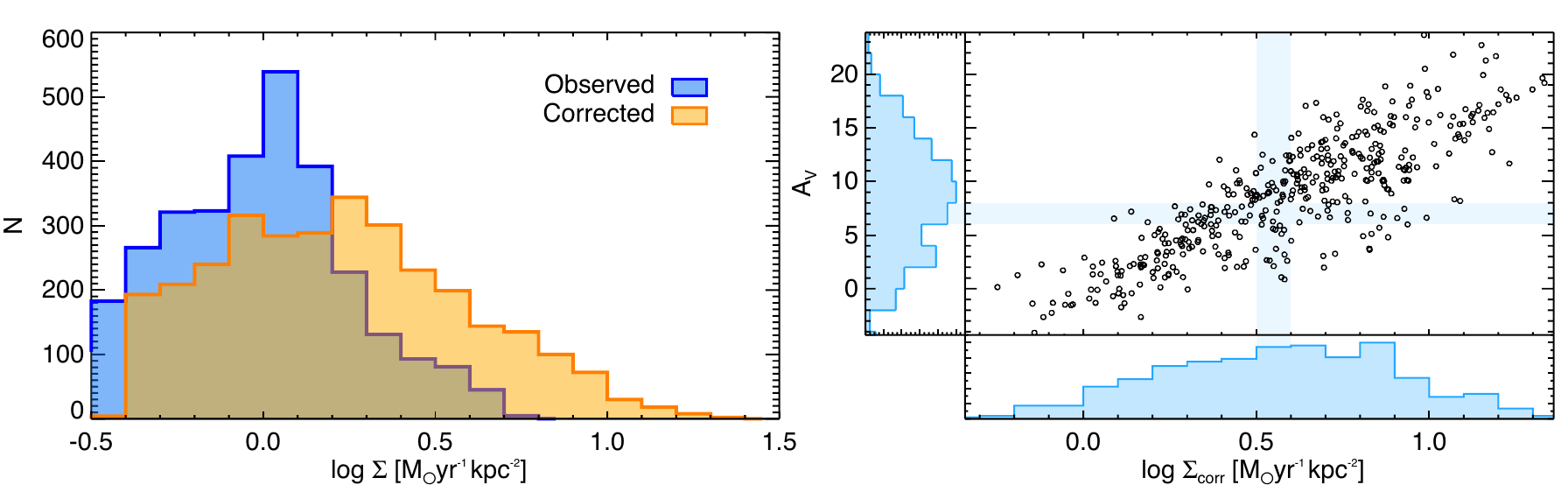} \\
\end{tabular}
\end{center}
\caption{\object{NGC 3110}. As for Fig.~\ref{figure:NGC2369} but for \object{NGC 3110}}
\label{figure:NGC3110}
\end{figure*}

\addtocounter{figure}{-1}
\addtocounter{subfig}{1}
\begin{figure*}[t]
\begin{center}
\begin{tabular}{c}
\includegraphics[angle=0, width=0.98\hsize]{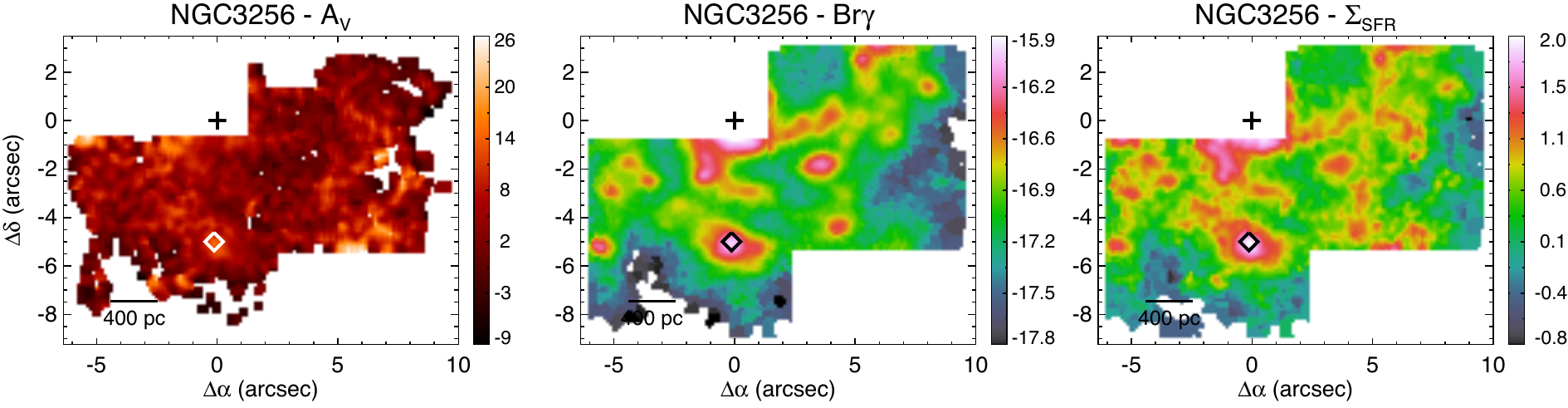} \\
\includegraphics[angle=0, height=.27\hsize]{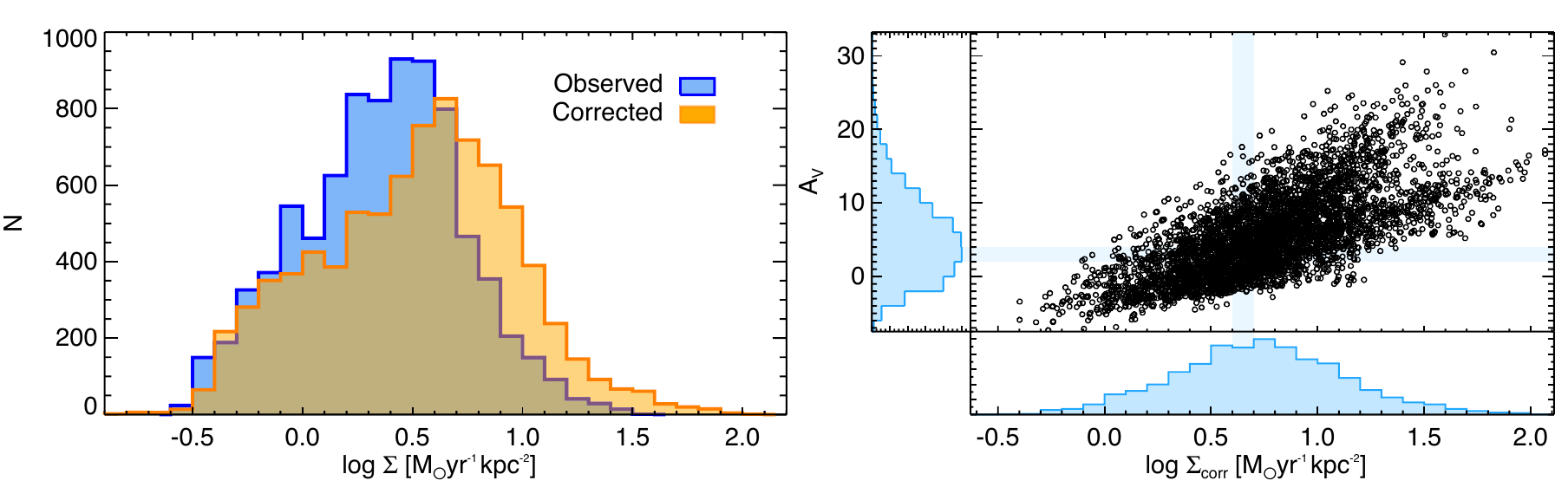} \\
\end{tabular}
\end{center}
\caption{\object{NGC 3256}. As for Fig.~\ref{figure:NGC2369} but for \object{NGC 3256}. Please note that the central spaxel lies outside the FoV since the nucleus was not observed in the K band. See \cite{Piqueras2012A&A546A} for further details.}
\label{figure:NGC3256}
\end{figure*}

\addtocounter{figure}{-1}
\addtocounter{subfig}{1}
\begin{figure*}[!hb]
\begin{center}
\begin{tabular}{c}
\includegraphics[angle=0, width=0.98\hsize]{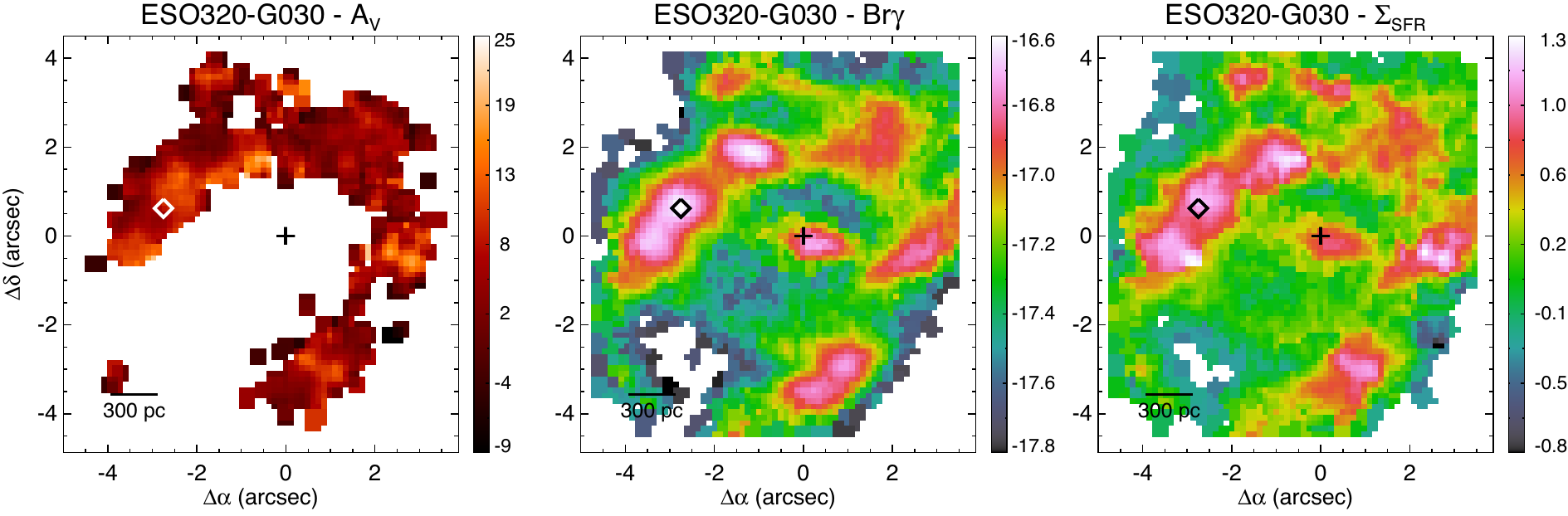} \\
\includegraphics[angle=0, height=.27\hsize]{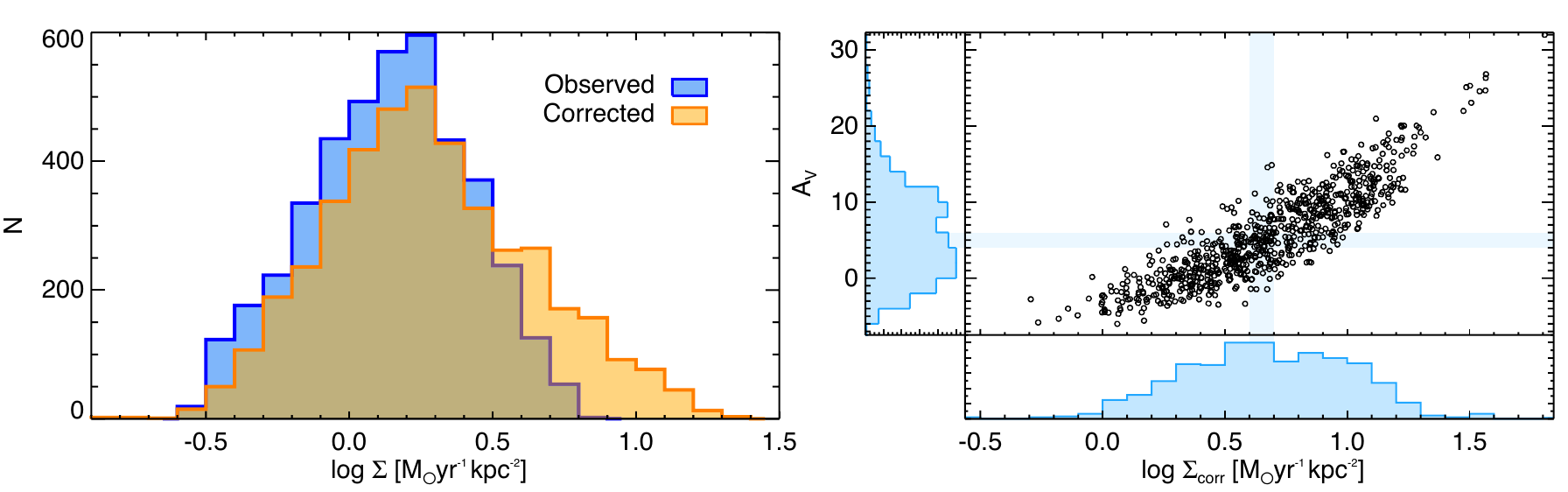} \\
\end{tabular}
\end{center}
\caption{\object{ESO 320-G030}. As for Fig.~\ref{figure:NGC2369} but for \object{ESO 320-G030}}
\label{figure:ESO320}
\end{figure*}

\addtocounter{figure}{-1}
\addtocounter{subfig}{1}
\begin{figure*}[t]
\begin{center}
\begin{tabular}{c}
\includegraphics[angle=0, width=0.98\hsize]{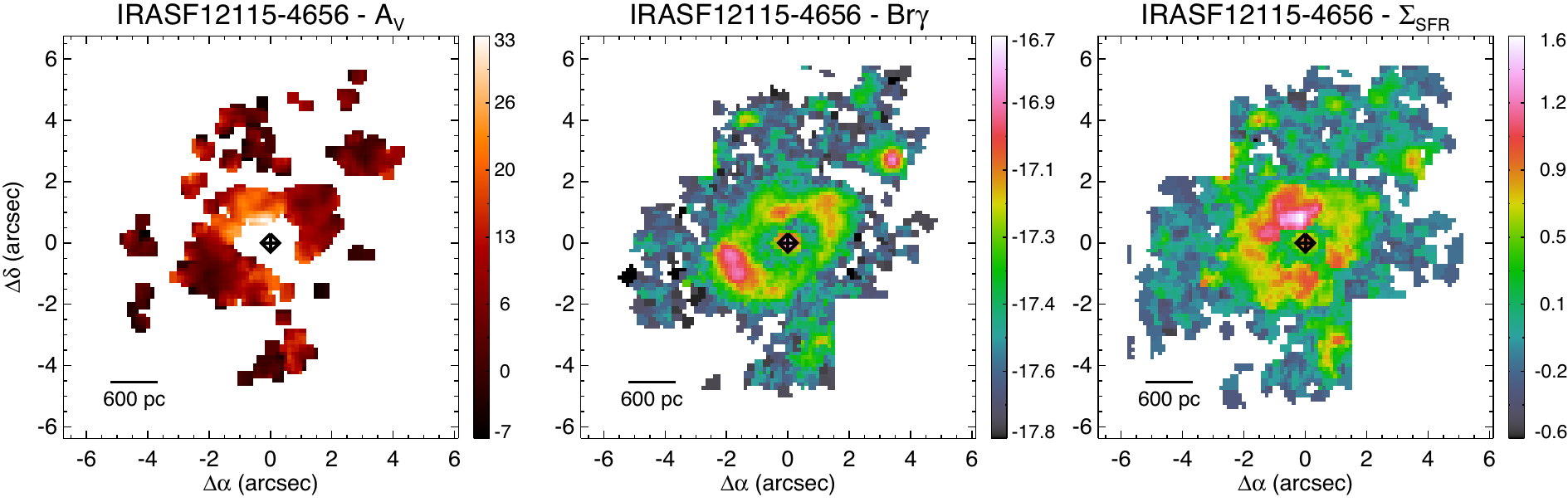} \\
\includegraphics[angle=0, height=.27\hsize]{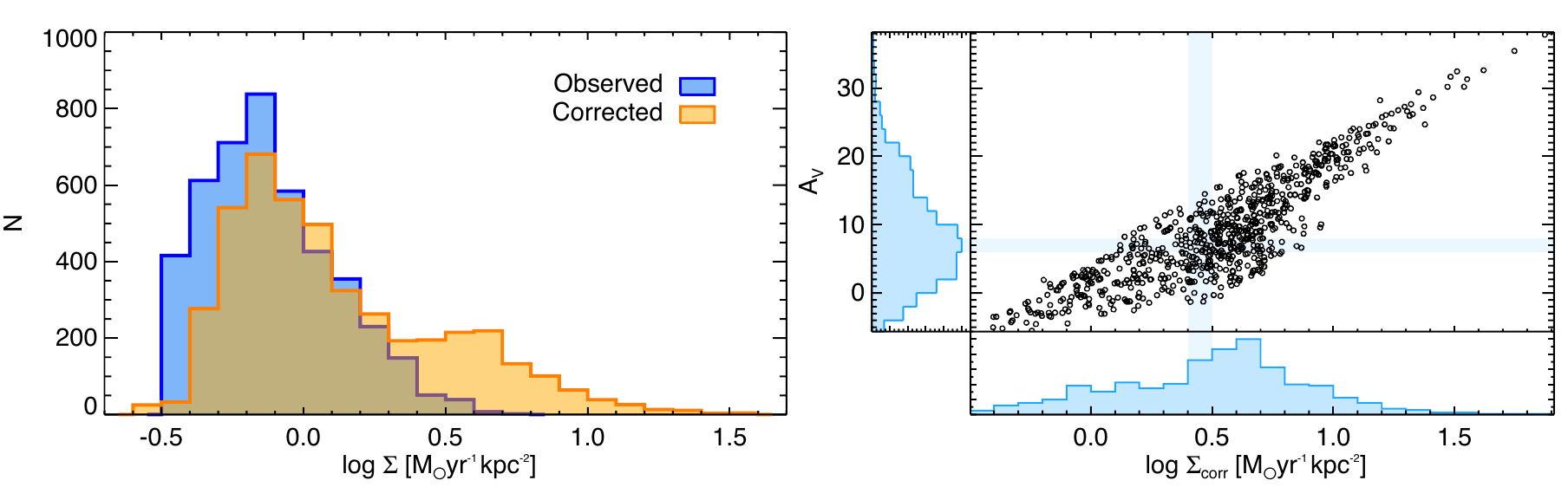} \\
\end{tabular}
\end{center}
\caption{\object{IRASF 12115-4656}. As for Fig.~\ref{figure:NGC2369} but for \object{IRASF 12115-4656}}
\label{figure:IRASF12115}
\end{figure*}

\addtocounter{figure}{-1}
\addtocounter{subfig}{1}
\begin{figure*}[!hb]
\begin{center}
\begin{tabular}{c}
\includegraphics[angle=0, width=0.98\hsize]{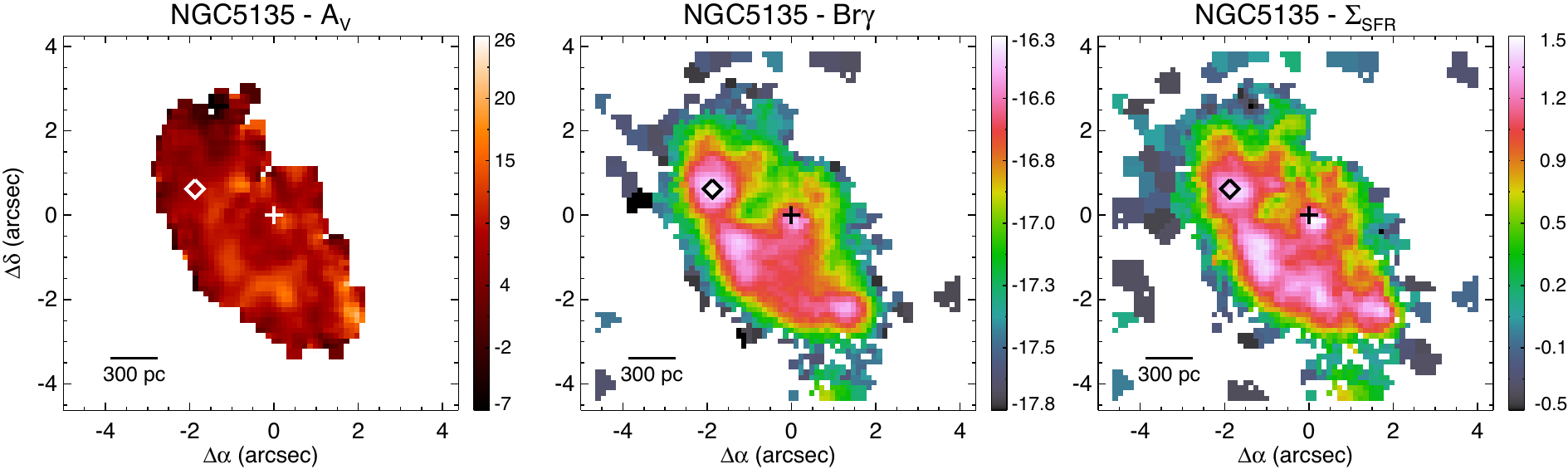} \\
\includegraphics[angle=0, height=.27\hsize]{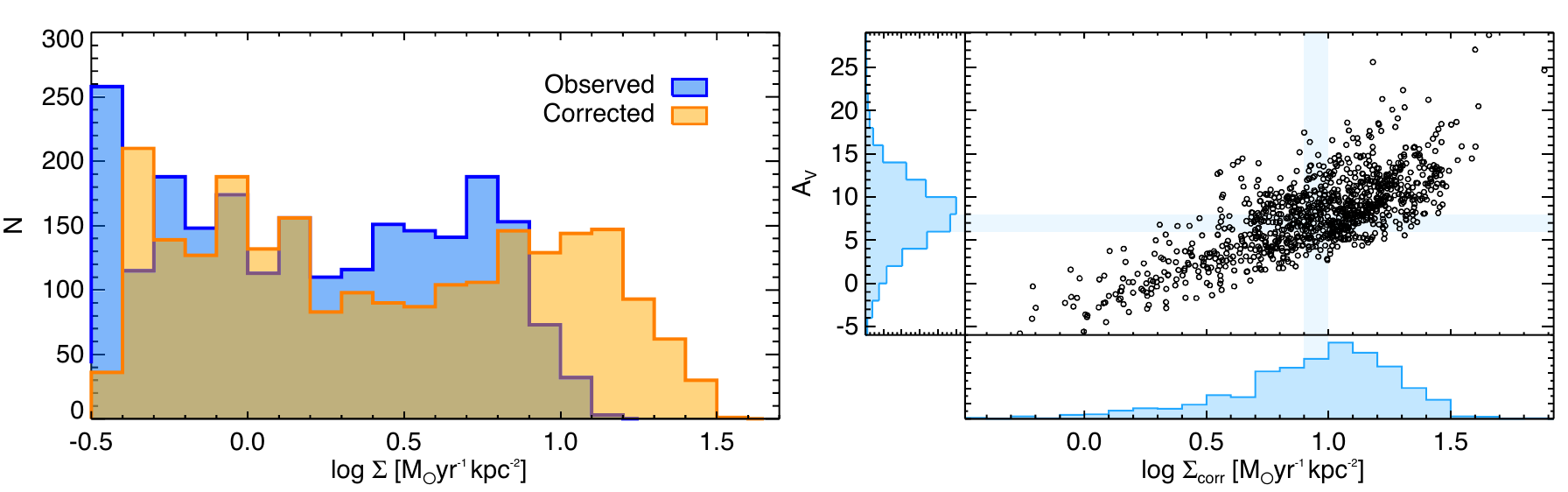} \\
\end{tabular}
\end{center}
\caption{\object{NGC 5135}. As for Fig.~\ref{figure:NGC2369} but for \object{NGC 5135}}
\label{figure:NGC5135}
\end{figure*}

\addtocounter{figure}{-1}
\addtocounter{subfig}{1}
\begin{figure*}[t]
\begin{center}
\begin{tabular}{c}
\includegraphics[angle=0, width=0.98\hsize]{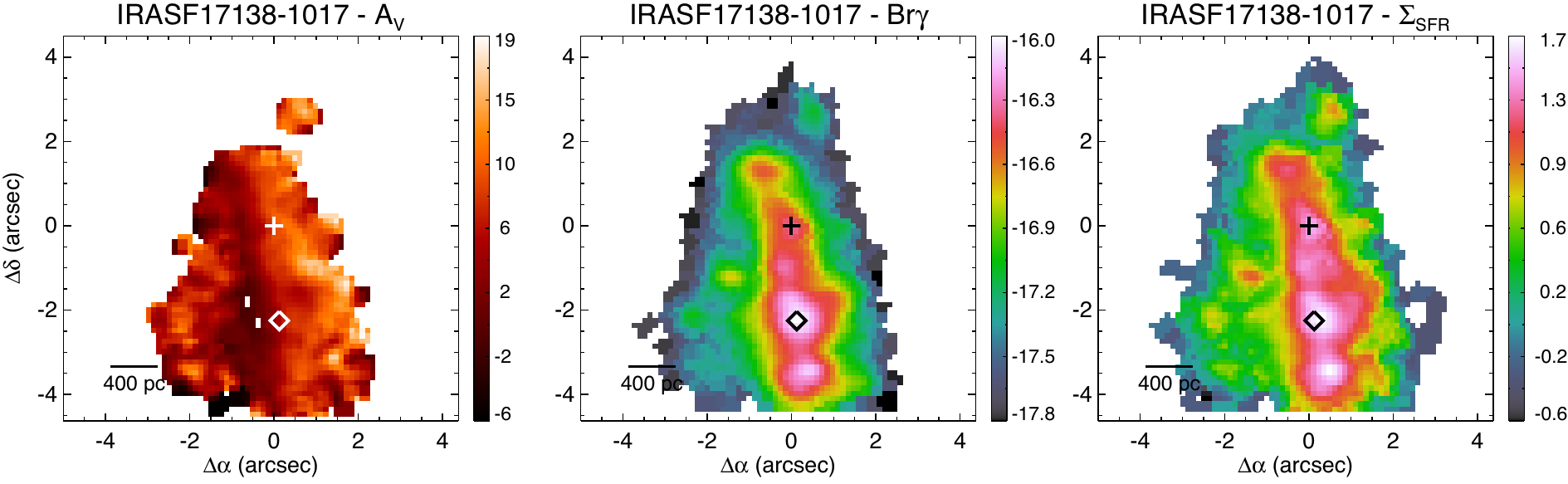} \\
\includegraphics[angle=0, height=.27\hsize]{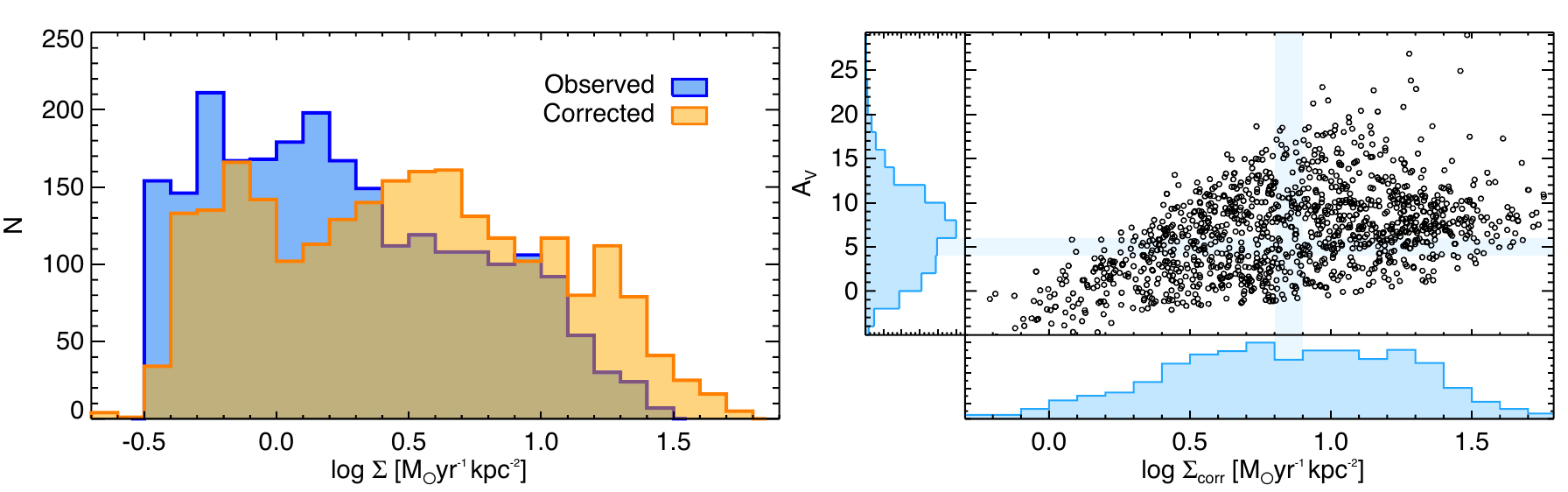} \\
\end{tabular}
\end{center}
\caption{\object{IRASF 17138-1017}. As for Fig.~\ref{figure:NGC2369} but for \object{IRASF 17138-1017}}
\label{figure:IRASF17138}
\end{figure*}

\addtocounter{figure}{-1}
\addtocounter{subfig}{1}
\begin{figure*}[!hb]
\begin{center}
\begin{tabular}{c}
\includegraphics[angle=0, width=0.98\hsize]{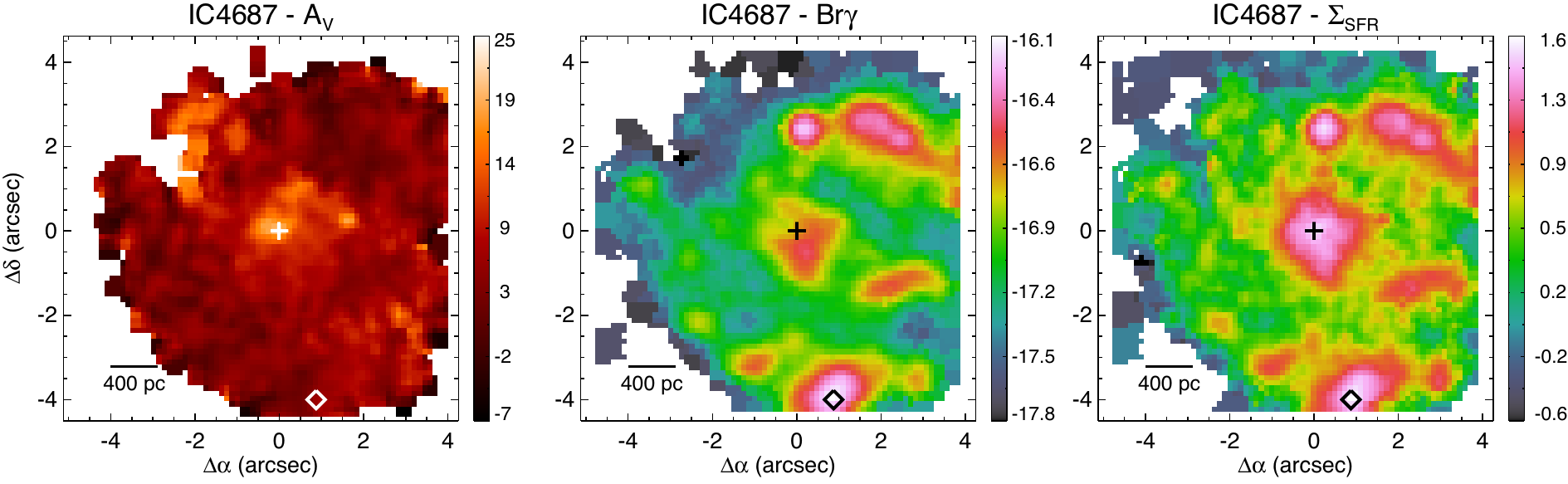} \\
\includegraphics[angle=0, height=.27\hsize]{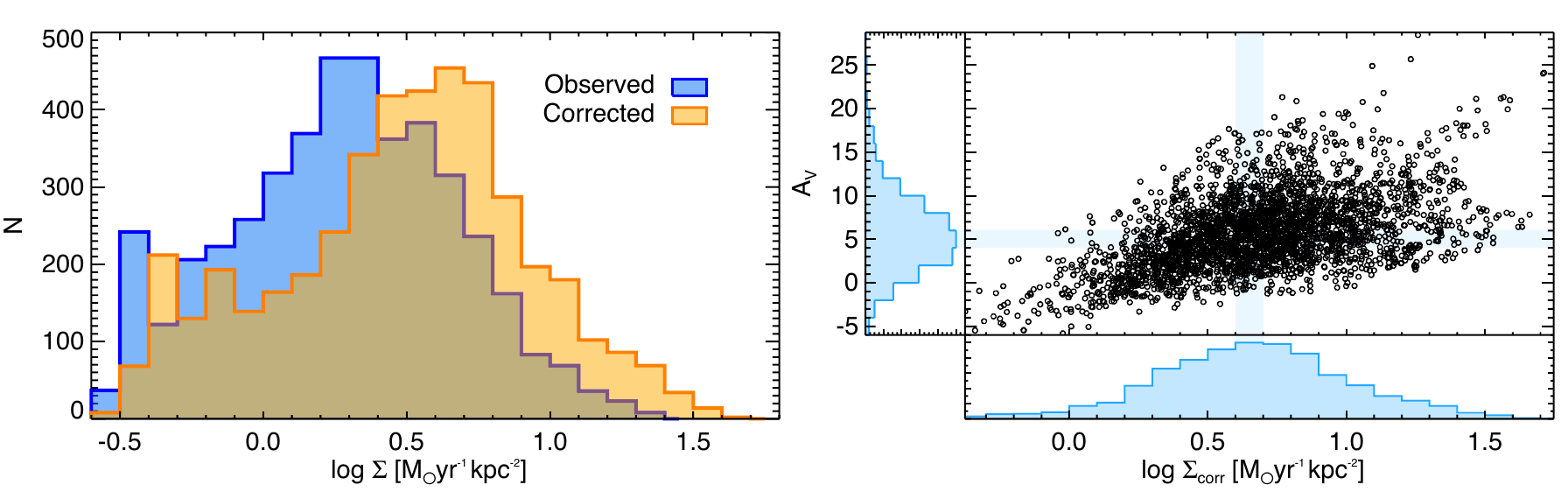} \\
\end{tabular}
\end{center}
\caption{\object{IC 4687}. As for Fig.~\ref{figure:NGC2369} but for \object{IC 4687}}
\label{figure:IC4687}
\end{figure*}

\addtocounter{figure}{-1}
\addtocounter{subfig}{1}
\begin{figure*}[t]
\begin{center}
\begin{tabular}{c}
\includegraphics[angle=0, width=0.80\hsize]{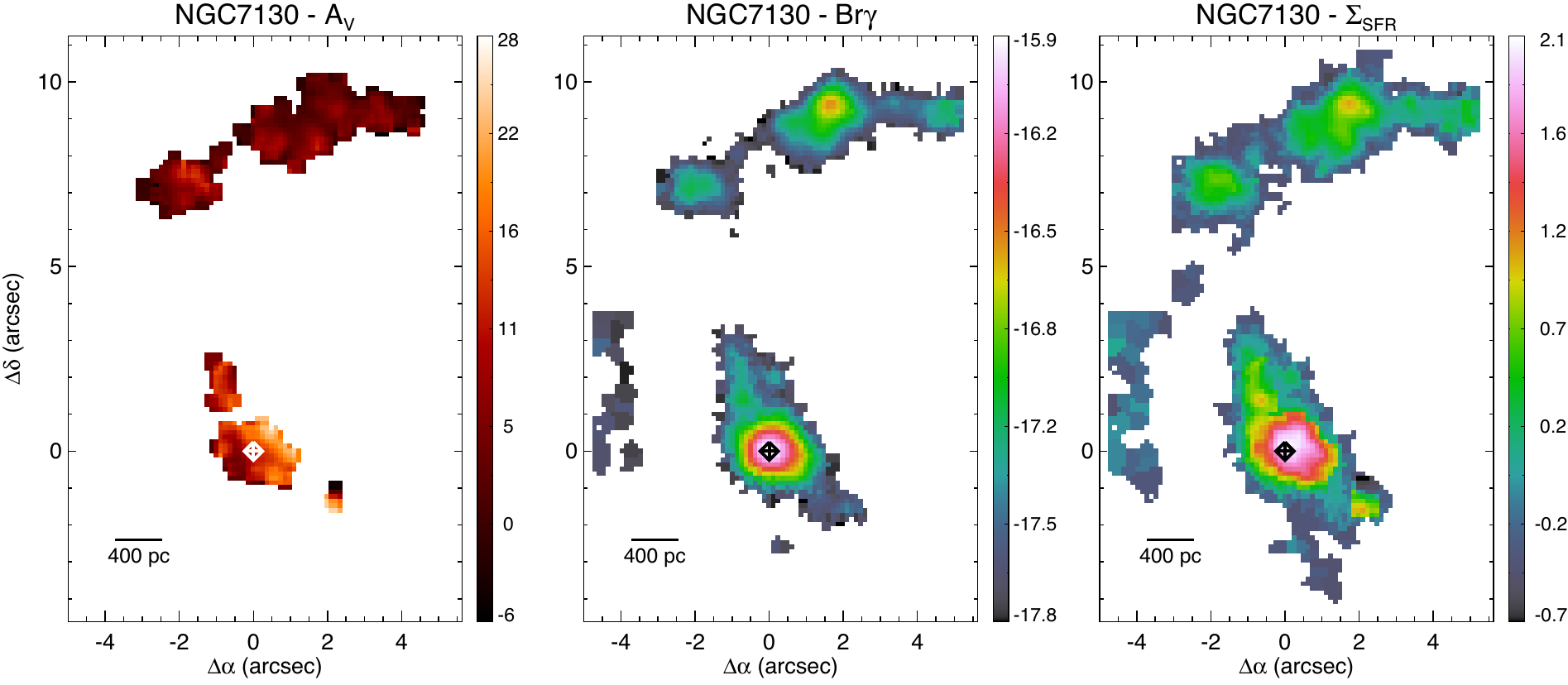} \\
\includegraphics[angle=0, height=.26\hsize]{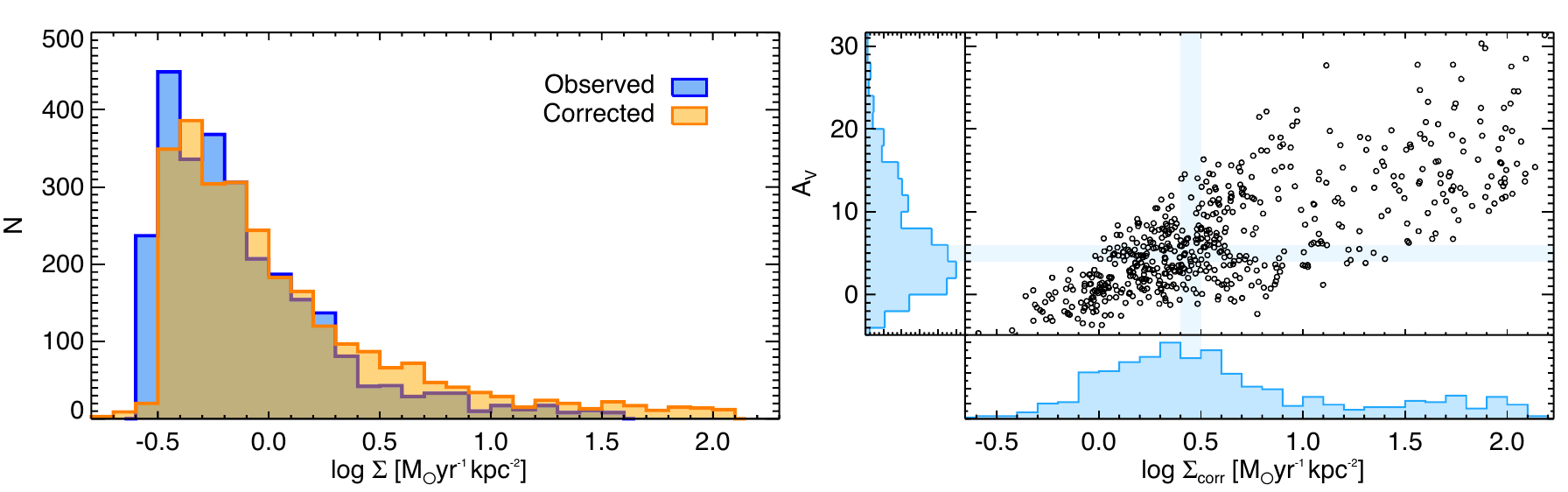} \\
\end{tabular}
\end{center}
\caption{\object{NGC 7130}. As for Fig.~\ref{figure:NGC2369} but for \object{NGC 7130}}
\label{figure:NGC7130}
\end{figure*}

\addtocounter{figure}{-1}
\addtocounter{subfig}{1}
\begin{figure*}[!hb]
\begin{center}
\begin{tabular}{c}
\includegraphics[angle=0, width=0.90\hsize]{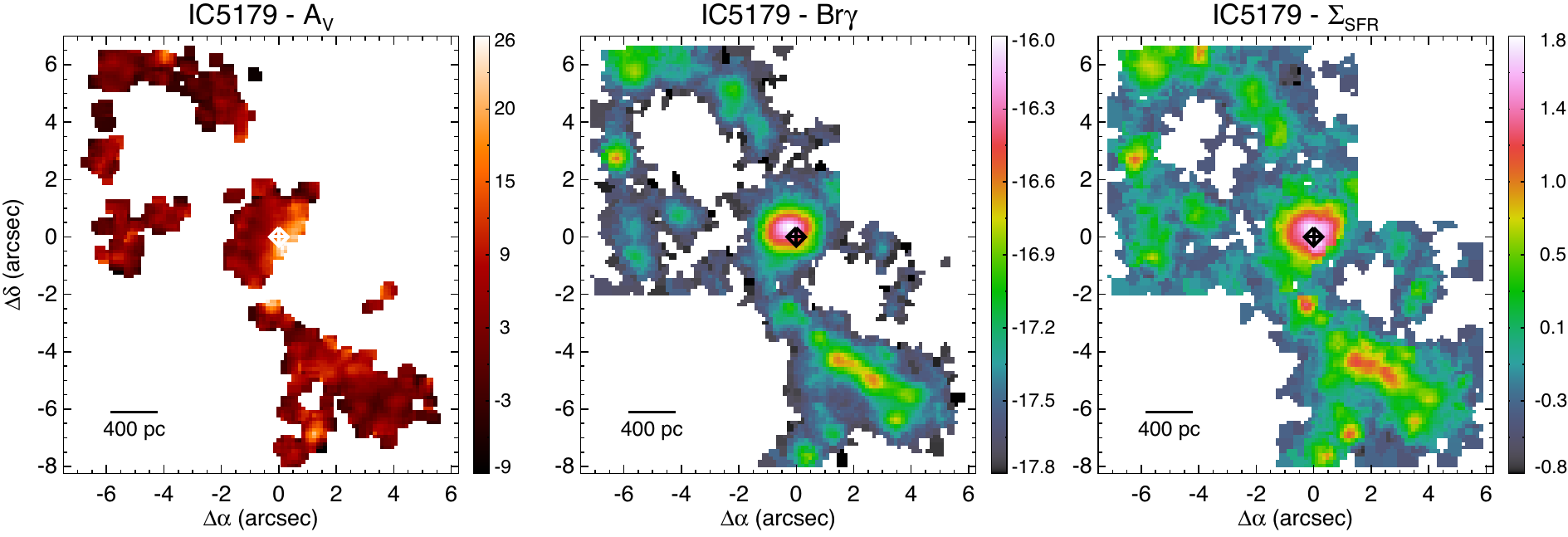} \\
\includegraphics[angle=0, height=.26\hsize]{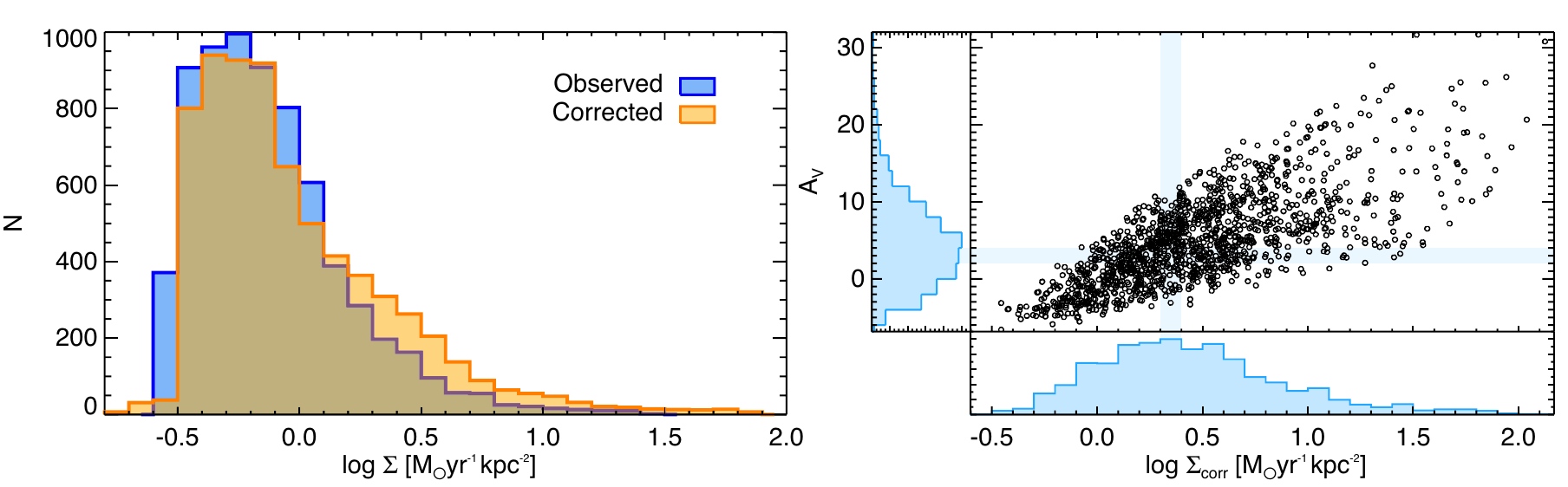} \\
\end{tabular}
\end{center}
\caption{\object{IC 5179}. As for Fig.~\ref{figure:NGC2369} but for \object{IC 5179}}
\label{figure:IC5179}
\end{figure*}


\renewcommand{\thefigure}{\Alph{section}.\arabic{figure}}
\refstepcounter{figure}

\label{figure:ULIRG}
\renewcommand{\thefigure}{\Alph{section}.\arabic{figure}\alph{subfig}}

\setcounter{figure}{1}
\setcounter{subfig}{1}
\begin{figure*}[t]
\begin{center}
\begin{tabular}{c}
\includegraphics[angle=0, width=0.94\hsize]{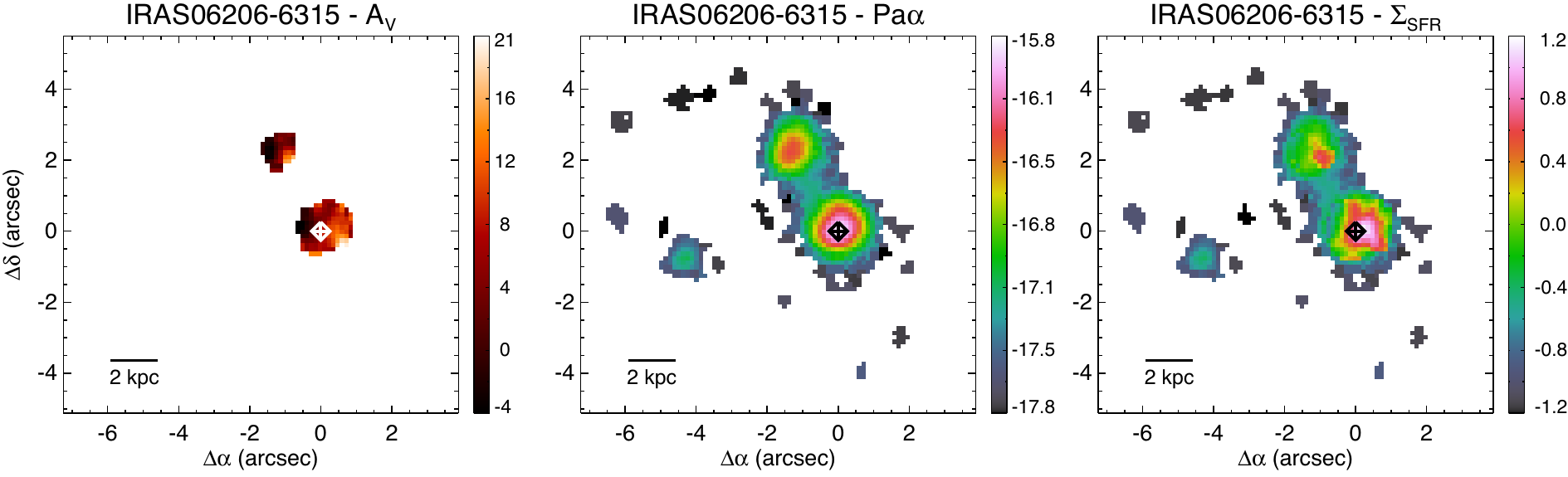} \\
\includegraphics[angle=0, height=.24\hsize]{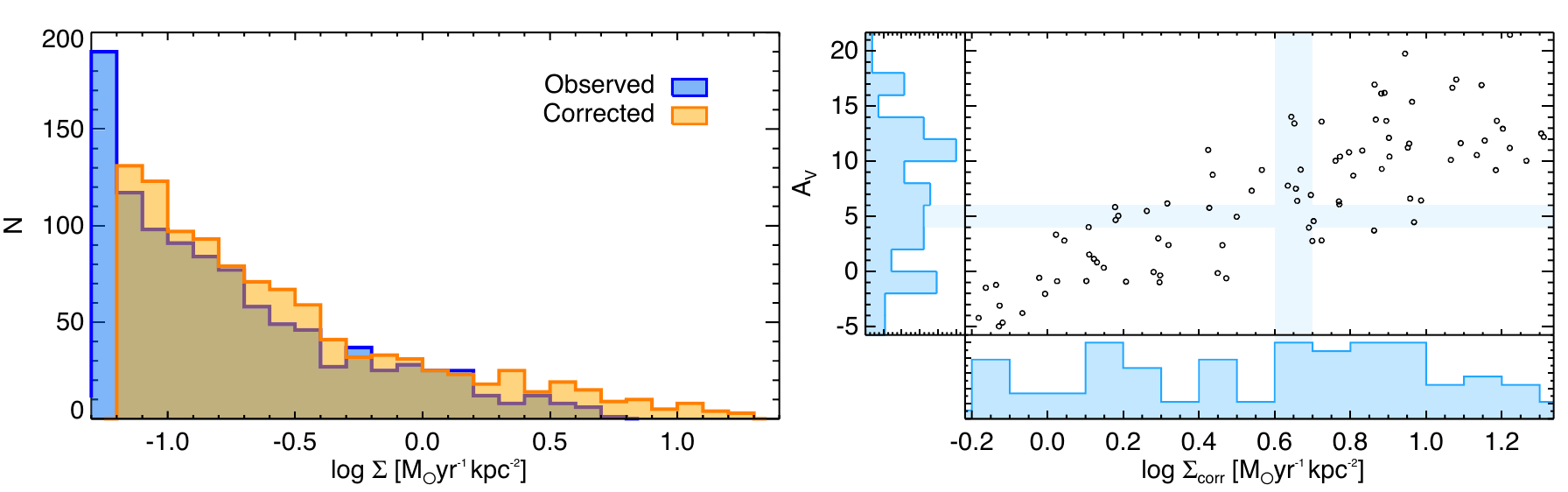} \\
\end{tabular}
\end{center}
\caption{\tiny \object{IRAS 06206-6315}. Top panels show the \Av\ map derived from the \Pal\ and \Brgl\ line ratio, the observed maps of the \Pa\ emission, together with the star formation rate surface density (\Si) map, corrected from extinction. Units are [mag], log[$\rm erg\,s^{-1}\,cm^{-2}$], and [$\rm M_{\sun}\,yr^{-1}\,kpc^{-2}$], respectively. The nucleus and \Pa\ peak are marked with a plus sign (+) and a diamond ($\Diamond$), respectively. The main nucleus is defined as the brightest spaxel of the SINFONI K-band image (Paper~I), and the \Brg\ (\Pa) peak corresponds to the brightest spaxel of the corresponding emission map. Bottom left panel shows the observed (blue histogram) and the corrected-from-extinction (yellow histogram) \Si\ spaxel-by-spaxel distributions. The relationship between the corrected \Si\ values and the \Av\ is shown in the bottom right panel only for those points with a spaxel-by-spaxel correction of the extinction. The blue histograms show the projected distribution onto each axis and are arbitrarily normalised, whereas the blue lines are the median of each distribution.}
\label{figure:IRAS06206}
\end{figure*}

\addtocounter{figure}{-1}
\addtocounter{subfig}{1}
\begin{figure*}[!hb]
\begin{center}
\begin{tabular}{c}
\includegraphics[angle=0, width=0.94\hsize]{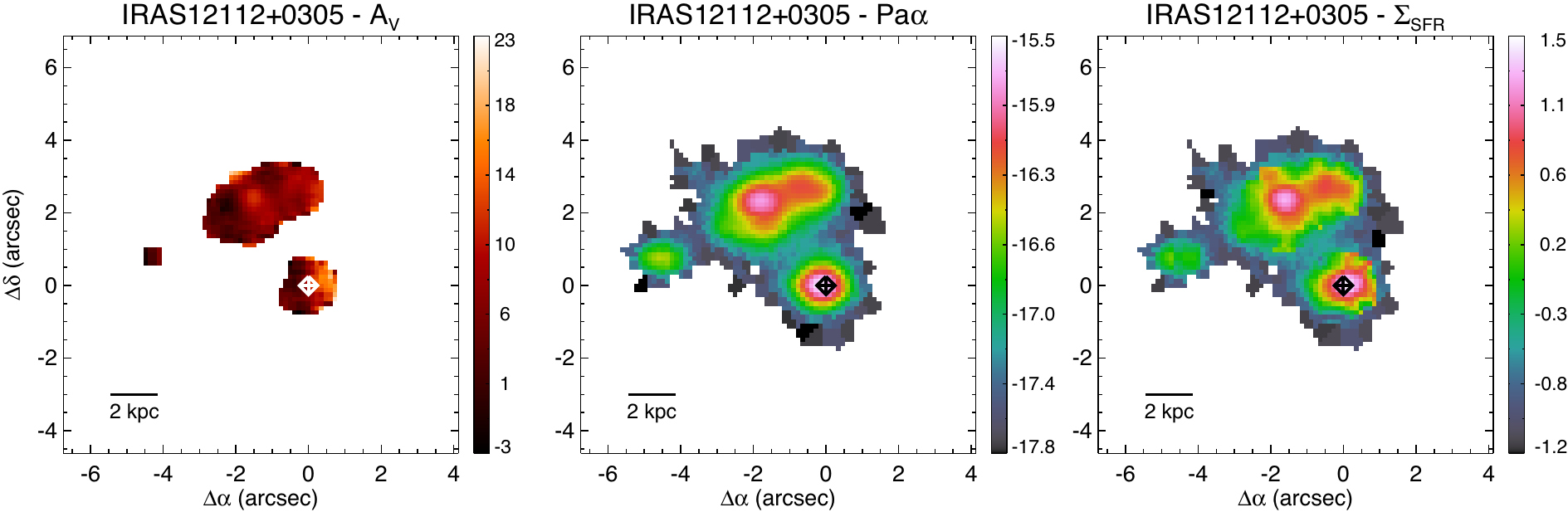} \\
\includegraphics[angle=0, height=.24\hsize]{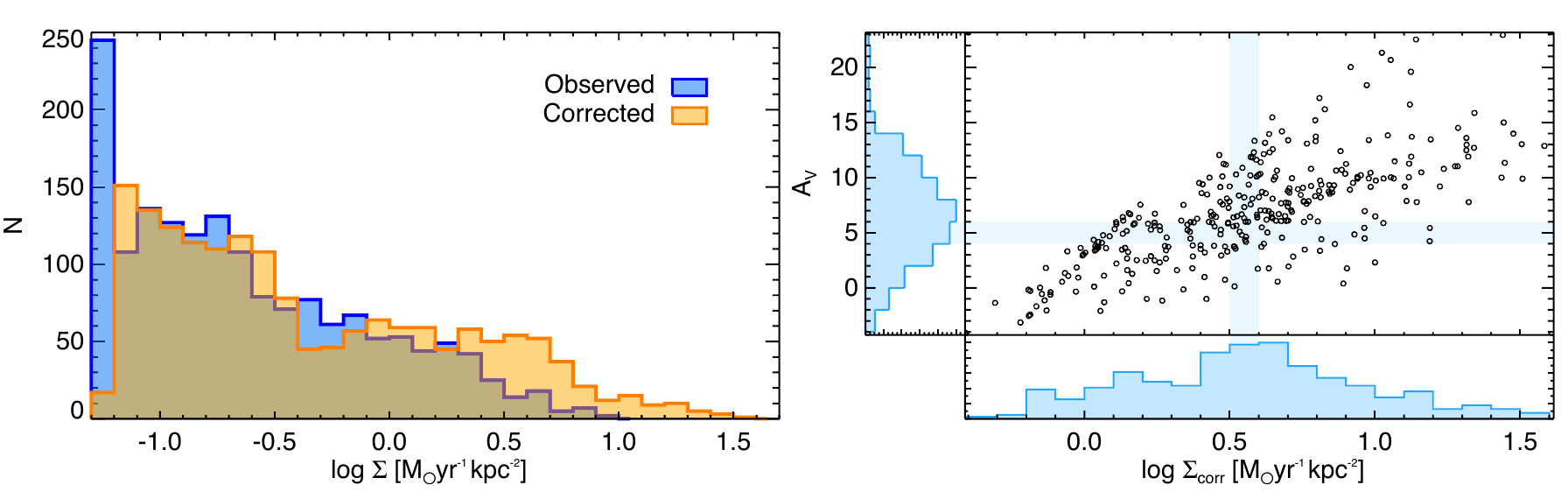} \\
\end{tabular}
\end{center}
\caption{As for Fig.~\ref{figure:IRAS06206} but for \object{IRAS 12112+0305}.}
\label{figure:IRAS12112}
\end{figure*}

\addtocounter{figure}{-1}
\addtocounter{subfig}{1}
\begin{figure*}[t]
\begin{center}
\begin{tabular}{c}
\includegraphics[angle=0, width=0.97\hsize]{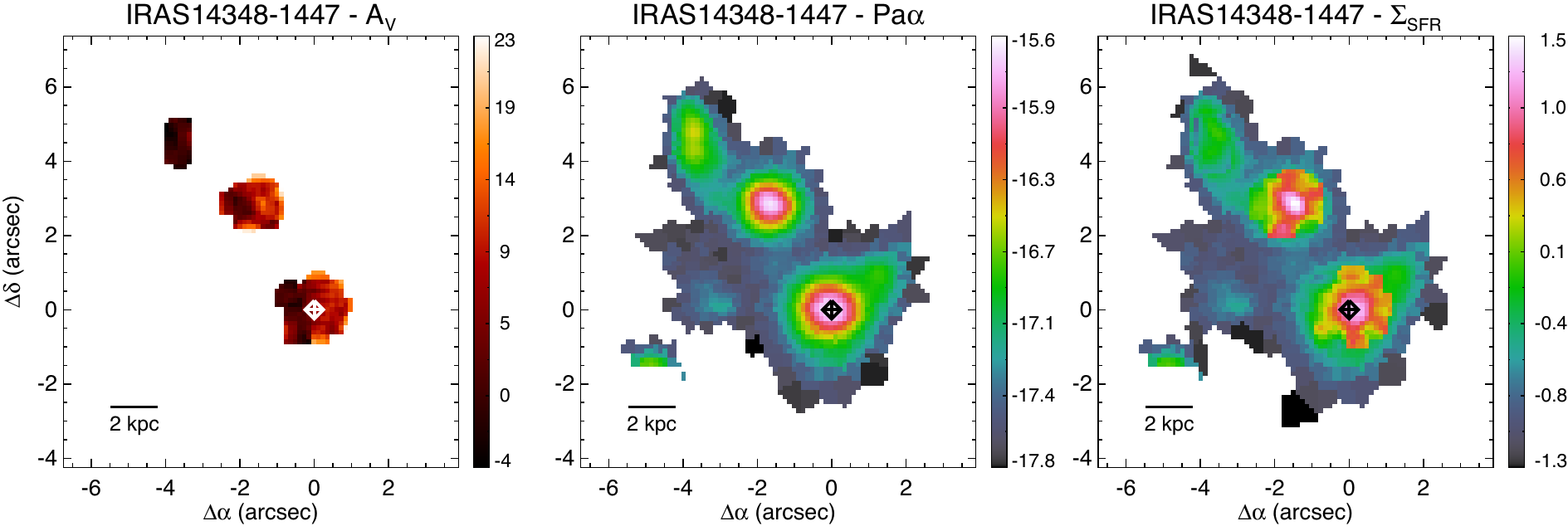} \\
\includegraphics[angle=0, height=.27\hsize]{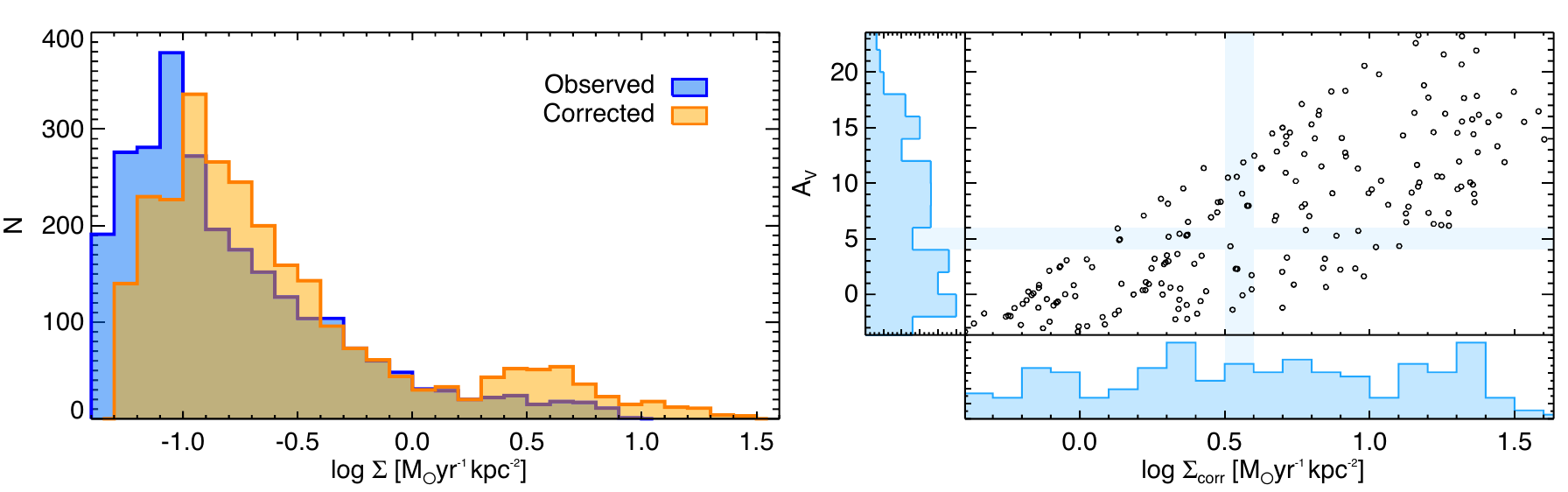} \\
\end{tabular}
\end{center}
\caption{As for Fig.~\ref{figure:IRAS06206} but for \object{IRAS 14348-1447}.}
\label{figure:IRAS14348}
\end{figure*}

\addtocounter{figure}{-1}
\addtocounter{subfig}{1}
\begin{figure*}[!hb]
\begin{center}
\begin{tabular}{c}
\includegraphics[angle=0, width=0.97\hsize]{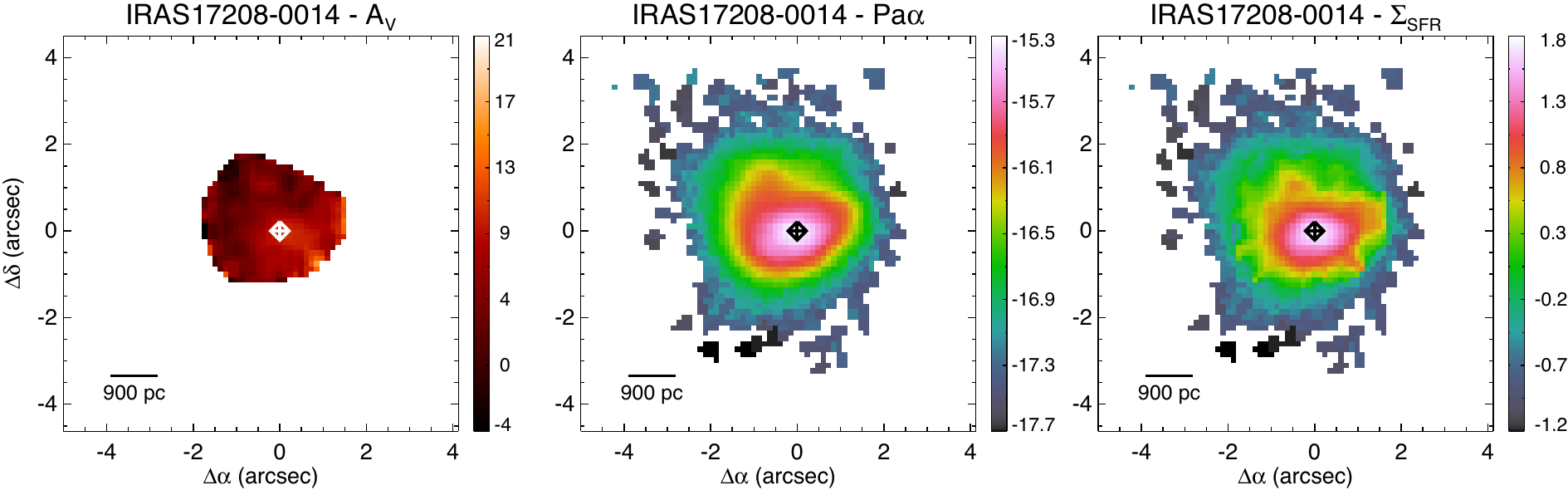} \\
\includegraphics[angle=0, height=.27\hsize]{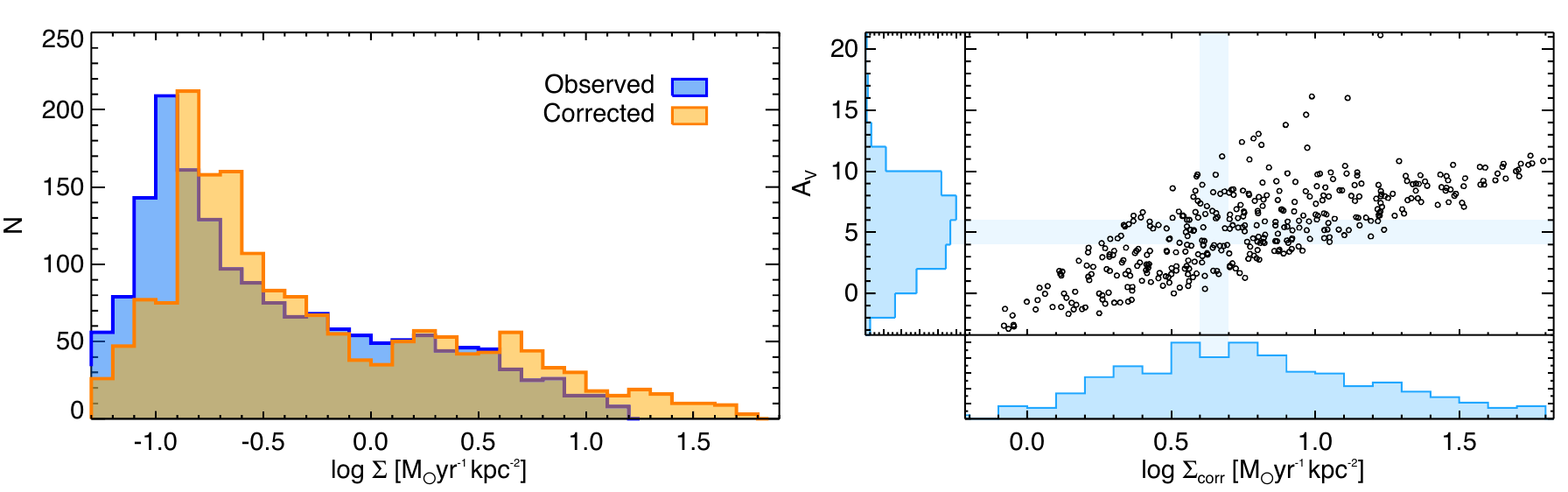} \\
\end{tabular}
\end{center}
\caption{As for Fig.~\ref{figure:IRAS06206} but for \object{IRAS 17208-0014}.}
\label{figure:IRAS17208}
\end{figure*}

\addtocounter{figure}{-1}
\addtocounter{subfig}{1}
\begin{figure*}[t]
\begin{center}
\begin{tabular}{c}
\includegraphics[angle=0, width=0.98\hsize]{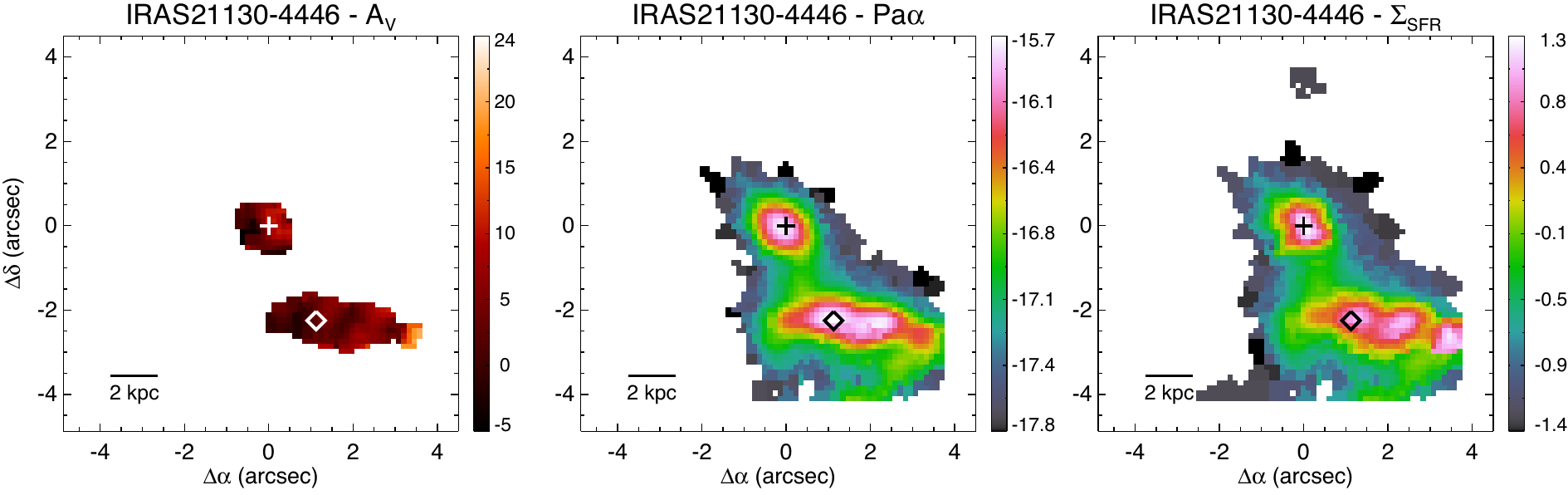} \\
\includegraphics[angle=0, height=.27\hsize]{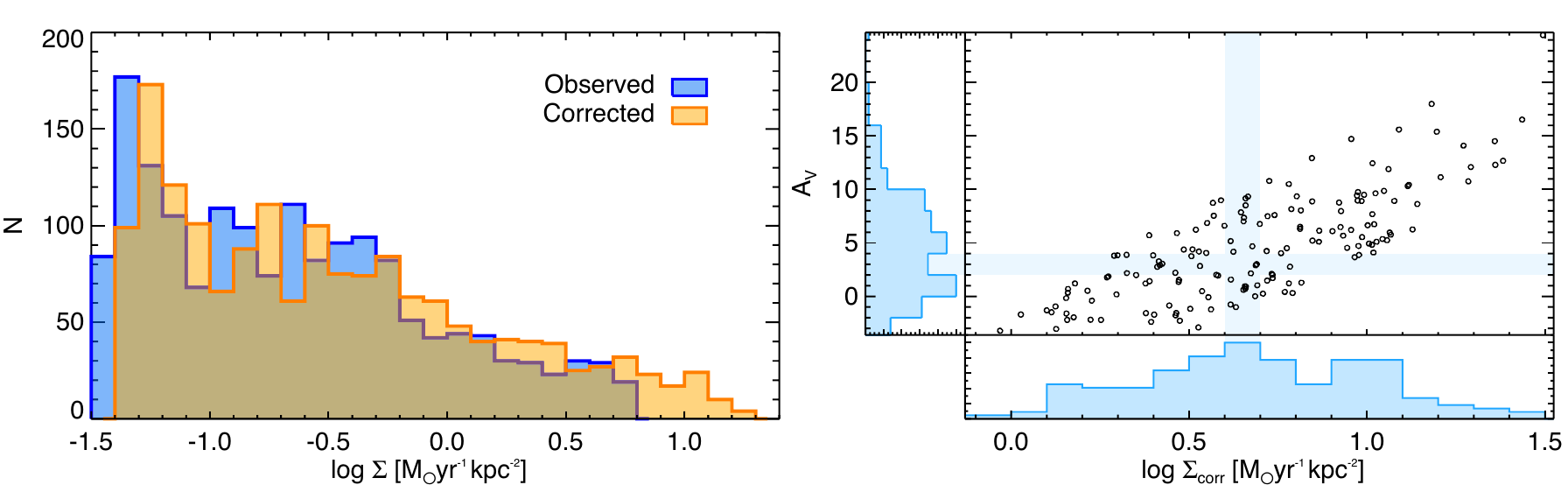} \\
\end{tabular}
\end{center}
\caption{As for Fig.~\ref{figure:IRAS06206} but for \object{IRAS 21130-4446}.}
\label{figure:IRAS21130}
\end{figure*}

\addtocounter{figure}{-1}
\addtocounter{subfig}{1}
\begin{figure*}[!hb]
\begin{center}
\begin{tabular}{c}
\includegraphics[angle=0, width=0.98\hsize]{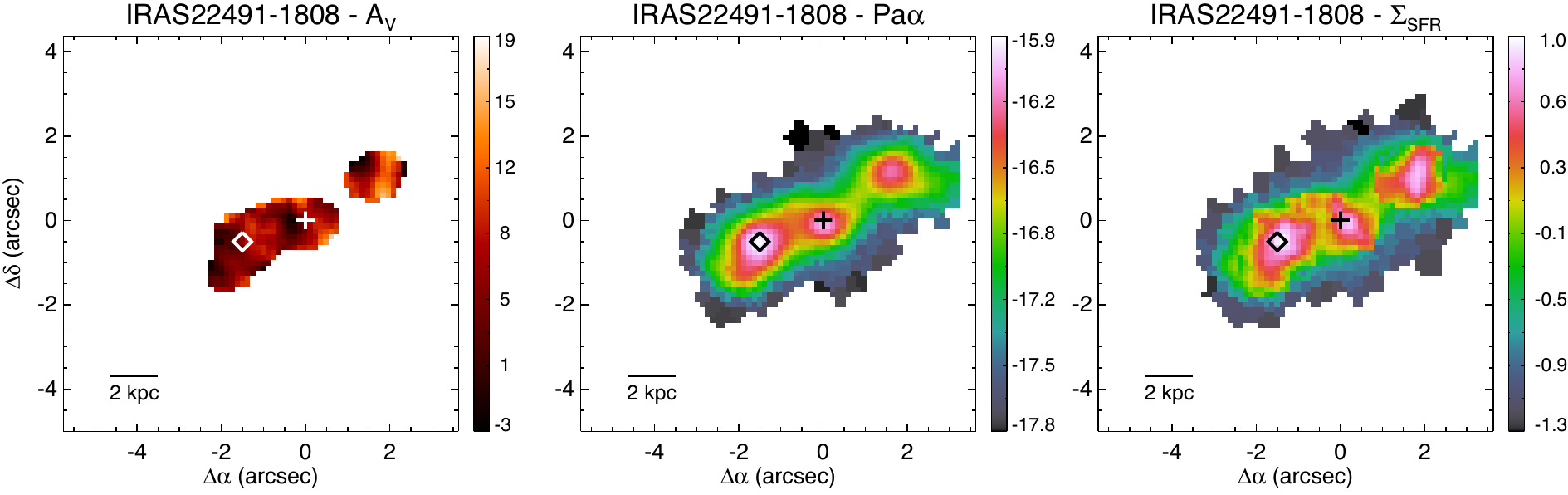} \\
\includegraphics[angle=0, height=.27\hsize]{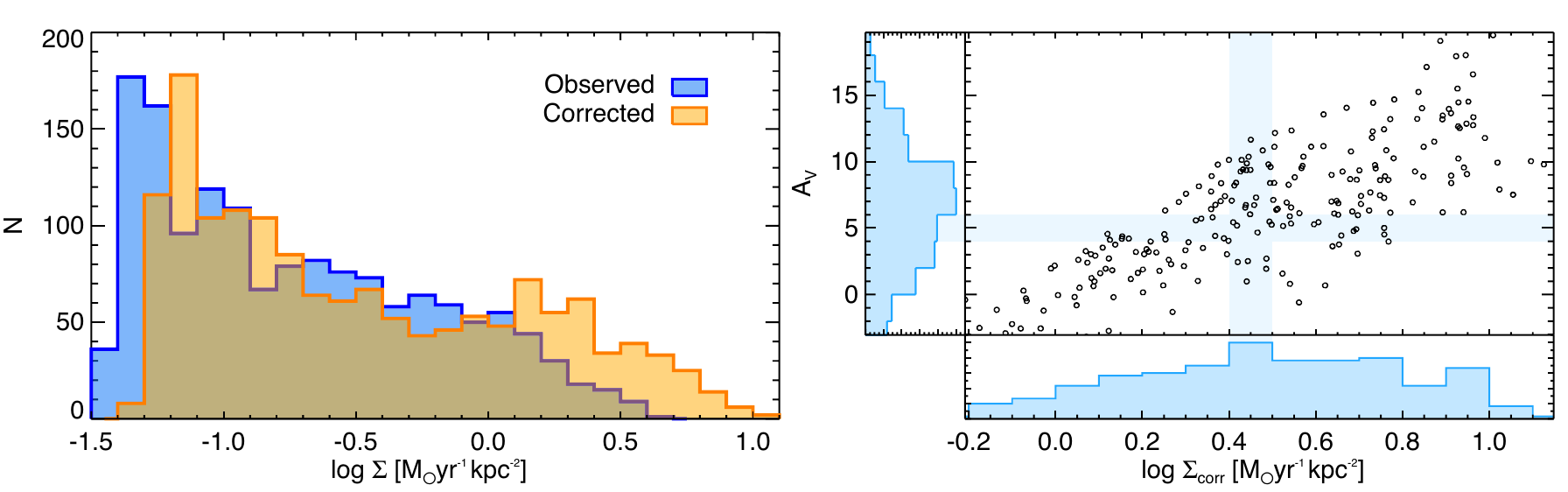} \\
\end{tabular}
\end{center}
\caption{As for Fig.~\ref{figure:IRAS06206} but for \object{IRAS 22491-1808}.}
\label{figure:IRAS22491}
\end{figure*}

\addtocounter{figure}{-1}
\addtocounter{subfig}{1}
\begin{figure*}[t]
\begin{center}
\begin{tabular}{c}
\includegraphics[angle=0, width=0.98\hsize]{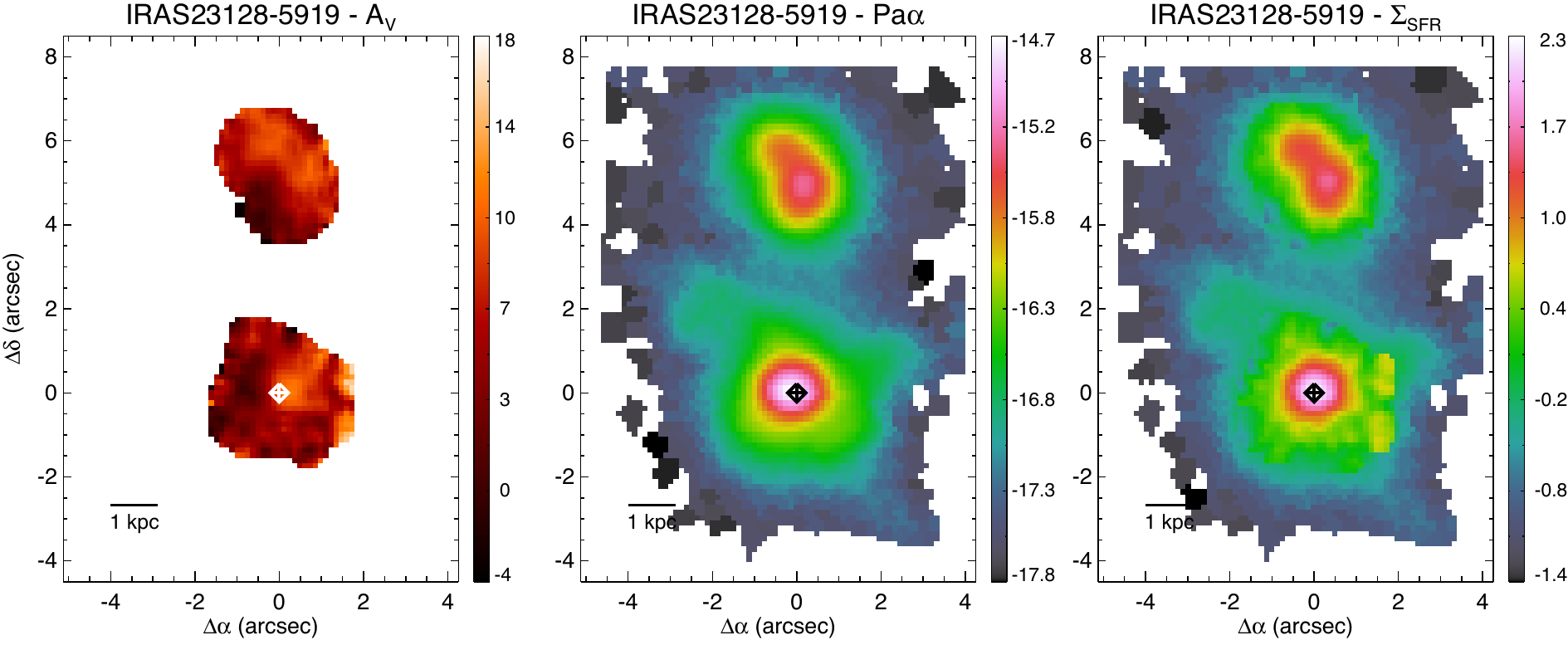} \\
\includegraphics[angle=0, height=.27\hsize]{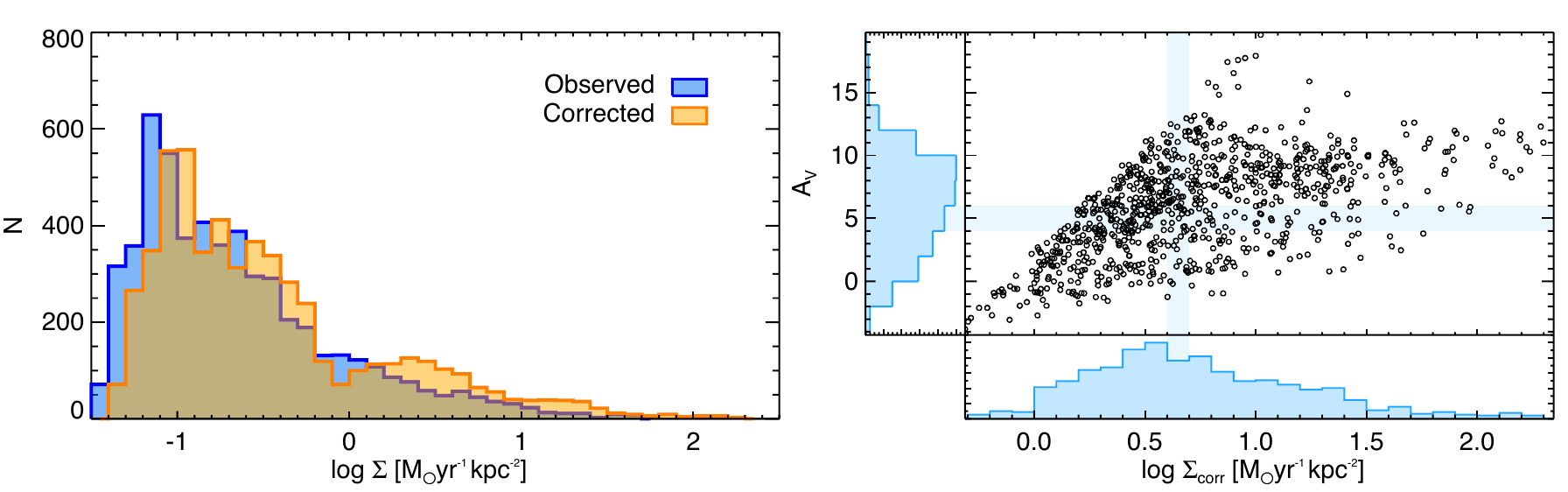} \\
\end{tabular}
\end{center}
\caption{As for Fig.~\ref{figure:IRAS06206} but for \object{IRAS 23128-5919}.}
\label{figure:IRAS23128}
\end{figure*}

\clearpage
\begin{landscape}
\begin{table}
{\tiny
\centering
\caption{Properties of individual star-forming clumps}
\label{table:regions_table}
\resizebox{1.3\textwidth}{!}{
{\setlength{\tabcolsep}{2pt}
\begin{tabular}{ccx{0.7cm}@{ $\pm$ }z{0.7cm}x{0.7cm}@{ $\pm$ }z{0.7cm}x{0.7cm}@{ $\pm$ }z{0.7cm}x{0.7cm}@{ $\pm$ }z{0.7cm}x{0.7cm}@{ $\pm$ }z{0.7cm}x{0.5cm}@{ $\pm$ }z{0.5cm}x{0.7cm}@{ $\pm$ }z{0.7cm}x{0.7cm}@{ $\pm$ }z{0.7cm}x{0.7cm}@{ $\pm$ }z{0.7cm}x{0.7cm}@{ $\pm$ }z{0.7cm}x{0.7cm}@{ $\pm$ }z{0.7cm}x{0.7cm}@{ $\pm$ }l}

\hline
\hline
\noalign{\smallskip}
  Object & Region &  \multicolumn{2}{c}{$\rm r_{eff}$} &  \multicolumn{2}{c}{$\rm L_{Pa\alpha}^{obs}$}& \multicolumn{2}{c}{$\rm L_{Pa\alpha}^{corr}$}& \multicolumn{2}{c}{$\rm \Sigma_{SFR}^{obs}$}& \multicolumn{2}{c}{$\rm \Sigma_{SFR}^{corr}$}&  \multicolumn{2}{c}{$\rm A_{V}$} &  \multicolumn{2}{c}{$\rm r_{core}$} &  \multicolumn{2}{c}{$\rm L_{Pa\alpha}^{obs}$}& \multicolumn{2}{c}{$\rm L_{Pa\alpha}^{corr}$}& \multicolumn{2}{c}{$\rm \Sigma_{SFR}^{obs}$}& \multicolumn{2}{c}{$\rm \Sigma_{SFR}^{corr}$}&  \multicolumn{2}{c}{$\rm A_{V}$} \\
  \noalign{\smallskip}
   (1) & (2) & \multicolumn{2}{c}{(3)} &  \multicolumn{2}{c}{(4)}& \multicolumn{2}{c}{(5)}& \multicolumn{2}{c}{(6)}& \multicolumn{2}{c}{(7)}&  \multicolumn{2}{c}{(8)} & \multicolumn{2}{c}{(9)} &  \multicolumn{2}{c}{(10)}& \multicolumn{2}{c}{(11)}& \multicolumn{2}{c}{(12)}& \multicolumn{2}{c}{(13)}&  \multicolumn{2}{c}{(14)} \\
   \noalign{\smallskip}
\hline
\noalign{\smallskip}
  IRAS06206-6315 &           Nucleus & 775 & 127 &   20 &   3 &   57 &   8 &   2.8 &   0.4 &   7.9 &   1.2 &  7.8 & 2.4& 512 &  29 &    10.5 &     1.6 &   31 &   4 &   3.5 &   0.5 &  10.3 &   1.6 &  8.1 & 2.5 \\
                 &                R1 & 912 &  74 &     9.3 &     1.4 &   15 &   2 &   0.9 &   0.1 &   1.5 &   0.3 &  4.0 & 1.4& 643 &  38 &     5.4 &     0.8 &     8.9 &     1.6 &   1.1 &   0.2 &   1.8 &   0.3 &  3.7 & 1.6 \\
                 &                R2 & 644 &  52 &    0.81 &    0.12 &    1.30 &    0.22 &  0.16 &  0.02 &  0.25 &  0.04 &  3.5 & 1.3& 579 &  66 &    0.69 &    0.11 &    1.10 &    0.19 &  0.17 &  0.03 &  0.27 &  0.05 &  3.5 & 1.3 \\
\hline
\noalign{\smallskip}
         NGC2369 &           Nucleus & 201 &  13 &     3.3 &     0.5 &   24 &   3 &   6.6 &   1.0 &   47 &   7 & 22.4 & 6.7& 133 &   8 &     1.9 &     0.3 &    12.0 &     1.9 &   8.6 &   1.3 &   54 &   8 & 21.0 & 6.3 \\
                 &                R1 & 210 &   8 &     2.3 &     0.3 &   21 &   3 &   4.2 &   0.6 &   38 &   5 & 25.2 & 7.6& 135 &  11 &    1.02 &    0.16 &    11.3 &     1.8 &   4.7 &   0.7 &   52 &   8 & 27.2 & 8.2 \\
                 &                R2 & 138 &   4 &     1.8 &     0.3 &     7.1 &     1.1 &   7.7 &   1.2 &   31 &   4 & 15.8 & 4.8& 181 &   9 &     2.6 &     0.4 &    11.3 &     1.7 &   6.9 &   1.0 &   29 &   4 & 16.4 & 5.0 \\
                 &                R3 & 161 &   7 &     1.6 &     0.3 &     7.0 &     1.1 &   5.3 &   0.8 &   22 &   3 & 16.3 & 4.9& 108 &  22 &    0.88 &    0.13 &     3.7 &     0.6 &   6.2 &   0.9 &   25 &   3 & 16.1 & 4.9 \\
\hline
\noalign{\smallskip}
         NGC3110 &           Nucleus & 351 &   1 &     4.4 &     0.7 &    11.0 &     1.7 &   2.9 &   0.4 &   7.4 &   1.1 & 10.4 & 3.2& 268 &  17 &     3.0 &     0.4 &     7.5 &     1.1 &   3.5 &   0.5 &   8.8 &   1.3 & 10.6 & 3.2 \\
                 &                R1 & 274 &  19 &     2.2 &     0.3 &     4.8 &     0.7 &   2.5 &   0.4 &   5.5 &   0.8 &  8.8 & 2.7& 146 &   9 &    0.92 &    0.14 &     2.0 &     0.3 &   3.8 &   0.6 &   8.2 &   1.3 &  8.6 & 2.6 \\
                 &                R2 & 282 &   5 &    1.32 &    0.20 &     3.0 &     0.5 &   1.4 &   0.2 &   3.1 &   0.5 &  9.3 & 2.8& 237 &  24 &    0.68 &    0.10 &     1.7 &     0.3 &   1.6 &   0.2 &   4.1 &   0.6 & 10.6 & 3.2 \\
                 &                R3 & 299 &   5 &    1.23 &    0.19 &     2.5 &     0.4 &   1.1 &   0.2 &   2.3 &   0.3 &  7.9 & 2.4& 397 &  27 &     1.9 &     0.3 &     3.6 &     0.6 &   1.1 &   0.2 &   2.0 &   0.3 &  7.0 & 2.2 \\
\hline
\noalign{\smallskip}
         NGC3256 &           Nucleus & 182 &   7 &     4.1 &     0.6 &    10.0 &     1.5 &  10.2 &   1.5 &   24 &   3 & 10.0 & 3.0&  86 &   5 &    1.50 &    0.23 &     4.6 &     0.7 &   16 &   2 &   51 &   7 & 12.6 & 3.8 \\
                 &                R1 & 151 &   2 &     2.9 &     0.4 &     5.9 &     0.9 &  10.3 &   1.6 &   21 &   3 &  8.2 & 2.5& 101 &   5 &    1.41 &    0.21 &     3.0 &     0.5 &  11.8 &   1.8 &   25 &   3 &  8.7 & 2.6 \\
                 &                R2 & 205 &   2 &     2.1 &     0.3 &     3.5 &     0.5 &   4.0 &   0.6 &   6.8 &   1.0 &  6.0 & 1.8&  75 &   3 &    0.52 &    0.08 &    0.69 &    0.11 &   7.6 &   1.2 &  10.0 &   1.5 &  3.1 & 1.0 \\
                 &                R3 & 120 &   3 &    1.49 &    0.23 &    1.60 &    0.24 &   8.3 &   1.3 &   8.9 &   1.4 &  0.8 & 0.3&  76 &   4 &    0.74 &    0.11 &    0.77 &    0.12 &  11.5 &   1.7 &  12.0 &   1.8 &  0.5 & 0.3 \\
                 &                R4 & 135 &   9 &    1.47 &    0.23 &     1.9 &     0.3 &   6.7 &   1.0 &   8.8 &   1.4 &  3.1 & 1.0&  90 &   6 &    0.80 &    0.12 &    1.09 &    0.17 &   9.0 &   1.4 &  12.3 &   2.0 &  3.5 & 1.2 \\
                 &                R5 & 151 &   2 &    1.20 &    0.18 &     1.8 &     0.3 &   4.4 &   0.7 &   6.6 &   1.0 &  4.6 & 1.4& 118 &   7 &    0.68 &    0.10 &    1.08 &    0.16 &   5.3 &   0.8 &   8.4 &   1.3 &  5.2 & 1.6 \\
                 &                R6 & 153 &  10 &    1.10 &    0.17 &     2.3 &     0.3 &   3.8 &   0.6 &   8.0 &   1.2 &  8.3 & 2.5& 134 &   8 &    0.81 &    0.12 &     1.7 &     0.3 &   4.1 &   0.6 &   8.5 &   1.3 &  8.3 & 2.5 \\
                 &                R7 & 116 &   2 &    0.93 &    0.14 &    1.27 &    0.19 &   5.7 &   0.9 &   7.8 &   1.2 &  3.5 & 1.1& 118 &   7 &    0.79 &    0.12 &    1.04 &    0.16 &   6.3 &   1.0 &   8.4 &   1.3 &  3.2 & 1.0 \\
                 &                R8 & 135 &   6 &    0.84 &    0.13 &    1.00 &    0.15 &   3.7 &   0.6 &   4.4 &   0.7 &  1.9 & 0.6&  70 &   4 &    0.32 &    0.05 &    0.40 &    0.06 &   5.4 &   0.8 &   6.6 &   1.0 &  2.4 & 0.9 \\
                 &                R9 & 101 &   8 &    0.87 &    0.13 &    1.20 &    0.19 &   6.9 &   1.0 &   9.5 &   1.5 &  3.6 & 1.3&  50 &   2 &    0.26 &    0.04 &    0.37 &    0.07 &   9.7 &   1.5 &  14.1 &   2.5 &  4.1 & 1.9 \\
                 &               R10 & 107 &  19 &    0.80 &    0.12 &    0.94 &    0.14 &   5.9 &   0.9 &   7.0 &   1.1 &  1.9 & 0.6&  91 &   5 &    0.57 &    0.09 &    0.68 &    0.10 &   6.4 &   1.0 &   7.7 &   1.2 &  2.0 & 0.6 \\
                 &               R11 & 120 &   3 &    0.69 &    0.10 &    1.48 &    0.23 &   3.9 &   0.6 &   8.4 &   1.3 &  8.6 & 2.6& 100 &   7 &    0.52 &    0.08 &    1.13 &    0.17 &   4.2 &   0.6 &   9.1 &   1.4 &  8.8 & 2.7 \\
                 &               R12 & 101 &   4 &    0.69 &    0.10 &    1.22 &    0.18 &   5.5 &   0.8 &   9.8 &   1.5 &  6.5 & 2.0&  70 &   4 &    0.37 &    0.06 &    0.67 &    0.10 &   6.1 &   0.9 &  11.1 &   1.7 &  6.8 & 2.1 \\
                 &               R13 &  96 &  14 &    0.57 &    0.09 &    1.01 &    0.15 &   5.0 &   0.8 &   8.9 &   1.4 &  6.4 & 2.0&  63 &   3 &    0.28 &    0.04 &    0.51 &    0.08 &   6.4 &   1.0 &  11.6 &   1.8 &  6.8 & 2.1 \\
                 &               R14 & 108 &   6 &    0.54 &    0.08 &    0.78 &    0.12 &   3.7 &   0.6 &   5.4 &   0.8 &  4.1 & 1.3& 146 &   8 &    0.90 &    0.14 &    1.45 &    0.22 &   3.6 &   0.5 &   5.8 &   0.9 &  5.5 & 1.7 \\
                 &               R15 &  86 &  14 &    0.50 &    0.07 &    0.98 &    0.15 &   5.5 &   0.8 &  10.7 &   1.6 &  7.7 & 2.3&  62 &   4 &    0.28 &    0.04 &    0.61 &    0.09 &   6.0 &   0.9 &  13.0 &   2.0 &  8.7 & 2.7 \\
                 &               R16 &  75 &  15 &    0.32 &    0.05 &    0.53 &    0.08 &   4.6 &   0.7 &   7.7 &   1.2 &  5.8 & 1.8& 109 &   5 &    0.61 &    0.09 &    0.99 &    0.15 &   4.2 &   0.6 &   6.8 &   1.0 &  5.5 & 1.7 \\
                 &               R17 &  75 &  15 &    0.28 &    0.04 &    0.35 &    0.05 &   4.2 &   0.6 &   5.2 &   0.8 &  2.5 & 0.9&  59 &   5 &    0.19 &    0.03 &    0.23 &    0.04 &   4.4 &   0.7 &   5.4 &   0.8 &  2.3 & 0.9 \\
\hline
\noalign{\smallskip}
     ESO320-G030 &           Nucleus & 174 &  10 &    0.67 &    0.11 &    0.98 &    0.16 &   1.9 &   0.3 &   2.7 &   0.4 &  4.3 & 1.4&  93 &   6 &    0.28 &    0.04 &    0.43 &    0.07 &   3.1 &   0.5 &   4.7 &   0.8 &  4.7 & 1.8 \\
                 &                R1 & 227 &   9 &     2.1 &     0.3 &     4.2 &     0.6 &   3.3 &   0.5 &   6.7 &   1.0 &  8.1 & 2.5& 167 &   9 &    1.18 &    0.18 &     2.4 &     0.4 &   4.3 &   0.6 &   8.7 &   1.3 &  8.1 & 2.5 \\
                 &                R2 & 246 &   2 &    1.61 &    0.24 &     2.4 &     0.4 &   2.2 &   0.3 &   3.2 &   0.5 &  4.3 & 1.3& 211 &  17 &    1.25 &    0.19 &     1.9 &     0.3 &   2.3 &   0.3 &   3.5 &   0.5 &  4.7 & 1.4 \\
                 &                R3 & 246 &  14 &     1.7 &     0.3 &     2.7 &     0.4 &   2.3 &   0.3 &   3.8 &   0.6 &  5.7 & 1.7& 155 &   8 &    0.79 &    0.12 &    1.47 &    0.22 &   3.4 &   0.5 &   6.3 &   1.0 &  7.0 & 2.2 \\
                 &                R4 & 210 &  13 &    1.22 &    0.18 &     2.7 &     0.4 &   2.3 &   0.3 &   5.0 &   0.8 &  8.9 & 2.7& 149 &   9 &    0.65 &    0.10 &     1.7 &     0.3 &   3.1 &   0.5 &   7.8 &   1.2 & 10.6 & 3.2 \\
                 &                R5 & 123 &  21 &    0.70 &    0.11 &    1.50 &    0.23 &   3.8 &   0.6 &   8.1 &   1.2 &  8.6 & 2.6&  92 &   6 &    0.46 &    0.07 &    0.92 &    0.14 &   4.4 &   0.7 &   9.0 &   1.4 &  7.9 & 2.5 \\
                 &                R6 & 161 &   7 &    0.49 &    0.08 &    0.78 &    0.12 &   1.6 &   0.2 &   2.5 &   0.4 &  5.3 & 1.7&  99 &   6 &    0.22 &    0.03 &    0.44 &    0.07 &   2.1 &   0.3 &   4.2 &   0.7 &  7.7 & 2.5 \\
\hline
\noalign{\smallskip}
  IRAS12112+0305 &           Nucleus & 551 & 115 &    16.5 &     2.5 &   52 &   7 &   4.4 &   0.7 &  14.1 &   2.1 &  8.7 & 2.6& 367 &  20 &     8.9 &     1.3 &   29 &   4 &   5.9 &   0.9 &   19 &   2 &  9.1 & 2.8 \\
                 &                R1 &1232 &  36 &   33 &   5 &   85 &  12 &   1.8 &   0.3 &   4.7 &   0.7 &  7.0 & 2.1& 531 &  32 &    10.5 &     1.6 &   36 &   5 &   3.3 &   0.5 &  11.5 &   1.7 &  9.3 & 2.8 \\
                 &                R2 & 628 & 103 &     2.5 &     0.5 &     3.4 &     0.7 &   0.5 &   0.1 &   0.7 &   0.1 &  2.2 & 0.8& 498 &  29 &     1.8 &     0.4 &     2.3 &     0.5 &   0.6 &   0.1 &   0.7 &   0.2 &  1.9 & 0.8 \\
\hline
\noalign{\smallskip}
 IRASF12115-4656 &  Nucleus$^{\dag}$ & 187 &   8 &    0.53 &    0.09 &    1.21 &    0.22 &   1.2 &   0.2 &   2.7 &   0.5 &  9.4 & 3.1&  85 &   6 &    0.14 &    0.03 &    0.27 &    0.06 &   1.6 &   0.3 &   2.9 &   0.7 &  7.2 & 3.3 \\
                 &                R1 & 398 &   4 &     3.7 &     0.6 &     7.9 &     1.2 &   1.9 &   0.3 &   4.1 &   0.6 &  8.5 & 2.6& 213 &  12 &    1.37 &    0.21 &     2.3 &     0.3 &   2.7 &   0.4 &   4.5 &   0.7 &  5.7 & 1.8 \\
                 &                R2 & 357 &  13 &     2.1 &     0.3 &    13.2 &     2.0 &   1.4 &   0.2 &   8.6 &   1.3 & 20.7 & 6.2& 157 &  12 &    0.53 &    0.08 &     3.8 &     0.6 &   2.0 &   0.3 &  13.9 &   2.1 & 22.2 & 6.7 \\
                 &                R3 & 338 &   4 &     2.0 &     0.3 &     4.4 &     0.7 &   1.5 &   0.2 &   3.1 &   0.5 &  8.6 & 2.6& 320 &  31 &     1.8 &     0.3 &     3.7 &     0.6 &   1.5 &   0.2 &   3.0 &   0.5 &  8.0 & 2.4 \\
                 &                R4 & 252 &  20 &    1.14 &    0.18 &     1.8 &     0.3 &   1.4 &   0.2 &   2.3 &   0.4 &  5.2 & 1.6& 108 &   6 &    0.40 &    0.06 &    0.63 &    0.10 &   2.6 &   0.4 &   4.1 &   0.6 &  5.0 & 1.6 \\
\hline
\noalign{\smallskip}
         NGC5135 &  Nucleus$^{\dag}$ & 221 &   4 &     3.1 &     0.5 &     6.6 &     1.0 &   5.3 &   0.8 &  11.1 &   1.7 &  8.4 & 2.5& 155 &  11 &     1.8 &     0.3 &     3.9 &     0.6 &   6.1 &   0.9 &  13.5 &   2.1 &  9.0 & 2.7 \\
                 &                R1 & 212 &  12 &     4.0 &     0.6 &     7.5 &     1.1 &   7.2 &   1.1 &  13.6 &   2.1 &  7.2 & 2.2& 133 &   7 &     2.0 &     0.3 &     3.9 &     0.6 &   9.3 &   1.4 &   17 &   2 &  7.5 & 2.3 \\
                 &                R2 & 221 &  12 &     3.9 &     0.6 &     9.1 &     1.4 &   6.7 &   1.0 &  15.6 &   2.4 &  9.5 & 2.9& 153 &   8 &     2.1 &     0.3 &     5.2 &     0.8 &   8.0 &   1.2 &   19 &   2 & 10.0 & 3.0 \\
                 &                R3 & 212 &  12 &     2.7 &     0.4 &     6.4 &     1.0 &   4.9 &   0.7 &  11.6 &   1.8 &  9.7 & 2.9& 143 &   8 &    1.50 &    0.23 &     3.3 &     0.5 &   6.5 &   1.0 &  14.5 &   2.2 &  9.1 & 2.8 \\
\hline
\hline
\end{tabular}}}
}
\end{table}

\clearpage
\addtocounter{table}{-1}
\begin{table}
{\tiny
\caption{Properties of individual star-forming clumps (cont.)}
\centering
\resizebox{1.3\textwidth}{!}{
{\setlength{\tabcolsep}{2pt}
\begin{tabular}{ccx{0.7cm}@{ $\pm$ }z{0.7cm}x{0.7cm}@{ $\pm$ }z{0.7cm}x{0.7cm}@{ $\pm$ }z{0.7cm}x{0.7cm}@{ $\pm$ }z{0.7cm}x{0.7cm}@{ $\pm$ }z{0.7cm}x{0.5cm}@{ $\pm$ }z{0.5cm}x{0.7cm}@{ $\pm$ }z{0.7cm}x{0.7cm}@{ $\pm$ }z{0.7cm}x{0.7cm}@{ $\pm$ }z{0.7cm}x{0.7cm}@{ $\pm$ }z{0.7cm}x{0.7cm}@{ $\pm$ }z{0.7cm}x{0.7cm}@{ $\pm$ }l}

\hline
\hline
\noalign{\smallskip}
  Object & Region &  \multicolumn{2}{c}{$\rm r_{eff}$} &  \multicolumn{2}{c}{$\rm L_{Pa\alpha}^{obs}$}& \multicolumn{2}{c}{$\rm L_{Pa\alpha}^{corr}$}& \multicolumn{2}{c}{$\rm \Sigma_{SFR}^{obs}$}& \multicolumn{2}{c}{$\rm \Sigma_{SFR}^{corr}$}&  \multicolumn{2}{c}{$\rm A_{V}$} &  \multicolumn{2}{c}{$\rm r_{core}$} &  \multicolumn{2}{c}{$\rm L_{Pa\alpha}^{obs}$}& \multicolumn{2}{c}{$\rm L_{Pa\alpha}^{corr}$}& \multicolumn{2}{c}{$\rm \Sigma_{SFR}^{obs}$}& \multicolumn{2}{c}{$\rm \Sigma_{SFR}^{corr}$}&  \multicolumn{2}{c}{$\rm A_{V}$} \\
  \noalign{\smallskip}
   (1) & (2) & \multicolumn{2}{c}{(3)} &  \multicolumn{2}{c}{(4)}& \multicolumn{2}{c}{(5)}& \multicolumn{2}{c}{(6)}& \multicolumn{2}{c}{(7)}&  \multicolumn{2}{c}{(8)} & \multicolumn{2}{c}{(9)} &  \multicolumn{2}{c}{(10)}& \multicolumn{2}{c}{(11)}& \multicolumn{2}{c}{(12)}& \multicolumn{2}{c}{(13)}&  \multicolumn{2}{c}{(14)} \\
   \noalign{\smallskip}
  \hline
  \noalign{\smallskip}
  IRAS14348-1447 &           Nucleus & 872 & 154 &   36 &   5 &   87 &  13 &   4.0 &   0.6 &   9.6 &   1.5 &  6.5 & 2.0& 515 &  27 &   18 &   2 &   44 &   6 &   5.9 &   0.9 &  14.5 &   2.2 &  6.8 & 2.1 \\
                 &                R1 & 703 & 115 &   20 &   3 &   66 &  10 &   3.5 &   0.5 &  11.1 &   1.7 &  8.7 & 2.6& 399 &  23 &    10.3 &     1.6 &   36 &   5 &   5.0 &   0.8 &   18 &   2 &  9.6 & 2.9 \\
                 &                R2 &1050 &  79 &     6.3 &     1.0 &     6.7 &     1.1 &   0.5 &   0.1 &   0.5 &   0.1 &  0.4 & 0.4& 731 &  40 &     3.3 &     0.5 &     3.7 &     0.6 &   0.6 &   0.1 &   0.7 &   0.1 &  0.9 & 0.6 \\
                 &                R3 &1573 &  53 &     2.8 &     0.4 &     4.1 &     0.6 &  0.09 &  0.01 &  0.14 &  0.02 &  2.8 & 0.9& 842 &  54 &    1.06 &    0.16 &     1.6 &     0.3 &  0.13 &  0.02 &  0.19 &  0.03 &  3.0 & 1.0 \\
\hline
\noalign{\smallskip}
 IRASF17138-1017 &           Nucleus & 181 &   7 &     3.4 &     0.5 &     7.4 &     1.1 &   8.4 &   1.3 &   18 &   2 &  8.9 & 2.7& 188 &  24 &     3.5 &     0.5 &     7.9 &     1.2 &   8.3 &   1.3 &   18 &   2 &  9.1 & 2.7 \\
                 &                R1 & 220 &   6 &     9.9 &     1.5 &    16.5 &     2.5 &   17 &   2 &   28 &   4 &  5.7 & 1.7& 151 &   8 &     5.2 &     0.8 &     9.2 &     1.4 &   20 &   3 &   36 &   5 &  6.4 & 1.9 \\
                 &                R2 & 237 &  18 &     6.2 &     0.9 &    11.5 &     1.7 &   9.2 &   1.4 &   17 &   2 &  7.0 & 2.1& 225 &  20 &     5.9 &     0.9 &    11.0 &     1.7 &   9.4 &   1.4 &   17 &   2 &  7.0 & 2.1 \\
                 &                R3 & 197 &  34 &     5.7 &     0.9 &    11.2 &     1.7 &  12.7 &   1.9 &   25 &   3 &  7.7 & 2.3& 138 &   9 &     3.6 &     0.5 &     7.4 &     1.1 &  15.1 &   2.3 &   31 &   4 &  8.1 & 2.4 \\
                 &                R4 & 158 &  26 &     1.7 &     0.3 &     3.3 &     0.5 &   5.7 &   0.9 &  10.6 &   1.6 &  7.2 & 2.2& 135 &  10 &    1.37 &    0.21 &     2.6 &     0.4 &   6.1 &   0.9 &  11.6 &   1.8 &  7.2 & 2.2 \\
                 &                R5 & 158 &  26 &    0.90 &    0.14 &     1.7 &     0.3 &   2.9 &   0.4 &   5.5 &   0.8 &  7.0 & 2.1&  86 &   5 &    0.32 &    0.05 &    0.66 &    0.10 &   3.9 &   0.6 &   8.0 &   1.2 &  8.3 & 2.6 \\
                 &                R6 & 158 &  29 &    0.51 &    0.08 &    1.08 &    0.17 &   1.7 &   0.3 &   3.5 &   0.5 &  8.5 & 2.6& 117 &   8 &    0.31 &    0.05 &    0.70 &    0.11 &   1.8 &   0.3 &   4.1 &   0.6 &  9.0 & 2.8 \\
                 &                R7 & 176 &   8 &    0.48 &    0.07 &    1.31 &    0.20 &   1.3 &   0.2 &   3.6 &   0.5 & 11.4 & 3.5& 129 &  10 &    0.30 &    0.05 &    0.92 &    0.14 &   1.5 &   0.2 &   4.7 &   0.7 & 12.7 & 3.9 \\
\hline
\noalign{\smallskip}
  IRAS17208-0014 &           Nucleus & 675 &  48 &   36 &   5 &  105 &  15 &   6.6 &   1.0 &   18 &   2 &  7.9 & 2.4& 347 &  21 &    14.3 &     2.2 &   47 &   7 &  10.1 &   1.5 &   33 &   5 &  8.9 & 2.7 \\
\hline
\noalign{\smallskip}
          IC4687 &           Nucleus & 355 &  12 &     8.0 &     1.2 &   22 &   3 &   5.2 &   0.8 &  14.8 &   2.2 & 11.8 & 3.5& 224 &  15 &     4.0 &     0.6 &    11.7 &     1.8 &   6.6 &   1.0 &   19 &   2 & 12.1 & 3.6 \\
                 &                R1 & 363 &  12 &    10.5 &     1.6 &    14.6 &     2.2 &   6.6 &   1.0 &   9.2 &   1.4 &  3.7 & 1.1& 221 &  12 &     4.2 &     0.6 &     6.3 &     1.0 &  10.0 &   1.5 &  14.8 &   2.2 &  4.5 & 1.4 \\
                 &                R2 & 268 &   5 &     9.3 &     1.4 &    14.4 &     2.2 &  10.7 &   1.6 &   16 &   2 &  5.0 & 1.5& 174 &   7 &     5.5 &     0.8 &     8.8 &     1.3 &  15.7 &   2.4 &   24 &   3 &  5.3 & 1.6 \\
                 &                R3 & 311 &   9 &     4.7 &     0.7 &     7.8 &     1.2 &   4.1 &   0.6 &   6.7 &   1.0 &  5.6 & 1.7& 178 &  10 &    1.65 &    0.25 &     2.5 &     0.4 &   6.1 &   0.9 &   9.4 &   1.4 &  4.8 & 1.4 \\
                 &                R4 & 257 &  11 &     2.9 &     0.4 &     4.2 &     0.6 &   3.7 &   0.6 &   5.3 &   0.8 &  4.0 & 1.2& 136 &   8 &    1.09 &    0.16 &    1.57 &    0.24 &   5.4 &   0.8 &   7.8 &   1.2 &  4.2 & 1.3 \\
                 &                R5 & 158 &  26 &     3.0 &     0.4 &     5.1 &     0.8 &  10.0 &   1.5 &   16 &   2 &  6.0 & 1.8&  94 &   5 &    1.32 &    0.20 &     2.4 &     0.4 &  12.6 &   1.9 &   23 &   3 &  6.9 & 2.1 \\
                 &                R6 & 257 &  11 &    1.48 &    0.22 &     2.0 &     0.3 &   1.9 &   0.3 &   2.6 &   0.4 &  3.6 & 1.1& 145 &   9 &    0.57 &    0.09 &    0.77 &    0.12 &   2.2 &   0.3 &   3.0 &   0.5 &  3.5 & 1.1 \\
                 &                R7 & 176 &   8 &    0.96 &    0.15 &    1.35 &    0.21 &   2.6 &   0.4 &   3.7 &   0.6 &  3.9 & 1.2& 131 &   9 &    0.62 &    0.09 &    0.89 &    0.14 &   2.9 &   0.4 &   4.3 &   0.7 &  4.2 & 1.3 \\
                 &                R8 & 197 &  34 &    0.94 &    0.14 &    1.40 &    0.21 &   2.0 &   0.3 &   2.9 &   0.4 &  4.5 & 1.4& 138 &  11 &    0.52 &    0.08 &    0.83 &    0.13 &   2.3 &   0.3 &   3.6 &   0.6 &  5.2 & 1.6 \\
\hline
\noalign{\smallskip}
  IRAS21130-4446 &           Nucleus & 680 & 143 &   18 &   2 &   49 &   7 &   3.5 &   0.5 &   9.6 &   1.5 &  7.5 & 2.3& 490 &  29 &    11.5 &     1.8 &   34 &   5 &   4.0 &   0.6 &  12.2 &   1.9 &  8.3 & 2.6 \\
                 &                R1 &1454 &  11 &   58 &   8 &   94 &  14 &   2.3 &   0.4 &   3.7 &   0.6 &  3.5 & 1.1&1441 &  57 &   38 &   5 &   62 &   9 &   3.6 &   0.5 &   5.8 &   0.9 &  3.6 & 1.1 \\
\hline
\noalign{\smallskip}
         NGC7130 &  Nucleus$^{\dag}$ & 205 &   6 &     8.2 &     1.2 &   30 &   4 &  16.1 &   2.4 &   59 &   9 & 14.8 & 4.5& 114 &   6 &     3.6 &     0.6 &    13.6 &     2.1 &   23 &   3 &   87 &  13 & 14.9 & 4.5 \\
                 &                R1 & 239 &  10 &     2.1 &     0.3 &     2.8 &     0.4 &   3.0 &   0.5 &   3.9 &   0.6 &  2.9 & 0.9& 121 &   7 &    0.92 &    0.14 &    1.22 &    0.18 &   5.1 &   0.8 &   6.7 &   1.0 &  3.1 & 1.0 \\
                 &                R2 & 246 &   5 &    0.87 &    0.13 &     1.7 &     0.3 &   1.2 &   0.2 &   2.2 &   0.3 &  7.3 & 2.2& 161 &  11 &    0.48 &    0.07 &    0.99 &    0.15 &   1.5 &   0.2 &   3.2 &   0.5 &  8.3 & 2.5 \\
                 &                R3 & 239 &  10 &    0.66 &    0.10 &    0.76 &    0.12 &   0.9 &   0.1 &   1.0 &   0.2 &  1.6 & 0.6& 165 &  11 &    0.43 &    0.07 &    0.47 &    0.08 &   1.3 &   0.2 &   1.5 &   0.2 &  1.1 & 0.7 \\
\hline
\noalign{\smallskip}
          IC5179 &           Nucleus & 157 &  11 &     3.5 &     0.5 &     9.5 &     1.4 &  11.7 &   1.8 &   31 &   4 & 11.1 & 3.4&  81 &   4 &    1.46 &    0.22 &     3.7 &     0.6 &   18 &   2 &   46 &   7 & 10.5 & 3.2 \\
                 &                R1 & 204 &   3 &    0.96 &    0.15 &    1.19 &    0.19 &   1.9 &   0.3 &   2.3 &   0.4 &  2.5 & 0.9& 146 &   7 &    0.64 &    0.10 &    0.84 &    0.13 &   2.4 &   0.4 &   3.1 &   0.5 &  3.0 & 1.0 \\
                 &                R2 & 275 &   2 &    0.89 &    0.14 &    0.98 &    0.16 &   1.0 &   0.2 &   1.1 &   0.2 &  1.0 & 0.4& 120 &   8 &    0.27 &    0.04 &    0.31 &    0.05 &   1.7 &   0.3 &   2.0 &   0.3 &  1.9 & 0.8 \\
                 &                R3 & 157 &  12 &    0.64 &    0.10 &    0.93 &    0.15 &   2.1 &   0.3 &   3.0 &   0.5 &  4.2 & 1.3&  66 &   3 &    0.21 &    0.03 &    0.34 &    0.05 &   4.1 &   0.6 &   6.5 &   1.0 &  5.3 & 1.8 \\
                 &                R4 & 222 &   8 &    0.54 &    0.08 &    0.80 &    0.13 &   0.9 &   0.1 &   1.3 &   0.2 &  4.4 & 1.4& 128 &   9 &    0.20 &    0.03 &    0.33 &    0.05 &   1.2 &   0.2 &   1.9 &   0.3 &  5.5 & 1.8 \\
                 &                R5 &  92 &  19 &    0.47 &    0.07 &    0.94 &    0.14 &   4.4 &   0.7 &   8.7 &   1.3 &  7.7 & 2.4&  85 &   5 &    0.42 &    0.06 &    0.80 &    0.12 &   4.7 &   0.7 &   9.1 &   1.4 &  7.4 & 2.3 \\
                 &                R6 &  87 &   7 &    0.38 &    0.06 &    0.69 &    0.10 &   4.0 &   0.6 &   7.2 &   1.1 &  6.6 & 2.0&  71 &   4 &    0.27 &    0.04 &    0.50 &    0.08 &   4.3 &   0.6 &   8.0 &   1.2 &  7.1 & 2.2 \\
                 &                R7 & 130 &  23 &    0.26 &    0.04 &    0.84 &    0.13 &   1.3 &   0.2 &   4.3 &   0.7 & 13.2 & 4.0&  87 &   6 &    0.14 &    0.02 &    0.56 &    0.09 &   1.7 &   0.3 &   6.5 &   1.0 & 15.3 & 4.7 \\
                 &                R8 & 148 &  12 &    0.27 &    0.04 &    0.40 &    0.06 &   1.0 &   0.2 &   1.5 &   0.2 &  4.6 & 1.5&  91 &   7 &    0.13 &    0.02 &    0.20 &    0.03 &   1.3 &   0.2 &   2.0 &   0.3 &  5.1 & 1.7 \\
                 &                R9 &  87 &   7 &    0.24 &    0.04 &    0.29 &    0.04 &   2.5 &   0.4 &   3.0 &   0.5 &  1.9 & 0.7&  76 &   6 &    0.17 &    0.03 &    0.20 &    0.03 &   2.8 &   0.4 &   3.2 &   0.5 &  1.7 & 0.7 \\
                 &               R10 &  87 &   7 &    0.23 &    0.04 &    0.29 &    0.04 &   2.6 &   0.4 &   3.2 &   0.5 &  2.4 & 0.9&  60 &   4 &    0.11 &    0.02 &    0.13 &    0.02 &   2.8 &   0.4 &   3.4 &   0.5 &  2.1 & 0.9 \\
                 &               R11 &  87 &   7 &    0.17 &    0.03 &    0.22 &    0.03 &   2.0 &   0.3 &   2.6 &   0.4 &  2.6 & 1.0&  70 &   6 &    0.13 &    0.02 &    0.16 &    0.03 &   2.2 &   0.3 &   2.8 &   0.4 &  2.7 & 1.0 \\
\hline
\noalign{\smallskip}
  IRAS22491-1808 &           Nucleus & 758 &  32 &     9.6 &     1.4 &   22 &   3 &   1.4 &   0.2 &   3.3 &   0.5 &  6.5 & 2.0& 439 &  26 &     4.5 &     0.7 &    11.0 &     1.7 &   2.0 &   0.3 &   5.0 &   0.8 &  6.7 & 2.1 \\
                 &                R1 & 780 &  32 &    16.6 &     2.5 &   38 &   5 &   2.2 &   0.3 &   5.1 &   0.8 &  6.2 & 1.9& 501 &  29 &     8.8 &     1.3 &   21 &   3 &   3.0 &   0.4 &   7.2 &   1.1 &  6.7 & 2.1 \\
                 &                R2 & 919 &  27 &     9.3 &     1.4 &   31 &   4 &   0.9 &   0.1 &   3.1 &   0.5 &  9.2 & 2.8& 500 &  27 &     4.1 &     0.6 &    13.8 &     2.1 &   1.4 &   0.2 &   4.6 &   0.7 &  9.0 & 2.7 \\
\hline
\noalign{\smallskip}
  IRAS23128-5919 &  Nucleus$^{\dag}$ & 490 &  86 &   60 &   9 &  186 &  28 &   20 &   3 &   62 &   9 &  8.4 & 2.5& 257 &  14 &   24 &   3 &   83 &  12 &   30 &   4 &  105 &  16 &  9.2 & 2.8 \\
                 &                R1 & 693 &  12 &   35 &   5 &   88 &  13 &   6.2 &   0.9 &  15.7 &   2.4 &  6.9 & 2.1& 385 &  20 &   16 &   2 &   43 &   6 &   9.3 &   1.4 &   23 &   3 &  7.1 & 2.1 \\
\hline
\hline
\end{tabular}}}
\tablefoot{\tiny Col. (2): Region label (see Fig.~\ref{sfr_maps}). Col (3):  Effective radius in [pc]. Cols (4) and (5): Observed (4) and extinction-corrected (5) \Pa\ luminosities measured within \reff, expressed in [$\rm \times10^{6}$\,\Lsolar]. The \Pa\ luminosities for the LIRGs are obtained from the \Brg\ fluxes using the case B recombination factor at T$=10,000$\,K and n$\rm_{\rm e}=10^4\,cm^{-3}$ \citep{Osterbrock:2006AGN2}. Cols. (6) and (7): Observed (6) and extinction-corrected (7) \Si\ in [$\rm M_{\sun}\,yr^{-1}\,kpc^{-2}$]. Col. (8): \Av\ in magnitudes. Col. (9): Core radius in [pc] (see text for definition). Cols. (10) to (14): Same as Cols. (4) to (8) but measured within \rcore. All the uncertainties are calculated by a \emph{bootstrap} method of $\rm N=300$ simulations, added in quadrature to the 15\% systematic error from the absolute flux calibration.
$^{\dag}$ Regions that host AGN.}
}
\end{table}
\end{landscape}

\begin{acknowledgements}
We thank the anonymous referee for his/her useful comments and suggestions that significantly improved the final content of this paper.
This work was supported by the Ministerio de Econom\'ia y Competitividad of Spain under grants BES-2008-007516, ESP2007-65475-C02-01, AYA2010-21161-C02-01, AYA2012-32295-C02-01, and AYA2012-39408-C02-01.
Based on observations collected at the European Organisation for Astronomical Research in the Southern Hemisphere, Chile, programmes 077.B-0151A, 078.B-0066A, and 081.B-0042A.
This research made use of the NASA/IPAC Extragalactic Database (NED), which is operated by the Jet Propulsion Laboratory, California Institute of Technology, under contract with the National Aeronautics and Space Administration. 
\end{acknowledgements}

\bibliographystyle{aa}
\bibliography{$HOME/Dropbox/aa}
\end{document}